\documentclass[a4paper,11pt]{article}
\pdfoutput=1 

\usepackage{jheppub} 

\usepackage[T1]{fontenc} 

\usepackage{times}
\usepackage{graphicx}
\usepackage{amssymb}
\usepackage{xspace}

\def\be{\begin{eqnarray}}
\def\ee{\end{eqnarray}}

\newcommand{\mt}[1]{\textrm{\tiny #1}}

\def\pt{{p_\mt{T}}}

\def\Raa{R_{\mt{AA}}}

\def\aC{{{\kappa}_{\rm coll}}}

\def\aR{{\kappa_{\rm rad}}}

\def\aSC{{\kappa_{\rm sc}}}

\def\w{{\bf w}}
\def\v{{\bf v}}

\def\x{{\bf x}}

\newcommand{\pythia}{{\sc Pythia}\xspace}

\title{\boldmath Predictions for Boson-Jet Observables and Fragmentation Function Ratios 
from a Hybrid Strong/Weak Coupling Model for Jet Quenching}

\author[a]{Jorge Casalderrey-Solana,}
\author[b]{Doga Can Gulhan,}
\author[c,d]{Jos\'e Guilherme Milhano,}
\author[a]{Daniel Pablos,}
\author[b,e]{Krishna Rajagopal}

\affiliation[a]{
Departament d'Estructura i Constituents
de la Mat\`eria and Institut de Ci\`encies del Cosmos (ICCUB),
Universitat de Barcelona, Mart\'\i \ i Franqu\`es 1, 08028 Barcelona, Spain
}

\affiliation[b]{Laboratory for Nuclear Science and Department of Physics, Massachusetts Institute of Technology (MIT), Cambridge, MA 02139 USA}

\affiliation[c]{CENTRA, Instituto Superior T\'ecnico, Universidade de Lisboa, Av.~Rovisco Pais, P-1049-001 Lisboa, Portugal}
\affiliation[d]{Physics Department, Theory Unit, CERN, CH-1211 Gen\`eve 23, Switzerland}

\affiliation[e]{
Center for Theoretical Physics, MIT, Cambridge, MA 02139, USA
}
\emailAdd{jorge.casalderrey@ub.edu}

\emailAdd{dgulhan@mit.edu}

\emailAdd{guilherme.milhano@tecnico.ulisboa.pt}

\emailAdd{dpablos@ecm.ub.es}

\emailAdd{krishna@mit.edu}

\preprint{{\footnotesize ICCUB-15-017, CERN-PH-TH-2015-184, MIT-CTP-4701}}

\abstract{
We have previously introduced a hybrid strong/weak coupling model for jet quenching in
heavy ion collisions in which we describe the production and fragmentation of jets at weak coupling, using \pythia,
and describe the rate at which each parton in the jet shower loses energy as it propagates
through the strongly coupled plasma, $dE/dx$, using an expression computed holographically at strong coupling. 
The model has a single free parameter that we fit to a single experimental measurement.
We then confront our model with experimental data on many other jet observables, focusing
in this paper on boson-jet observables, finding that it provides a good description of present jet data.
Next, we provide the predictions of our hybrid model for many measurements to come, including
those for inclusive jet, dijet, photon-jet  and Z-jet observables in heavy ion collisions with
energy $\sqrt{s}=5.02$~ATeV coming soon at the LHC.  As the statistical uncertainties on
near-future measurements of photon-jet observables are expected to be much smaller than
those in present data, with about an order of magnitude more photon-jet events expected, 
predictions for these observables are particularly important.
We find that most of our pre- and post-dictions do not depend 
sensitively on the form we choose for the
rate of energy loss  $dE/dx$ of the partons in the shower.  This gives our predictions considerable
robustness.  To better discriminate between possible forms for the rate of energy loss, though,
we must turn to intrajet observables.  Here, we focus on ratios of fragmentation functions.
We close with a suggestion for a particular ratio, between the fragmentation functions of
inclusive and associated jets with the same kinematics  in the same collisions, 
which is particularly
sensitive
to the $x$- and $E$-dependence of $dE/dx$, and hence may be used to learn 
which mechanism of parton energy loss best describes the quenching of jets.
}

\begin{document} 
\maketitle
\flushbottom

\section{Introduction}
\label{sec:intro}

The LHC has ushered in a new era in the exploration of the properties of matter under extreme conditions. 
By colliding Pb ions at center of mass energies  in the multi-TeV regime,
the LHC has provided us with droplets of the hottest matter ever produced in the laboratory
and a diverse suite of copious high energy probes with which to explore the microscopic properties
of the strongly coupled, liquid, quark-gluon plasma (QGP) discovered at RHIC~\cite{Adcox:2004mh,Arsene:2004fa,Back:2004je,Adams:2005dq}.
For jet probes in particular, the results of the successful LHC Run 1 have shown that many of the properties of these fundamental QCD objects 
are substantially modified when the jets are produced in Pb-Pb collisions as compared to when they are produced in p-p 
collisions~\cite{Aad:2010bu,Chatrchyan:2011sx,Chatrchyan:2012nia,Chatrchyan:2012gt,Chatrchyan:2012gw,Aad:2012vca,Aad:2013sla,Chatrchyan:2013kwa,Abelev:2013kqa,Chatrchyan:2013exa,Chatrchyan:2014ava,Aad:2014wha,Aad:2014bxa,Adam:2015ewa,Adam:2015doa,Aad:2015bsa}. 
 These modifications are results of the final state interaction between the jets and the droplets of hot QCD matter formed in heavy ion collisions. 
 As the principal underlying effect is the energy loss suffered by each of the components of the jet showers on their way out of the hot matter, the various modifications of jet properties observed in heavy ion collisions are referred to, in sum, as jet quenching.
 The phenomenon of jet quenching was first discovered without reconstructing individual jets via the strong reduction
 in the number of intermediate-$\pt$ hadrons in heavy ion collisions at RHIC~\cite{Adcox:2001jp,Adler:2002xw}.
  Precisely because varied modifications of jet properties can now be well measured, this suite of probes has the potential to provide us with unique and important information about the properties of QGP and about the interaction between energetic partons and this strongly coupled liquid. 
 The imminent start of the LHC heavy ion Run 2, which will increase both the center of mass energy and luminosity for heavy ion collisions and hence
 will substantially increase the production rates of jets and all other hard probes including high energy photons and Z-bosons,
 makes many much more quantitative analyses of the striking phenomena observed in the first LHC run imminent.
 Turning this opportunity into precise extractions of QGP properties from the experimental data to come 
requires a diverse suite of theoretical tools.  In this paper, we shall present substantial advances in
the development of one such tool, introduced in Ref.~\cite{Casalderrey-Solana:2014bpa}.

It is the discovery that the QGP of QCD is a strongly coupled liquid, with intense collective phenomena, no apparent
quasiparticle structure, and a very rich phenomenology that makes it of such interest.
But, at the same time, it makes the theoretical description of its properties and dynamics 
much more challenging, since many of the perturbative tools available to describe weakly coupled hard QCD processes
become inapplicable in the strongly coupled liquid plasma produced in the range of temperatures accessible to colliders.
Fortunately, in recent years gauge/gravity duality has emerged as a tool with which to analyze the dynamics of the strongly coupled
plasma in various non-Abelian gauge theories.
Although it is not yet known whether QCD itself has a dual gravitational description, calculations done using these
new holographic methods (where one maps a question in the gauge theory plasma onto a calculation done in
its dual gravitational description) have yielded many qualitative insights into the properties and dynamics of the QGP in QCD,
as produced and probed in heavy ion collisions.
(For a review see Ref.~\cite{CasalderreySolana:2011us}.)

Although we now have a toolkit that includes perturbative methods valid for weakly coupled hard processes
and holographic calculations that are valuable guides to strongly coupled dynamics,
the description of the interactions between jets and the plasma remains challenging because
physics at both hard and soft scales enters in central ways.
Jets in the 100 GeV range are produced in hard processes with large momentum transfers which must be
described by perturbative QCD.  Furthermore, the energetic, virtual, partons produced in a hard scattering
immediately experience a rapid sequence of branching processes which reduce their large initial virtuality,
ultimately down to the hadronization scale.
The evolution of this branching process is governed by the DGLAP equations and is described well 
by perturbation theory, since the initial virtualities of the partons in the shower that results from the branching
are large.
Nevertheless, when the jet fragments interact with the plasma created in Pb-Pb collisions, soft exchanges of energy and momentum 
with the plasma constituents are governed by physics at scales of order the temperature of the medium, where a strongly coupled description
is mandated.  
And, it is these soft interactions
that are critical to understanding the modification of jets produced in Pb-Pb collisions caused by their interaction with the plasma.
This multi-scale nature of the physics of jet probes and their modification 
means that neither approaches that seek to treat the both the jets and their interaction with the plasma perturbatively~\cite{Baier:1996kr,Baier:1998kq,Gyulassy:2000er,Wiedemann:2000za,Wang:2001ifa,Arnold:2002ja,Jeon:2003gi,Jacobs:2004qv,Turbide:2005fk,Wicks:2005gt,Qin:2007rn,CasalderreySolana:2007zz,Majumder:2010qh,D'Eramo:2010ak,MehtarTani:2010ma,MehtarTani:2011tz,Ovanesyan:2011xy,CasalderreySolana:2011rz,Dai:2012am,D'Eramo:2012jh,Blaizot:2013hx,Mehtar-Tani:2013pia,Wang:2013cia,Armesto:2013fca,Blaizot:2013vha,Burke:2013yra,Mehtar-Tani:2014yea,Kang:2014xsa,Chien:2014nsa,Kurkela:2014tla,Blaizot:2014ula,Apolinario:2014csa,Blaizot:2014rla,Blaizot:2015jea,Ghiglieri:2015zma}
nor those that seek to address the entire dynamics of the quenching of 
energetic probes at strong coupling via gauge/gravity duality~\cite{Herzog:2006gh,CasalderreySolana:2006rq,Gubser:2006bz,Liu:2006nn,Gubser:2008as,Hatta:2008tx,Chesler:2008wd,Chesler:2008uy,Arnold:2010ir,Arnold:2011qi,Arnold:2011vv,Chernicoff:2011xv,Chesler:2011nc,Arnold:2012uc,Arnold:2012qg,Ficnar:2013wba,Ficnar:2013qxa,Chesler:2014jva,Rougemont:2015wca} can capture the physics at all the relevant scales correctly.

To face this challenge, in Ref.~\cite{Casalderrey-Solana:2014bpa} we introduced a phenomenological model specifically
designed to treat the different dynamics arising on different scales differently, using a perturbative description where it is appropriate
and deploying insights from holographic calculations where soft momentum transfers and strongly coupled physics enter.
While there have been other attempts to combine results obtained from weak and strong 
coupling~\cite{Horowitz:2008ig,Marquet:2009eq,Horowitz:2011gd,Betz:2011jp,Betz:2014cza,Iancu:2014ava,Betz:2015oia},
our approach
is distinct since it focusses on using two different calculational frameworks at the different energy scales where each is
appropriate.
In a nutshell, our model combines a perturbative description of the creation and evolution of jets as implemented in modern Monte Carlo codes, in particular in
\pythia~\cite{Sjostrand:2007gs}, with a prescription for the mechanism by which, and consequently the rate $dE/dx$ at which,
each parton in the \pythia shower loses energy that we infer from the holographic 
computation~\cite{Chesler:2014jva} of how energetic light quarks lose energy in the strongly coupled plasma of a gauge theory with
a dual gravitational description. 
This hybrid model contains a single parameter which controls the stopping distance of fast partons in the QCD plasma. We fitted this parameter
 to data in Ref.~\cite{Casalderrey-Solana:2014bpa}, and
 then confronted this hybrid strong/weak coupling model 
with data on  several single-jet and dijet observables, obtaining very satisfactory agreement. 
Nonetheless, our analysis also showed that this same set of data could be described almost as well
by different assumptions for the energy loss rate $dE/dx$ 
of particles in plasma. This indicates that the observables considered in Ref.~\cite{Casalderrey-Solana:2014bpa}
are not very sensitive to the microscopic dynamics governing the interaction between an energetic probe
and the strongly coupled plasma.

In this paper, we continue our exploration of the predictions of the hybrid model that we introduced in Ref.~\cite{Casalderrey-Solana:2014bpa},
and whose construction and main aspects we review briefly in
Section~\ref{sec:hybridmodel}.
In Section~\ref{Sec:FlowEffects} and Appendix~\ref{BI-Eloss}, we improve
some aspects of the implementation of the model:
we incorporate the effects of the motion of the hydrodynamic fluid on the rate of energy loss,
and we  embed the jets in our model in the 
viscous hydrodynamic simulations of Refs.~\cite{Shen:2014vra,Shen:2014nfa}, rather than the inviscid hydrodynamic simulations from Ref.~\cite{hiranoLHC} that
we used in  Ref.~\cite{Casalderrey-Solana:2014bpa}. 
We discuss the implications of these improvements 
for the extraction of the value of the single  parameter in our model in Sections~\ref{sec:effects_flow} and  \ref{sec:species_dep}. 

After laying this groundwork, in Section~\ref{sec:BJCorr} we use our model to analyze photon-jet and Z-jet
correlations (generically, boson-jet correlations)  in heavy ion collisions.
These observables provide a different and complementary set of data with which to confront our model; doing
so is the first of our principal goals.
After specifying the details of how we generate and quench our samples of boson-jet events in Section~\ref{sec:MC},  
in Section~\ref{sec:photonjet_results} we compare the predictions of our model to presently 
available photon-jet data from LHC heavy ion collisions. We find that our model
is just as successful here as it was at describing inclusive jet observables and dijet observables 
in Ref.~\cite{Casalderrey-Solana:2014bpa} and Appendix~\ref{update}.\footnote{In Appendix~\ref{update}, we update our model calculations of the observables from Ref.~\cite{Casalderrey-Solana:2014bpa}
to include the improved model implementation introduced in Section~\ref{Sec:FlowEffects}.}
These many successes give us confidence in our approach.

In Section~\ref{sec:BJCorr} and Appendix~\ref{Model_dep}, we find that, within current experimental and theoretical uncertainties, 
two different control hypotheses for the form of the rate of energy loss $dE/dx$ yield almost as good
descriptions of the single-jet, dijet, and photon-jet observables that we have investigated as we obtain from
our hybrid strong/weak coupling model.
Similarly, the gamma-jet data we consider in this work can also be described well by the perturbative treatment of 
 Ref.~\cite{Wang:2013cia}.

%
Therefore, these observables are not particularly effective at discriminating between rather different assumptions
about the interactions between the components of the jets and the strongly coupled medium. (This
observation was first made for the specific case of the dijet asymmetry observable in Ref.~\cite{Renk:2012cx}.)
 However, we have observed various small differences  among the 
 predictions of our hybrid strong/weak coupling model and those of our two control models, which
 opens the possibility that the higher energy and higher statistics data from the coming LHC heavy ion Run 2
 will yield more precise measurements that can be used to constrain the microscopic dynamics of how
 jets interact with the plasma and, ultimately, the microscopic properties of the plasma itself.
With these goals in mind, and to facilitate direct comparisons to data that is coming soon,  
in Section~\ref{sec:photonjet_results}
we provide predictions for  photon-jet observables in 
Pb-Pb collisions at $\sqrt{s}=5.02$~ATeV and in Section~\ref{sec:zjet_predictions} we do the same for Z-jet observables. 
We include the predictions of our model for a number of single-jet and dijet observables
at $\sqrt{s}=5.02$~ATeV in Appendix~\ref{DijetPredictions}.  We hope that near-future high-statistics measurements will
make the myriad predictions of our hybrid model for heavy ion collisions at $\sqrt{s}=5.02$~ATeV 
among the most incisive results of this paper.

With the goal of finding observables that are more discriminating, in the sense that they are more
sensitive to different assumptions about the rate of energy loss $dE/dx$ and hence are better
able to give us information about the microscopic dynamics behind jet quenching, in Section~\ref{Sec:FF}
we turn to exploring more differential observables.  Before adding further physics (and further parameters)
to our hybrid model, we cannot use it to describe the modification of the angular shapes of jets.  We can,
however, analyze fragmentation functions.
In Section \ref{sec:FFBJ} we study the fragmentation functions of the jets produced in association
with either a photon or a Z-boson, providing predictions for the ratios of these fragmentation functions
in Pb-Pb collisions with $\sqrt{s}=5.02$~ATeV to the same fragmentation functions in p-p collisions.  These predictions are qualitative at best, however,
because they are sensitive to differences between the hadronization of jets in Pb-Pb and p-p collisions that
are not under theoretical control.  And, these observables turn out not to be particularly discriminating.
In Section~\ref{sec:ff_dijet} we use the tools that we have constructed to identify a new observable
that {\it is} discriminating. This observable is the ratio between the fragmentation functions of
inclusive jets in Pb-Pb collisions and the fragmentation functions of the associated (lower energy)
jets from a dijet pair. Much of the uncertainty
associated with hadronization should cancel in this ratio, since we nowhere need p-p reference data
and since we compare the fragmentation functions of associated jets and inclusive 
jets with the same range of energies,
in the same Pb-Pb collisions.  Furthermore, by explicit calculation we find a clear and substantial separation between the predictions
of our hybrid strong/weak coupling model for this ratio and the predictions that we obtain
from our two control models.  And, we find a good qualitative understanding of why
this separation occurs, which is to say of why this ratio is such a discriminating observable.
We look forward to measurements of this observable in the coming LHC heavy ion Run 2
as these data should tell us much about the dynamics of how jets interact with the strongly coupled plasma.

In Section \ref{conclusions} we discuss the main conclusions of this study and look ahead.

\section{Description of the Model and its Implementation}
\label{sec:Section2}

\subsection{The Hybrid Model Approach}
\label{sec:hybridmodel}

In this Section, we briefly review the main aspects of the hybrid model that
we introduced in our previous publication~\cite{Casalderrey-Solana:2014bpa}. 
Our model is a response to the challenge of
addressing the wide range of energy scales involved in the production of energetic
quarks, gluons, photons and Z-bosons in hard processes followed by the interaction
of the quarks and gluons with the strongly coupled matter formed in ultrarelativistic
heavy ion collisions.
Our hybrid model treats the weakly coupled short distance dynamics involved in the creation and
hard evolution of jets perturbatively while at the same time using insights from holographic
calculations that assume strong coupling in the treatment of the soft long distance interactions
of the jet with the strongly coupled fluid created in the collision.
The former processes occur at momentum scales set by the large virtuality of the elementary hard partonic interaction, which we shall
denote $Q$.  The latter processes involve momenta at or close to the typical scales that characterize the strongly coupled fluid, 
which we take to be of order the temperature $T$. 
For the energetic processes involved in jet production in Pb-Pb collisions at the LHC, $Q\gg T$.
We use this separation of scales to justify our explicit separation of the treatment of the relevant dynamics for each regime.
We are, however, certainly not doing a systematic expansion order by order in powers of the ratio of these scales. Instead,
we construct a model in which the qualitatively different 
physics at the widely separated scales is each treated in the most appropriate language, in so doing introducing
a single model parameter that at the present stage we must fit to data.

In our model, we assume that the soft in-medium processes
cannot alter either the short distance production of hard partons  in elementary partonic collisions or the hard branching processes by which the large initial virtuality relaxes. 
In p-p collisions in vacuum, those two processes are both described well 
by perturbative QCD; in particular, the radiative branching processes are incorporated into evolution equations which are in turn
the basis of the different high energy Monte Carlo event generators like \pythia~\cite{Sjostrand:2007gs}, which we shall employ.
The probabilistic implementation of this evolution describes 
the formation of a shower of partons initiated from the initial hard process, with splitting probabilities dictated by perturbative 
QCD. 
In the environment created in a heavy ion collision,
as this parton shower develops its constituents are continuously
exchanging momentum with the matter produced in the collision.
In our model, we assume that
the splitting probabilities are not modified by these soft exchanges. 
This does not mean, though, that the shower remains unchanged. 
As a consequence of their soft exchanges of momenta with the medium,
the particles in the shower lose energy and pick up momentum transverse
to their original direction as they propagate through the strongly coupled liquid.
In this publication, as in Ref.~\cite{Casalderrey-Solana:2014bpa}, we focus entirely
on the loss of energy of the partons in the shower, as we can do so in a model
with only a single free parameter.  In a future publication we shall introduce the transverse
momentum kicks, at the expense of adding at least one more parameter. Here, as in
Ref.~\cite{Casalderrey-Solana:2014bpa}, we shall only describe experimental observables
that are sensitive to the loss of energy by the partons in the shower and insensitive to
the transverse momentum kicks that they also experience.
Since the momentum transfers between the shower partons and the medium
are not large, the physics of energy loss must be described at strong coupling.
After assigning a life-time to each parton in the shower 
according to a formation time argument~\cite{CasalderreySolana:2011gx,Casalderrey-Solana:2014bpa},
we know the points in spacetime where the parton on each branch of the shower
is formed, and splits.  We model the energy loss of each parton in the shower
as a continuous process,
supplementing the in-medium evolution with an explicit energy loss rate $dE/dx$
that models the strongly coupled dynamics of parton energy loss.
We do not track what becomes of the energy lost by each parton in the shower,
implicitly assuming that the lost energy is incorporated into the strongly coupled fluid,
ultimately becoming soft hadrons with momenta of order $T$ that we do not model.
The form that we assume for the rate of energy loss $dE/dx$ therefore fully encodes all the strongly coupled
 in-medium dynamics incorporated in our model.   

In our model, we explore the consequences of
an energy loss rate $dE/dx$ whose form is that appropriate for the rate of energy loss of an
energetic massless quark (excitation in the fundamental color representation) traversing 
a slab of plasma with temperature $T$ and thickness $x$ in the strongly coupled plasma of
$\mathcal {N}=4$ supersymmetric Yang-Mills (SYM) theory~\cite{Chesler:2014jva}, 
\be
\label{CR_rate}
\left. \frac{d E}{dx}\right|_{\rm strongly~coupled}= - \frac{4}{\pi} E_{\rm in} \frac{x^2}{x_{\rm stop}^2} \frac{1}{\sqrt{x_{\rm stop}^2-x^2}}\,, 
\quad \quad x_{\rm stop}= \frac{1}{2\aSC}\frac{E_{\rm in}^{1/3}}{T^{4/3}}\ ,
\ee
obtained via the gauge/gravity duality. Here, $E_{\rm in}$ is the initial energy that the massless quark has
before it enters the plasma, $E(x)$ is the energy that it has after traversing the slab of thickness $x$,
and $x_{\rm stop}$ 
is the stopping distance of the high energy excitation --- the smallest slab thickness that results in the energetic
excitation losing all of its energy within the slab of plasma.
In ${\cal N}=4$ SYM theory,
the dimensionless
constant $\aSC$ appearing in the expression for $x_{\rm stop}$ is determined explicitly in terms of the
\'{}t Hooft coupling $\lambda$ and is 
$\aSC=1.05 \lambda^{1/6}$~\cite{Chesler:2008uy,Chesler:2011nc,Ficnar:2013wba}. 
The premise of our hybrid model is that the form of $dE/dx$ in the strongly coupled quark-gluon plasma
of QCD is the same as in (\ref{CR_rate}); we shall see that this hypothesis is uncontradicted by many
and varied sets of data.  However, there is no reason at all to expect that the relationship between
$\aSC$ and $\lambda$ should be the same in QCD and ${\cal N}=4$ SYM theory, as the strongly coupled plasmas
of the two theories have different, and differently many, microscopic degrees of freedom.
Furthermore, there are ambiguities in the definition of jets in $\mathcal{N}=4$ SYM theory: since hard processes in this
theory do not produce jets~\cite{Hatta:2008tx,Hofman:2008ar}, different theoretical calculations have been developed in which highly energetic colored
excitations are formed in different ways -- no one of which is preferred over others as a model for jets in QCD since
none is  model for jet production in QCD. 
And, the proportionality
constant between $E_{\rm in}^{1/3}/T^{4/3}$ and $x_{\rm stop}$ can depend on details of the particular way
in which a highly energetic colored excitation is formed.
For both these reasons, and as discussed in more detail in  Ref.~\cite{Casalderrey-Solana:2014bpa}, in our model we will assume that 
any differences between $dE/dx$ in the strongly coupled plasmas of QCD and $\mathcal{N}=4$ SYM theory 
can be absorbed in the value of $\aSC$, which we therefore take as a free parameter whose value
must be fixed by fitting to data.   We will refer to the form  (\ref{CR_rate}) for $dE/dx$
as strongly coupled energy loss. Our hybrid model constitutes applying this prescription for energy loss
branch-by-branch to the partons in a shower that described in vacuum by \pythia.  We shall specify
the implementation of our hybrid model fully in subsequent subsections.

In order to have some other benchmarks against which to compare 
the success of our hybrid model, as in Ref.~\cite{Casalderrey-Solana:2014bpa} we will also explore 
two other quite
different forms for the energy loss rate $dE/dx$, one inspired by perturbative calculations of radiative
energy loss and the other by perturbative calculations of collisional energy loss.
These expressions are given by 
\be
\label{Eloss_equations}
\left. \frac{d E}{dx}\right|_{\rm radiative} = -\aR T^3 x\,, \quad \quad \left. \frac{d E}{dx}\right|_{\rm collisional} = - \aC T^2
\ee
where we shall again treat $\aR$ and $\aC$ as  parameters to be fixed by fitting to data. 
This
oversimplified treatment of radiative and collisional energy loss is not meant to supersede other much more sophisticated analysis of these 
mechanisms~\cite{Renk:2008pp,Schenke:2009gb,Armesto:2009fj,Lokhtin:2011qq,Zapp:2013vla,Burke:2013yra,Zapp:2013zya}. Our only goal is to use them as benchmark expressions for $dE/dx$ with very different
dependence (or lack thereof) on $x$ and $E_{\rm in}$ to that found in (\ref{CR_rate}).

Note that we shall take (\ref{CR_rate}) and (\ref{Eloss_equations}) as rates of energy loss
for quarks in the parton shower; gluons lose more energy. In the case of (\ref{Eloss_equations}), gluons
lose more energy by a factor of $C_A/C_F=9/4$ and we shall take $\aR$ and $\aC$ to be larger by
this factor for gluons in the parton shower.  In the case of (\ref{CR_rate}), $x_{\rm stop}$ is
shorter for gluons than quarks by a factor of $(C_A/C_F)^{1/3}$~\cite{Gubser:2008as}, and as in Ref.~\cite{Casalderrey-Solana:2014bpa}
we shall therefore
take $\aSC$ to be larger for gluons in the parton shower than for quarks by a factor of $(9/4)^{1/3}$.

The energy loss rates Eqs.~(\ref{CR_rate}) and (\ref{Eloss_equations}) were derived to 
describe the degradation of the energy of partons traversing a static plasma with some constant temperature $T$. 
In reality, the strongly coupled liquid created in a heavy ion collision is finite in spatial extent and 
expands and cools rapidly, meaning that $T$
depends strongly on both position and time.  We will nevertheless use these expressions for the instantaneous $dE/dx$, 
to describe the energy lost by the partons in a shower as they traverse the expanding cooling strongly
coupled matter and ultimately the energy lost by the jets that emerge from the debris of the heavy ion collision.
We describe our implementation of this approach in the next Section.

\subsection{The Effects of Flow on the Rate of Energy Loss}
\label{Sec:FlowEffects}

In the previous Section, we have specified the rate of energy loss for an energetic
parton traversing static plasma with some constant temperature $T$.  In order to
use these expressions in the description of how an energetic parton loses energy
as it traverses the expanding, flowing, cooling, plasma created in a heavy ion
collision, we will exploit the fact that the expansion, flow and cooling of this fluid
are described well by nearly inviscid hydrodynamics.
This implies that the dynamic medium can be described as a collection of fluid cells that are each close to 
thermal equilibrium, locally.  As is standard in fluid mechanics, local thermal equilibrium should be understood 
from a coarse-grained point of view: at every fluid cell there is a macroscopic system of size much larger than 
any microscopic scale, such as the inverse temperature of the cell. 
From this coarse-grained perspective,  the temperature and the velocity of the fluid cell change from 
point to point, and from time to time.

Accounting for the variation in the temperature of the fluid is straightforward,
and was already incorporated in our previous publication~\cite{Casalderrey-Solana:2014bpa}.
We assume that the temperature $T$ appearing in the formulae (\ref{CR_rate}) or (\ref{Eloss_equations})
for $dE/dx$ varies in space and time, and at each point in spacetime passed by the energetic
parton is given by the temperature of the fluid, in the local fluid rest frame,
at that point in spacetime.
The basic assumption behind this adiabatic prescription is that the length scale on which 
an infinitesimal energy loss occurs is small compared to the length scale over which
$T$ changes.  In Ref.~\cite{Casalderrey-Solana:2014bpa}, we took the variation in temperature into account
by integrating (\ref{CR_rate}) or (\ref{Eloss_equations}) along the trajectory of each parton
in the shower in the collision center-of-mass frame, taking $T$ at each point along
the trajectory from the hydrodynamic solution describing the bulk fluid.  

The prescription employed in Ref.~\cite{Casalderrey-Solana:2014bpa}
does not take the velocity of the hydrodynamic fluid fully into
consideration, as in this prescription  the velocity of the moving fluid only affects
$dE/dx$ in so far as it results in changes in the temperature of the fluid.
There is an additional effect, that we chose to neglect for simplicity in Ref.~\cite{Casalderrey-Solana:2014bpa}
but shall incorporate here.  It is seen most simply by considering a fluid with a constant temperature 
flowing with a uniform velocity.   It is clear from their derivations that (\ref{CR_rate}) or (\ref{Eloss_equations}) describe the
rate of energy loss of an energetic parton moving through this fluid {\it in the local fluid rest frame.}
If this is not the collision center-of-mass frame, $dE/dx$ in that frame must be obtained via
a Lorentz transformation.  
In this paper we will
incorporate this effect  into
our model description and, in so doing, will improve upon our previous treatment of the
effects of the medium dynamics on the loss of energy of partons in a jet.
Not surprisingly, we will find that incorporating the effects of fluid flow on the rate of energy
loss has significant effects on our results at large rapidity, since it is at large rapidity that the
boost between the local fluid rest frame and the collision center-of-mass frame becomes large.

Let us denote the rate of energy loss in the local fluid rest frame by 
\be
\frac{d E_F}{dx_F}= \mathcal{F}_F(x_F, E^F_{\rm in})\,,
\label{dEdxFluidFrame}
\ee 
where 
the function $\mathcal{F}_F(x_F, E^F_{\rm in})$ is given by the right-hand side of (\ref{CR_rate}) or (\ref{Eloss_equations}) and
where we have highlighted in the notation that $dE_F/dx_F$ depends upon the distance $x_F$ that the parton has
travelled in the local fluid rest frame and, in the case of (\ref{CR_rate}), upon the initial energy $E^F_{\rm in}$ that the parton had when it was produced
at a splitting point in the shower, again as evaluated in the local fluid rest frame. 
In making this statement, we have assumed that the effects of the spatial and temporal gradients in the fluid
on $dE_F/dx_F$ can be neglected.\footnote{The effects of spatial and temporal gradients in the fluid
on the rate of energy loss of an infinitely heavy quark moving through strongly coupled plasma
have been computed, to lowest order in fluid gradients~\cite{Chesler:2013urd,Lekaveckas:2013lha,Rajagopal:2015roa}.  They can be significant early in a collision, before hydrodynamization.  Once
the fluid is hydrodynamic, these effects are small.  We expect that the same is true for the effects
of fluid gradients on the rate of energy loss of a massless parton
moving through the plasma, but we defer checking this by explicit calculation to future work.}
We now Lorentz transform the rate of energy loss (\ref{dEdxFluidFrame}) back to the collision center-of-mass frame,
obtaining a result that we shall denote by
\be
\frac{d E}{dx}= \mathcal{F}(t, E_{\rm in}) \, ,
\ee 
where $t$ is the time in the collision center-of-mass frame since the parton
was produced and $E_{\rm in}$ is the energy that the parton had, in that frame,
when it was produced.  We can change from $x$ to $t$ at will because throughout our treatment
we are assuming that the energetic partons in the shower move at (very close to) the speed of light.
The functions $\mathcal{F}_F$ and $\mathcal{F}$ are related explicitly by a Lorentz transformation
that we perform in Appendix~\ref{BI-Eloss}. 
The result takes on the surprisingly simple form
\be
\label{eq:biform}
\mathcal{F}(x, E_{\rm in}) =\mathcal{F}_F(x_F, E^F_{\rm in}(E)) )
\ee
where $E^F_{\rm in}$ and $x_F$ are the 
initial energy and the path length in the local fluid rest frame. 
These are related to quantities defined in the collision center-of-mass frame by
\be
\label{eq:EF}
E^F_{\rm in}&=& E_{\rm in} \,\gamma_F \left(1-\w \v\right) \, ,\\
\label{eq:xF}
x_F(t)&=&\int_{t_0}^{t} d t \sqrt{\left[\w^2 + \gamma_F^2\left(\v^2 - 2 \v \w + (\v \w)^2\right) \right]} \, ,
\ee
where ${\bf w}\equiv{\bf p}/E$ is the parton velocity, ${\bf v}$ and $\gamma_F$ are the fluid velocity and Lorentz factor, $t_0$ the time the parton 
was produced
and $t$ is the observation time, all in the collision center-of-mass frame. The derivations of these expressions
are also given in Appendix \ref{BI-Eloss}.

The result (\ref{eq:biform}) implies that if the rate of energy loss does not depend
explicitly on the energy of the parton or the distance that the parton has travelled
through the medium, as in the case of the collisional rate in (\ref{Eloss_equations}),
the fluid velocity will have no effect on $dE/dx$.  In this case, our treatment
is equivalent to the simpler treatment of
Ref.~\cite{Casalderrey-Solana:2014bpa}.
In the case of the radiative energy loss rate in (\ref{Eloss_equations}), or for
the strongly coupled rate of energy loss (\ref{CR_rate}) that we employ in
our hybrid model, we expect that including the effects of fluid flow on the
energy degradation of jets will be particularly important for jets at large rapidity.

As a simple but illustrative example,  let us consider the energy lost by an energetic parton propagating
through a fluid that is experiencing Bjorken flow, namely boost-invariant longitudinal expansion with no transverse flow.
If the parton has a large rapidity, the fluid that it is propagating through has a large longitudinal
velocity meaning that there is a substantial boost between the local fluid rest frame and
the collision center-of-mass frame.
 Assuming that both the parton and the boost invariant fluid are produced at the same time, 
 the  longitudinal velocity of the parton coincides with the fluid velocity at its location.
If the parton travels a distance $L$ in the collision center-of-mass frame during a time $t$, then according to 
 Eq.~(\ref{eq:xF}) the distance that the parton travels through the fluid in the local fluid rest frame is
\be
x_F(t)=\int_0^{t} \left|\w_T \right| dt= \frac{L}{\cosh y} \, ,
\ee
where $\w_T$ and y are the transverse velocity of the parton its rapidity respectively. 
For particles with significant rapidity $y$, say $y>1$, we see that the distance they travel
in the local fluid rest frame is substantially less than the distance they travel in the
collision center-of-mass frame.  If $dE/dx$ grows with $x$, not taking this Lorentz
contraction effect into account, as in Ref.~\cite{Casalderrey-Solana:2014bpa},
results in an overestimate of the amount of energy the parton loses.  Conversely, incorporating
this effect, as we do in this paper, will reduce the energy loss of partons with significant
rapidity relative to that in Ref.~\cite{Casalderrey-Solana:2014bpa}.
If the rate of energy loss is given by the expression
in (\ref{Eloss_equations}) inspired by radiative energy loss, this is the principal effect of
flow on the  energy loss suffered by an energetic parton.  If the rate of energy
loss is given by the strongly coupled form (\ref{CR_rate}), as in our hybrid model,
this effect is important but it is also important to note that $E_{\rm in}$ is also different
in different frames.
 Using (\ref{eq:EF}),
\be
E^F_{\rm in }=\frac{E_{\rm in}}{\cosh y} \,
\ee
in this simple Bjorken flow.
The effect of this diminution in $E^F_{\rm in}$ on the rate of energy loss (\ref{CR_rate}) is complex.  On the one hand, reducing $E^F_{\rm in}$ reduces 
the rate of energy loss.  On the other hand, it reduces $x_{\rm stop}$ which  increases the rate of energy loss.
The net effect is not clear {\it a priori}.
In the next Section, we will address this issue by performing a complete simulation of jets in plasma and compare it to the 
results we obtained in our
previous publication \cite{Casalderrey-Solana:2014bpa}, where we took into account the changing temperature
along the parton trajectory but left out the effects of the fluid velocity.

\subsection{\label{sec:effects_flow}The Effects of Flow on Single-jet and Dijet observables}

We have re-analyzed the single jet and dijet observables studied in our previous publication~\cite{Casalderrey-Solana:2014bpa} to include flow effects as prescribed by Eq.~(\ref{eq:biform}). As already mentioned,  the effect of flow  for the different energy loss models we have studied is different, so
we have done this reanalysis for all three expressions for the parton energy loss in (\ref{CR_rate}) and (\ref{Eloss_equations}).

Following the same procedure as in  Ref.~\cite{Casalderrey-Solana:2014bpa}, we embedded dijets generated by \pythia 8.183~\cite{Sjostrand:2007gs} 
into  boost-invariant 
hydrodynamic simulations of the collision dynamics. 
Initially, as in \cite{Casalderrey-Solana:2014bpa}, we employed the ideal hydrodynamic simulations of Pb-Pb collisions with
center-of-mass energy $\sqrt{s}= 2.76$ AGeV computed by
Hirano, Huovinen and Nara~\cite{hiranoLHC}. These HHN simulations 
incorporate an equation of state obtained from the lattice calculations of Ref.~\cite{Bazavov:2009zn}. 
This equation of state has
a crossover transition in a range of temperatures $180<T_c<200$ MeV,
somewhat higher than the crossover transition temperatures obtained from more
recent lattice calculations.
The value of the crossover temperature $T_c$ has quantitative effects on our results because  in our model we quench jets by applying
an energy loss prescription, (\ref{CR_rate}) or (\ref{Eloss_equations}), to the partons
in a shower only as long as those partons are at points in spacetime where the medium
has a temperature greater than $T_c$.  When the medium temperature experienced by a given shower
parton drops below $T_c$, we turn energy loss off.  We make this sharp distinction for simplicity; it is of course unrealistic. We expect
that using expressions derived for QGP like (\ref{CR_rate}) or (\ref{Eloss_equations}) overestimates the energy
loss at temperatures just above $T_c$, since the rate of energy loss in a hadron gas is less than that in QGP
meaning that as the crossover temperature is approached from above the rate of energy loss must
drop in a way not captured by (\ref{CR_rate}) or (\ref{Eloss_equations}).
Of course, turning the energy loss off completely is certainly an underestimate at temperatures just below $T_c$.
If we turn energy loss off at a specified $T_c$ as we do, varying the value of $T_c$ gives us a way to estimate
the systematic errors introduced into our model results by our crude treatment of late-time energy loss.

\begin{table}[t]
 \begin{center}
 \begin{tabular}{|c||c|c||c|c||c|c|}
 \hline
& \multicolumn{2} {c ||} {HHN hydro}& \multicolumn{2} {c ||}  {HHN hydro}& \multicolumn{2} {c |}{SH Hydro} \\
Parameter  & \multicolumn{2} {c ||} {without flow effects}& \multicolumn{2} {c ||}  { with flow effects}& \multicolumn{2} {c |}{with flow effects} \\
\cline{2-7}
 & \multicolumn{2} {c ||} {$T_c$ range}& \multicolumn{2} {c ||}  {$T_c$ range} & \multicolumn{2} {c |}{$T_c$ range}\\
 & 180 MeV & 200 MeV & 180 MeV & 200 MeV & 145 MeV & 170 MeV 
\\
\hline
$\aSC$ & $0.26 -  0.31$ & $0.30 - 0.35$ & $0.39-  0.46$& $0.45 - 0.53$ &$0.32 - 0.37$ &$0.35 -  0.41$
\\
\hline
$\aR$ & $0.81 - 1.2$ & $1.0 - 1.6$ & $1.6 - 2.4$& $2.1 -  3.3$ &$0.97-1.5$ &$1.2-1.8$
\\
\hline
$\aC$ & $2.5 -  3.5$ & $2.9 - 4.2$ & $2.5- 3.5$& $2.9 - 4.2$ &$1.8-2.6$ &$2.2-3.0$\\
\hline
  \end{tabular}
\end{center}
 \caption{\label{alphatable} Values 
 of the fit parameters needed in the specification of $dE/dx$ in our three different energy loss models, Eqs.  \ref{CR_rate} and   \ref{Eloss_equations}. 
 The parameters are extracted 
 by comparing model predictions for $\Raa$ for jets with 100~GeV$<\pt <$110~GeV 
 in central
 Pb-Pb at $\sqrt{s}=2.76$  ATeV collisions at the LHC to experimental data~\cite{Raajet:HIN}. 
 The parameters are extracted by employing three different treatments of the bulk hydrodynamic fluid.
 In the first two columns, we use hydrodynamic simulations from Ref.~\cite{hiranoLHC}, that we denote by HHN.
 The first column repeats results from Ref.~\cite{Casalderrey-Solana:2014bpa}; the second column shows
 the effects of including the effects of the flow velocity on the rate of energy loss as described in Section 2.2.
  The different choices of $T_c$ denote the temperature below which we turn off parton energy loss in our model.
 In each cell in the table, the range of the fit parameter is determined by fitting to one experimentally
 measured $R_{\rm AA}$ data point, as described in the text.
 In the third column, we use hydrodynamic simulations from Ref.~\cite{Shen:2014vra} as described in the text that were
 provided to us by Shen and Heinz (SH), with choices of $T_c$ that span the range of crossover temperatures
 favored by current lattice calculations.
 }
\end{table}

In Ref.~\cite{Casalderrey-Solana:2014bpa} we presented all our results with both $T_c=180$~MeV
and $T_c=200$~MeV.  We do the same in the first two columns of Table \ref{alphatable}. The first column
repeats results from Ref.~\cite{Casalderrey-Solana:2014bpa} that include only the effects of variations in $T$ with no effects of the flow velocity
on the energy loss included. The second column comes from our present calculations, including the effects 
of flow as in Eq.~(\ref{eq:biform}).  In each cell in the table, we quote a range of values for the
parameter $\aSC$ or $\aR$ or $\aC$ in the expression for the rate of energy loss
$dE/dx$ of the partons in the shower that we select, (\ref{CR_rate}) or (\ref{Eloss_equations}).  The range
corresponds to fitting $R_{\rm AA}$ for jets with 100~GeV $< \pt <$ 110~GeV to the upper and lower limits
of the experimental uncertainty on the measured value of this quantity.  We then take the entire range 
spanned by varying $R_{\rm AA}$ for jets with 100~GeV $< \pt <$ 110~GeV and $-2<\eta<2$
over its experimentally allowed range~\cite{Raajet:HIN}
and by varying $T_c$ from 180 to 200 MeV as the allowed range for a given $\kappa$.  As described in
full in Ref.~\cite{Casalderrey-Solana:2014bpa} and reviewed in Appendix \ref{update}, we then compute
a variety of single jet and dijet observables over wide ranges in $\pt$ and centrality and obtain
good descriptions of the data, in particular using our hybrid model with $dE/dx$ given by (\ref{CR_rate}).
This does not change when we incorporate the effects of flow; what changes are the fitted values
of the $\kappa$ parameters, as shown in the first two columns of Table~\ref{alphatable}.
Note that $\aC$ does not change since, as we discussed in the previous Section,
flow does not affect $dE/dx$ if $dE/dx$ is given by the collisional expression from (\ref{Eloss_equations}).
Both $\aSC$ and $\aR$ increase: in these cases, incorporating the effects of flow reduces the
energy loss of partons with substantial rapidity and since we are fitting to $R_{\rm AA}$ for
jets with pseudorapidities over the range $-2 <\eta< 2$ we have to increase $\aSC$ and $\aR$.

Throughout the rest of this paper, we will embed jets and implement our hybrid model for their quenching
in a new set of hydrodynamic simulations of Pb-Pb collisions at $\sqrt{s}=2.76$~ATeV and 5.02~ATeV
provided to us by Shen and Heinz, based upon the codes developed by them and their
collaborators in Refs.~\cite{Shen:2014vra} and \cite{Shen:2014nfa}.  These 2+1-dimensional 
simulations, with boost-invariant longitudinal expansion,
incorporate an equation of state from Ref.~\cite{Huovinen:2009yb} (referred to there as the s95p-v1-PCE150 equation
of state) that incorporates results from lattice QCD calculations and from a hadron resonance gas
at low temperatures. With this equation of state, it is possible for us to set the $T_c$ below which we stop parton
energy loss within a range $145 \, {\rm MeV}< T_c< 170\, {\rm MeV}$ that reflects results for the range
of crossover temperatures from current lattice calculations, see for example Refs.~\cite{Aoki:2009sc,Bazavov:2014pvz}.  For simplicity, the hydrodynamic calculations
employ a temperature-independent $\eta/s=1/(4\pi)$. Also for simplicity, we employed
ensemble-averaged MC-Glauber initial conditions \cite{MCGlauber:rev}, neglecting all effects of
event-by-event fluctuations. Because we are not concerned
with the evolution below $T_c$, we evolved the initial conditions using 
viscous hydrodynamics alone, with no cascade afterburner.

As shown in the third column of Table~\ref{alphatable}, we have refitted the parameters $\aSC$, $\aR$ and $\aC$ in the
expressions (\ref{CR_rate}) and (\ref{Eloss_equations}) to the experimentally measured $R_{\rm AA}$
for jets with $-2<\eta<2$ and 100~GeV$<\pt<$110~GeV using the SH hydro simulations and setting $T_c$ to 145 MeV
or 170 MeV to yield some sense of the theoretical systematic uncertainty as before.  The values of
the parameters so obtained are smaller than those in the second column, reflecting the lower value of $T_c$
meaning that energy loss continues for the partons in the showers for a longer time. We have
checked that most of the difference between the parameter values in the second and third
columns is indeed due to the change in the value of $T_c$, as anticipated.  
Perhaps coincidentally, the parameter
values in the third column are quite similar to those from Ref.~\cite{Casalderrey-Solana:2014bpa} given in
the first column: the consequences of adding flow effects and lowering $T_c$ to a large degree cancel.

In Appendix \ref{update}, we update the results of Ref.~\cite{Casalderrey-Solana:2014bpa}
for the jet $R_{\rm AA}$ as a function of $\pt$ and centrality, and for the dijet imbalance in
Pb-Pb collisions at $\sqrt{s}=2.76$~ATeV, comparing them to CMS data~\cite{Chatrchyan:2011sx,Raajet:HIN}. 
We see that with SH hydro and a lower $T_c$, just as with HHN hydro and a higher $T_c$
in Ref.~\cite{Casalderrey-Solana:2014bpa}, when we fit the one parameter $\aSC$ in our
hybrid model to one data point the model does a very good job of describing these full
data sets.  In Appendix \ref{DijetPredictions}, we present results for $R_{\rm AA}$ and
the dijet imbalance for Pb-Pb collisions with $\sqrt{s}=5.02$~ATeV.  These constitute
predictions for the LHC heavy ion run coming late this year.

The improvements we have introduced in our description of the dynamics of the expanding droplet of QGP 
in going from the first column of Table 1 to the second column to the third column
have resulted in changes to the numerical values of $\aSC$, $\aR$ and $\aC$.  However, the conclusions
about the implications of the values of these parameters that we reached in Ref.~\cite{Casalderrey-Solana:2014bpa} 
all remain unchanged.
The extracted values  of $\aR$ and $\aC$ imply such large ``weak'' couplings or such large
logarithmic corrections to the leading order energy loss result that a perturbative analysis is called into question.\footnote{For
an analysis of the effects of large logarithmic corrections to medium parameters, see Refs.~\cite{Iancu:2014kga,Blaizot:2014bha}.} 
The extracted value of $\aSC$ in the strongly coupled expression (\ref{CR_rate})
for $dE/dx$ in our hybrid model corresponds to a stopping distance for energetic partons in the strongly
coupled QGP of QCD that is about three to four times longer than that in the strongly coupled plasma
of ${\cal N}=4$ SYM theory. And, as can be seen by comparing the plots in Appendix~\ref{update}
to those in Ref.~\cite{Casalderrey-Solana:2014bpa}, the hybrid model with the improvements
that we have introduced continues to provide a good
description of data on both single jet and dijet observables
over a wide range of jet $\pt$ and event centrality.

\subsection{\label{sec:species_dep}Species Dependence of Jet Suppression}

\begin{figure}[tbp]
\centering 
\begin{tabular}{cc}
\includegraphics[width=.5\textwidth]{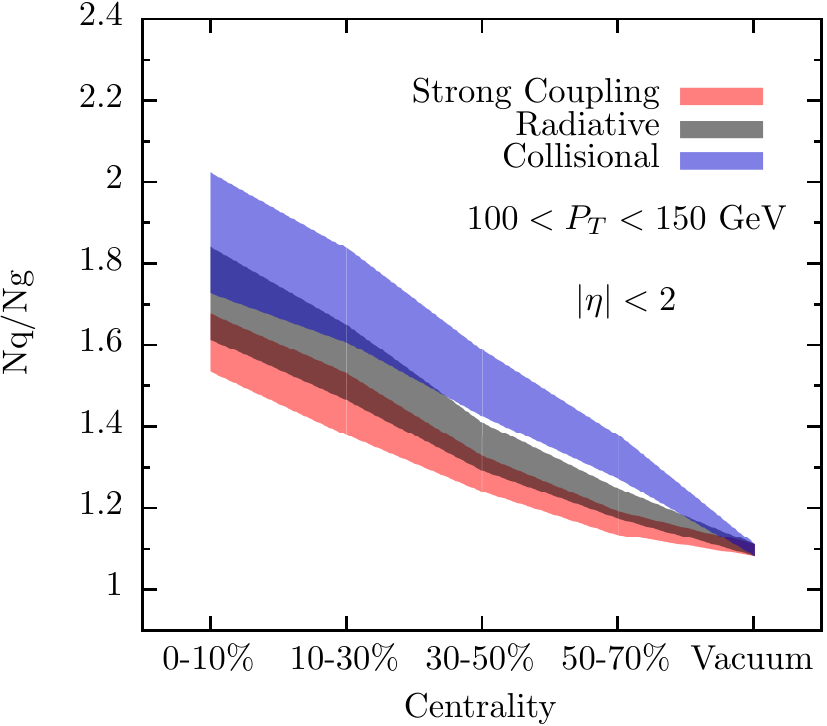}
\put(-170,40){\small{$\sqrt{s}=2.76~\rm{ATeV}$}}
&
\includegraphics[width=.5\textwidth]{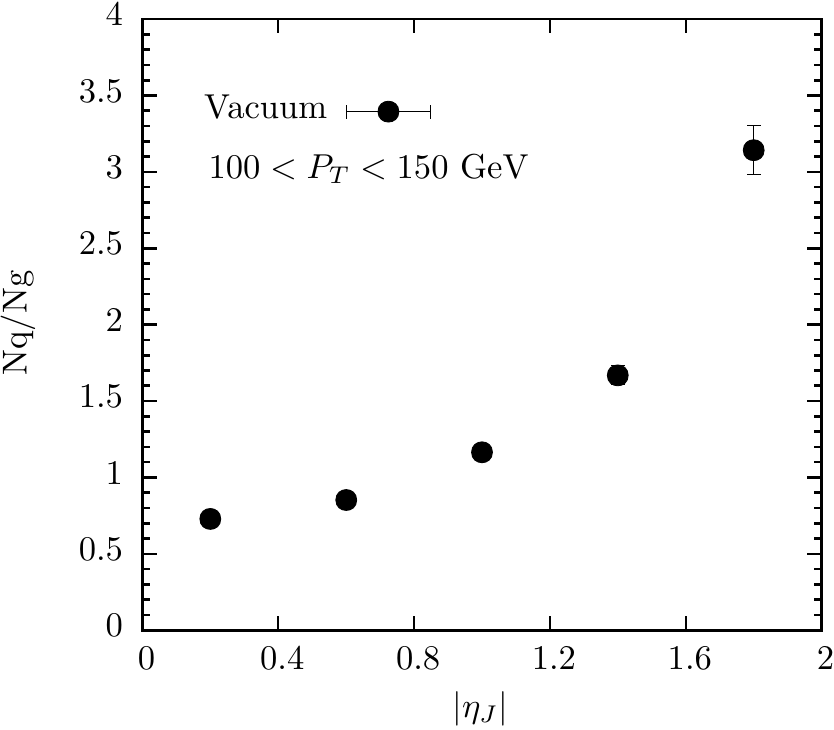}
\put(-90,40){\small{$\sqrt{s}=2.76~\rm{ATeV}$}}
\end{tabular}
\caption{\label{Fig:NQG}  Left panel: Ratio of quark-initiated jets to gluon-initiated jets with $|\eta|<2$ 
as a function of centrality for different 
models. 
Right panel: Ratio of quark-initiated jets to gluon-initiated jets as a function of jet pseudorapidity
for p-p collisions in vacuum, according to \pythia. 
}
\end{figure}

Although the central conclusions we draw from confronting our
hybrid model with the measured observables compiled in Appendix~\ref{update} are unchanged from
those we reached in Ref.~\cite{Casalderrey-Solana:2014bpa}, the conclusions that we drew there
about the species dependence of jet suppression {\it are} substantially affected by the inclusion
of the effects of flow on parton energy loss that we have described in the preceding Sections.
In the left panel of Fig.~\ref{Fig:NQG}, we repeat a calculation from Ref.~\cite{Casalderrey-Solana:2014bpa},
now with SH hydro and the lower $T_c$ range from the third column of Table~\ref{alphatable}, and reach
different conclusions.    
We show the ratio of the number of quark-initiated jets to the number of gluon-initiated jets with jet momenta in the range
$100< \pt <150$ ~GeV  and $\left| \eta \right| < 2$ as a function of centrality for the three models we have 
studied. As in \cite{Casalderrey-Solana:2014bpa},
we have defined the species of the jet-initiator from our \pythia ``data'' (in a way that is impossible
to do in experimental data) as the identity of the parton that initiated the DGLAP shower to which the
hardest particle in the jet belongs.
The inclusion of the effects of flow on parton energy loss has resulted in the bands corresponding 
to our three different models being much closer to each other in Fig.~\ref{Fig:NQG} than they
were in Ref.~\cite{Casalderrey-Solana:2014bpa}.  The power to discriminate between
models of energy loss implicit in Fig.~\ref{Fig:NQG} is much less than was the case
in Ref.~\cite{Casalderrey-Solana:2014bpa}.
 This change arises as a combination of two facts.  First, as we discussed in Section~\ref{Sec:FlowEffects} the inclusion
 of the effects of flow reduces the energy loss of partons (both quarks and gluons) at large rapidity if $dE/dx$ is given by the strong coupling
 expression (\ref{CR_rate}) or the radiative expression from (\ref{Eloss_equations}), but leaves things unchanged if
 the collisional expression from (\ref{Eloss_equations}) is employed.    Second, as shown in the right panel
 of Fig.~\ref{Fig:NQG}, without any energy loss there are many more quark-initiated jets at large
 rapidity than at mid-rapidity for jets with the same range of $\pt$ as in the left panel of the Figure.
 As the pseudorapidity increases, jets in this interval of transverse 
momentum become more and more quark dominated. 
The consequences of these two facts in concert is that the
 many quark-initiated jets at large rapidity lose less energy once flow effects are included than they
 did in the analysis of Ref.~\cite{Casalderrey-Solana:2014bpa}, if $dE/dx$ depends on flow as in either the strongly coupled or radiative 
 expressions.  In these cases, the effects of flow on the rate of energy loss serve to enhance
 the relative weight of the large rapidity jets in the final jet sample, and these large rapidity jets
 are predominantly quark-initiated.
 This means that,
 after summing over $|\eta|<2$ as in the left panel of Fig.~\ref{Fig:NQG}, the inclusion of the effects
 of flow pushes the strongly coupled and radiative $N_q/N_g$ bands in this Figure upwards, bringing both much closer
 to the collisional band.  This happens even though, as we mentioned in Section~\ref{sec:hybridmodel} and discussed in detail in Ref.~\cite{Casalderrey-Solana:2014bpa},
the Casimir-dependence of the rate of energy loss in (\ref{CR_rate}) is weaker than in (\ref{Eloss_equations}): this difference
in the Casimir dependence of $dE/dx$ pushes 
the strongly coupled $N_q/N_g$ band in the left panel of Fig.~\ref{Fig:NQG} downward relative to
the collisional band, but the effects of flow in concert with the
rapidity-dependence of the production of quark and gluon jets in p-p collisions brings it back up.
Consequently, our current results
render the species-dependence of quenching rather insensitive to the underlying energy loss dynamics.

\begin{figure}[tbp]
\centering 
\includegraphics[width=.49\textwidth]{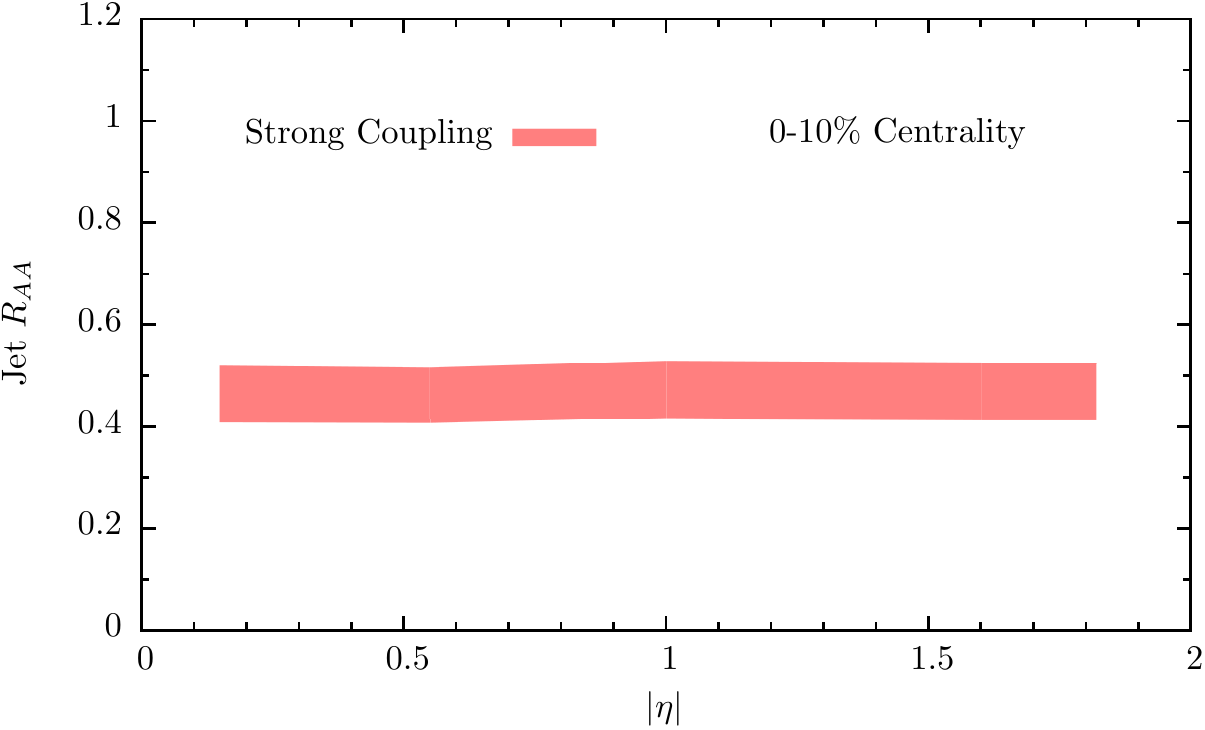}
\put(-170,90){\tiny{$\sqrt{s}=2.76~\rm{ATeV}$}}
\includegraphics[width=.49\textwidth]{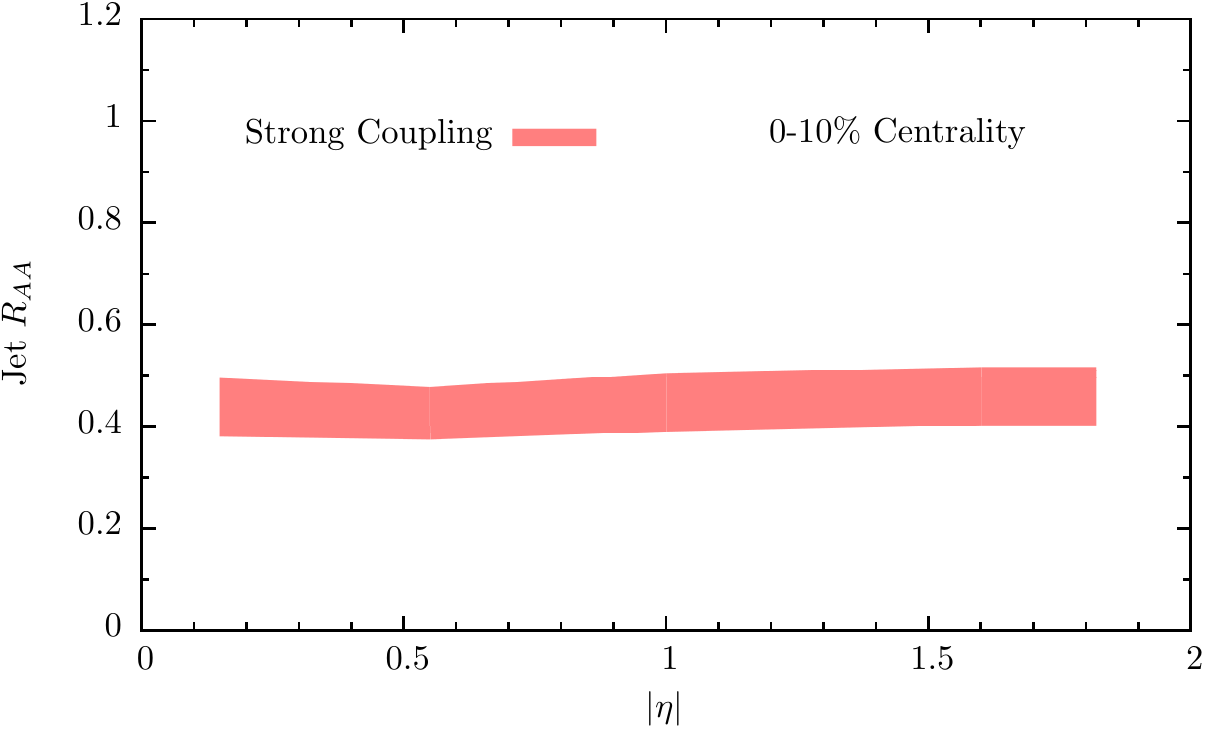}
\put(-170,90){\tiny{$\sqrt{s}=5.02~\rm{ATeV}$}}
\caption{\label{Fig:Rap}  
Pseudorapidity dependence of jet $R_{AA}$ in our hybrid model for $\sqrt{s}=2.76$~ATeV (left) and  $\sqrt{s}=5.02$~ATeV (right).
}
\end{figure}

We close this section by noting 
that despite the substantial rapidity dependence of the species-dependence 
of jet production, and of jet quenching, the resulting net jet $R_{\rm AA}$ is remarkably independent of rapidity in our hybrid model,
as shown in Fig.~\ref{Fig:Rap}.  This was noted previously in Ref.~\cite{Renk:2014gwa}, and we find
the same behavior in our control models in which we choose the dependence of $dE/dx$ according
to either one of the expressions (\ref{Eloss_equations}).
Although given
the strong species-dependence manifest in the right panel of Fig.~\ref{Fig:NQG} the flatness
of the rapidity dependence seen in Fig.~\ref{Fig:Rap} appears coincidental, it is in agreement
with experimental results from the ATLAS collaboration~\cite{Aad:2014bxa}.

\section{\label{sec:BJCorr}Boson-Jet Correlations, Including Predictions for $\sqrt{s}=5.02$~ATeV Collisions and for Z-jet Correlations}

The extensive exploration in Ref.~\cite{Casalderrey-Solana:2014bpa} demonstrated
that the hybrid model that we have developed describes currently available
inclusive jet and dijet data from LHC heavy ion collisions
rather successfully. In Appendix~\ref{update}, 
we compile updated versions of these results, using SH hydro and taking
flow effects into account as we described in the previous Section, and we
reconfirm the conclusions of Ref.~\cite{Casalderrey-Solana:2014bpa}.
This gives us considerable confidence in the model framework.
However, this class of observables proves not to be very sensitive to
whether we choose $dE/dx$ as in (\ref{CR_rate}) or (\ref{Eloss_equations})~\cite{Casalderrey-Solana:2014bpa}.
The distinctions between the strongly coupled form for $dE/dx$ and
our two control models provided by these observables are not sufficient to 
differentiate between these different hypotheses for the microscopic dynamics
of energy loss, at least with present uncertainties.   The strongly coupled form for $dE/dx$
does provide a better description of the dijet imbalance, but its predictions
are not sufficiently distinct from those of the control models.
We must, therefore, consider further observables.


In deciding how to go further, we face a choice.  We could start adding more physics to the model, which
would allow us to confront new classes of observables.  For example, if we were to add in the transverse
momentum kicks that the medium delivers to the shower partons passing through it, and the recoil that
the shower partons delivers to the medium, we could engage with jet shape observables.
This is an attractive prospect, but we defer it to future work.  Before adding to the model, and
in particular before adding a second free parameter, it is our responsibility to
first ask whether there are further observables that our present model, with its single parameter
already fixed as described in the previous Section, could reasonably be expected to describe.
This is our goal in this paper.

For most of the remainder of this paper, we turn to observables involving jets back-to-back with either
an energetic and isolated photon or a Z-boson.  We will keep the parameter $\aSC$ in our hybrid
model fixed exactly as in the previous Section, and confront our model with existing data on
three different observables describing $\gamma$-jet correlations 
in Pb-Pb collisions 
at $\sqrt{s}=2.76$~ATeV.  As before, our model describes
the data well, although it is fair to note in advance that the statistical error bars on these measurements
are substantial because to this point the data sets of $\gamma$-jet events have not been large.
Nevertheless, the confrontation of the hybrid model with these three new sets of data provides
a strong independent validation of our hybrid model, which has no further adjustable parameters.

Our central purpose in this Section is to use our hybrid model to provide predictions for
the measurements that the data sets with higher statistics by about an order of magnitude
that are anticipated late this year will
make possible.  To this end, we provide predictions for three $\gamma$-jet observables
and three Z-jet observables in Pb-Pb collisions 
at $\sqrt{s}=5.02$~ATeV.  In making these predictions, we will keep 
the parameter $\kappa_{\rm sc}$ in the hybrid model and the parameters $\aR$ and $\aC$ in the control models set to 
the same values that we obtained in the previous Section by fitting to one data point at $\sqrt{s}=2.76$~ATeV.  The principal
change in going from $\sqrt{s}=2.76$~ATeV to $\sqrt{s}=5.02$~ATeV is a (modest) increase
in the temperature of the plasma.  The principal effects of this increase arise from the explicit
temperature dependence in (\ref{CR_rate}) or (\ref{Eloss_equations}), and so are included in our analysis.
We are leaving out any small decrease in the values of the $\kappa$'s that may arise if the plasma
becomes slightly less strongly coupled, anticipating that this effect will likely be too small to be
resolved given other uncertainties, both theoretical and experimental.


Although the observables that we study in this Section are similar in some respects to
those  that describe dijet events,
since neither $\gamma$'s nor Z-bosons 
interact strongly with the plasma, when one
triggers on an energetic $\gamma$ or Z-boson
the jets produced in the same hard scattering
have points of origin that are distributed through the
collision volume differently than is the case for dijet pairs,
where selection bias favors points of origin closer to the
surface where at least one of the jets will suffer less energy loss.
This means that the jets produced in association 
with a $\gamma$ or a Z-boson sample 
a different path-length distribution than the jets in  dijet pairs. 
A second difference between boson-jet events and dijet events
is that both $\gamma$'s and Z-bosons are much more likely to 
be produced in association with a quark jet than with a gluon jet.
The third difference is that since the $\gamma$ or Z-boson suffers
no energy loss, as they do not interact with the medium, their
energy and direction are good proxies for the initial energy and
initial direction of the hard parton (usually,
the quark) going in the opposite direction.  
This is quite different than in the case of dijet events, where 
generically both jets should be expected to have lost some energy.
Of course, the hard parton opposite a $\gamma$ or Z-boson may split
into more than one jet; in the analysis of $\gamma$-jet and
Z-jet correlations, it cannot be assumed that the boson is back-to-back
with only one jet in the final state.

\subsection{\label{sec:MC}Generation and Selection of Monte Carlo Events}

For our study of photon-jet correlations 
we analyzed $10^5$ hard scattering processes, 
in \pythia 8.183~\cite{Sjostrand:2007gs} p-p collisions at $\sqrt{s}=2.76$~TeV
and another $10^5$ such events in \pythia p-p collisions at $\sqrt{s}=5.02$~TeV.
We require at least one photon with a transverse momentum above
a desired cut, typically choosing $\pt^\gamma>\pt^{\rm cut}=60$~GeV, and with a pseudorapidity
in the range $|\eta_\gamma|<1.44$.
We set the $\pt^{\rm min}$ parameter in \pythia  (basically, the minimum
momentum transfer in the hard processes that \pythia is sampling) safely
lower than $\pt^{\rm cut}$, for example choosing $\pt^{\rm min}=40$~GeV if we are recording
events that include a photon with $\pt^\gamma>60$~GeV, since \pythia does not
reliably reproduce the photon spectrum all the way down to its $\pt^{\rm min}$.\footnote{For one of our observables,
we wanted to include photons with $\pt^\gamma>\pt^{\rm cut}=40$~GeV in our analysis.  For this observable, we ran
$10^5$ events with $\pt^{\rm min}=20$~GeV.}
We placed each of these $10^5$ events at random in the transverse plane
of a Pb-Pb collision, with $\sqrt{s}=2.76$~ATeV or 5.02 ATeV, choosing
the location according to the probability distribution for the number of
binary collisions occurring at each transverse position according to an
optical Glauber model. We keep the direction of the photon (in azimuth
and in rapidity) as it was in the \pythia p-p collision from which we have
taken it.
We generate the hard processes in \pythia 
without underlying p-p events because we embed each hard process 
into a hydrodynamic description of the matter produced in the
heavy ion collision, meaning that including the underlying p-p event
from \pythia would be double-counting.  As we described
in the previous Section, we use
the  boost-invariant 2+1-dimensional 
hydrodynamical simulations of the expanding cooling droplet
of matter produced in the heavy ion collision from Ref.~\cite{Shen:2014vra}
to determine the local temperature and local velocity vector of the fluid at
the position of each of the partons in the jet shower described by \pythia,
as they propagate through the cooling hydrodynamic fluid as a function of time.
We follow each parton in the shower, and as long as a parton sees a local
temperature that is greater than $T_c$ we reduce its energy at a rate $dE/dx$
given by (\ref{CR_rate}) in our hybrid model, or by (\ref{Eloss_equations}) if we are investigating
one of our two control models.  
When a parton that has lost some energy branches, 
we start each of its daughters off with the same fractional energy loss as the parent   had when it branched.  The daughters
then lose further energy according to (\ref{CR_rate}) in our hybrid model or (\ref{Eloss_equations}) in our control models.
In this way, we compute the energy lost by each of the partons
in the showers produced by the hard scattering process, in so doing modifying the partons in the final state relative to 
what they were in the original \pythia event. 
As in Ref.~\cite{Casalderrey-Solana:2014bpa}, we neglect energy loss in the
pre-equilibrium stage of the collision before the hydrodynamic evolution
of Ref.~\cite{Shen:2014vra} starts at $\tau=0.6$~fm.
We have described these procedures in greater detail in Ref.~\cite{Casalderrey-Solana:2014bpa}.
As in the previous Section, we vary our choice of the temperature $T_c$ below which we 
turn off the energy loss over the range 
$145< T_c<170$~MeV.
The result of our analysis is $10^5$ hard scattering events (for each collision energy and
for each choice of $T_c$) in which the partons accompanying the photon have lost energy
as a result of their passage through the hydrodynamic medium.

Next, we analyze the $10^5$ events in order to ``measure'' each observable that is of interest to us.
In doing so, we only analyze the (modified) parton showers coming from the hard scattering. 
We do not include any of the particles that would be created as the hydrodynamic fluid
freezes out.  This means that we do not have any background particles and we therefore
do not perform any background subtraction in our calculation of observables.  When we compare
to data, we of course compare to measurements made after the experimentalists have done
their background subtraction.
For this reason, for those observables which have not been fully unfolded in the experimental analysis
we include a smearing procedure, to be describe in more detail later, to mimic resolution effects.

Although we of course know (from \pythia) exactly which photons in our events are the photons
coming from the originating hard process (prompt photons) and which photons are instead
produced in the parton shower (fragmentation photons), we do not use this information
in our computation of observables.  Instead, we perform an isolation analysis of
all high energy photons in our Monte Carlo data, patterned upon what experimentalists
must do.
This procedure allows us to study a sample of photons which is closer to that in experimental analyses, 
since in experiments prompt and fragmentation 
photons can only be distinguished via isolation cuts, with finite efficiencies. 
As it turns out, quenched jets associated with prompt photons constitute about 80\% of our
final ``data'' sets, with about 20\% coming from fragmentation of jets 
in events without a prompt photon. 

For readers who would like a little more detail, our procedure is as follows.
We select events containing at least one photon with $\pt^\gamma >\pt^{\rm cut}$
and $|\eta_{\gamma}|<1.44$. After quenching the parton showers in these events as 
described above, following the analysis of Refs.~\cite{Chatrchyan:2012gt,CMS-HIN-13-006} we consider a photon to be isolated
if the sum of the energy of all the particles within a cone around  this photon 
of radius $\Delta r\equiv \sqrt{\Delta \phi^2+ \Delta \theta^2}=0.4$ is below $5$~GeV. 
Contrary to what is done in the analysis of experimental data, we perform this isolation
cut at the partonic level, instead of at the hadronic level. 
In the unlikely situation that two or more isolated photons are found in a single event, 
we treat the one with the highest transverse momentum as the leading photon.
Next, we construct a sample of photon-jet pairs from our sample of isolated photons. 
We reconstruct jets using the anti-$k_T$ algorithm \cite{Cacciari:2011ma,Cacciari:2008gp}, typically choosing $R=0.3$, and keep
those events in which we find an associated
jet at an angular distance $\Delta \phi>7\pi/8 $ from the isolated 
photon. This angular cut suppresses the contribution of events in which the isolated photon
is a fragmentation photon and events with more than one associated jet 
to our final photon-jet sample.

We generate a sample of Z-jet events via an analogous procedure. We select $10^5$ \pythia
events in which a hard scattering process produces a Z-boson with $\pt^Z>\pt^{\rm cut}$,
again typically choosing $\pt^{\rm cut}=60$~GeV.
We need not worry about fragmentation Z-bosons: the large mass of the Z makes it very unlikely
that they are produced in the parton showers.  We therefore need not apply any isolation procedure:
any Z is a prompt Z.
For the sake of simplicity, we use the same $\pt$ and $\eta$ cuts applied to photon events. 
We reconstruct jets using the anti-$k_T$ algorithm as before, again typically choosing $R=0.3$,
and obtain our Z-jet sample by requiring an associated jet at an angular distance $\Delta \phi>7\pi/8 $ from the Z. 
In experimental measurements, Z-bosons are reconstructed through their di-muon decay.  Since high energy muons are
not modified at all by their passage through the plasma, and since the properties of the associated jets are the 
same no matter how the Z-boson decays, to increase the size of our sample we simply do not allow our Z-bosons to decay,
keeping all of them in our sample.

\subsection{\label{sec:photonjet_results}Photon-Jet Observables: Comparison with Experimental Results at $\sqrt{s}=2.76$~ATeV
and Predictions for $\sqrt{s}=5.02$~ATeV}

We use the Monte Carlo samples of photon-jet events, prepared and quenched as described above, 
to construct different observables which can be confronted with experimental data.
In this section, we describe these observables, we show the results obtained from
our model for collisions at $\sqrt{s}=2.76$~ATeV and compare them to data,
and we provide the predictions that we obtain from our model for collisions
at $\sqrt{s}=5.02$~ATeV.

We first construct the photon-jet imbalance.
This is analogous to the dijet imbalance, except that here
we have a photon instead of a leading jet.
In the case of dijets we define the associated jet 
as the jet in the pair that has less transverse momentum, meaning
that the ratio of the transverse momentum of the associated jet
to that of the leading jet is less than one by definition.
This is not so in the case of the photon-jet imbalance, since
the associated jet can have more or less transverse momentum
than the photon.  Defining
$x_{J\gamma}\equiv \pt^{\rm jet}/\pt^{\gamma}$, this observable
can be less than or greater than one.
For example, in p-p collisions in vacuum, an associated jet
can have more transverse momentum than the photon if there is a second
jet in the event, in the same hemisphere as the photon. In a heavy ion collision,
the energy loss experienced by the partons propagating
through the medium pushes $x_{J\gamma}$ downwards.
However, if the passing jet sweeps particles from the
plasma into the jet cone this can in principle push $x_{J\gamma}$ upwards,
but this effect is expected to be small at large energies.
Our hybrid model neglects this possibility, meaning that in every
event in our sample $x_{J\gamma}$ is less than (or equal to)
what it would have been in the absence of the medium.

\begin{figure}[tbp]
\centering 
\begin{tabular}{cc}
\includegraphics[width=.5\textwidth]{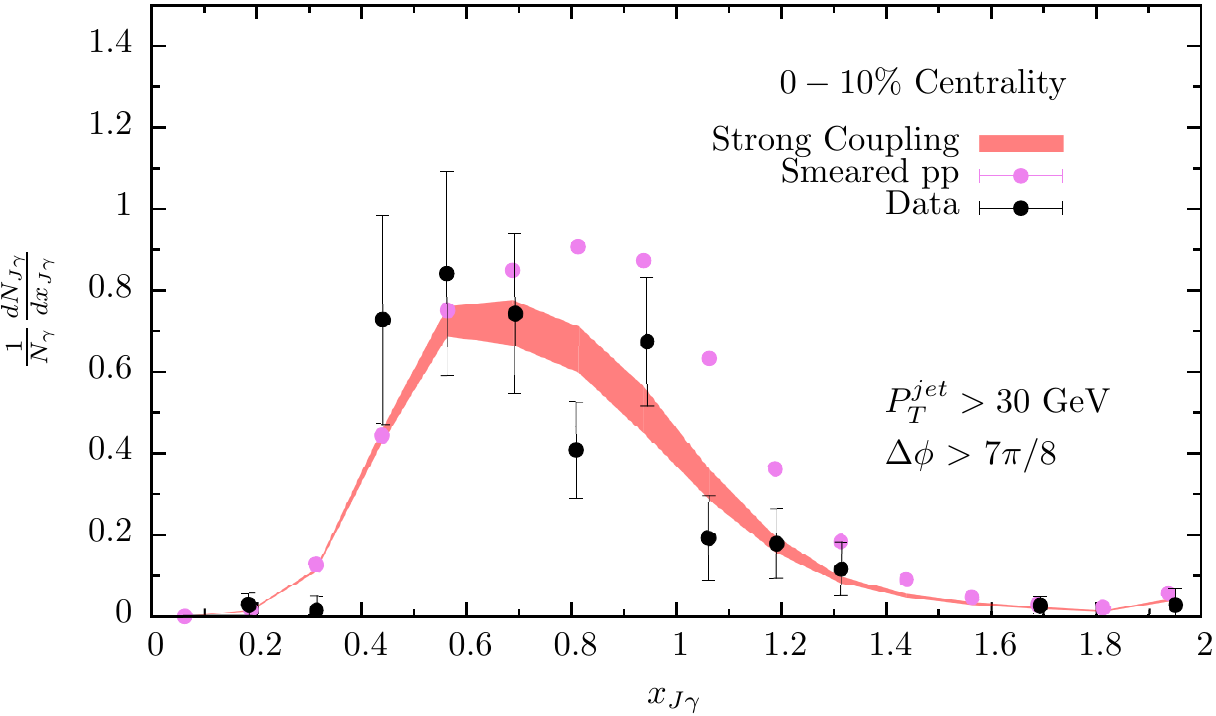}
\put(-180,110){\tiny{$\sqrt{s}=2.76~\rm{ATeV}$}}
&
\includegraphics[width=.5\textwidth]{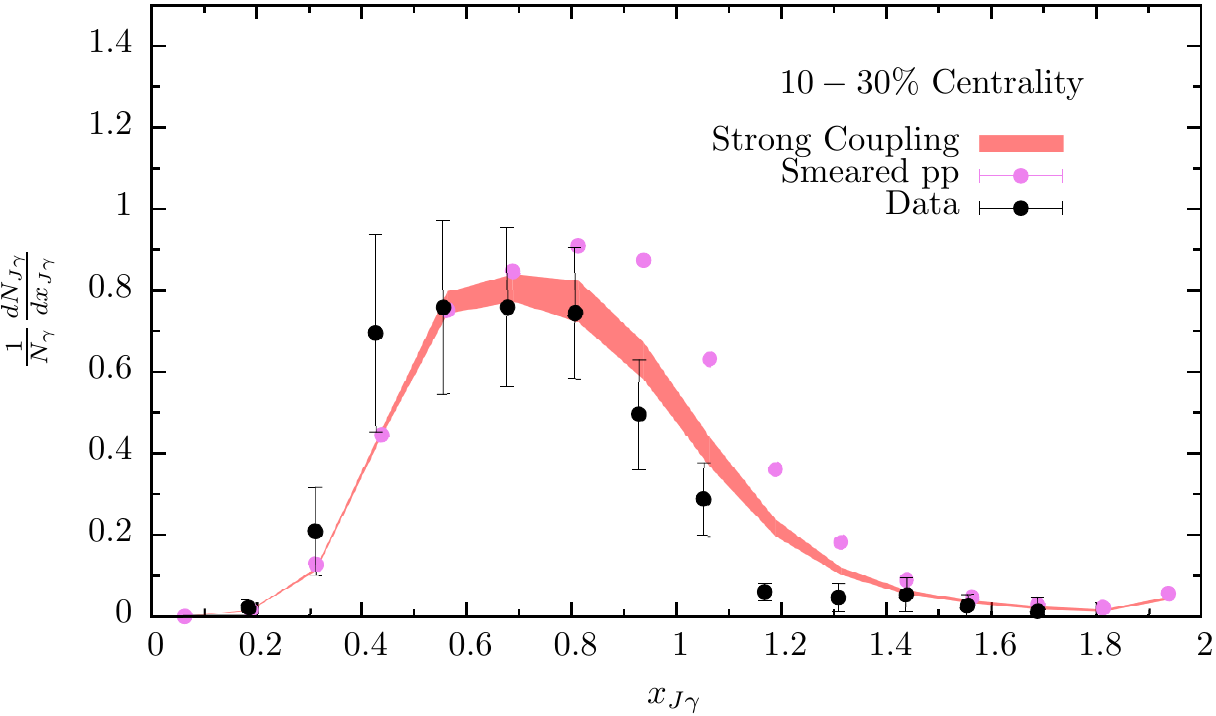}
\put(-180,110){\tiny{$\sqrt{s}=2.76~\rm{ATeV}$}}
\\
\includegraphics[width=.5\textwidth]{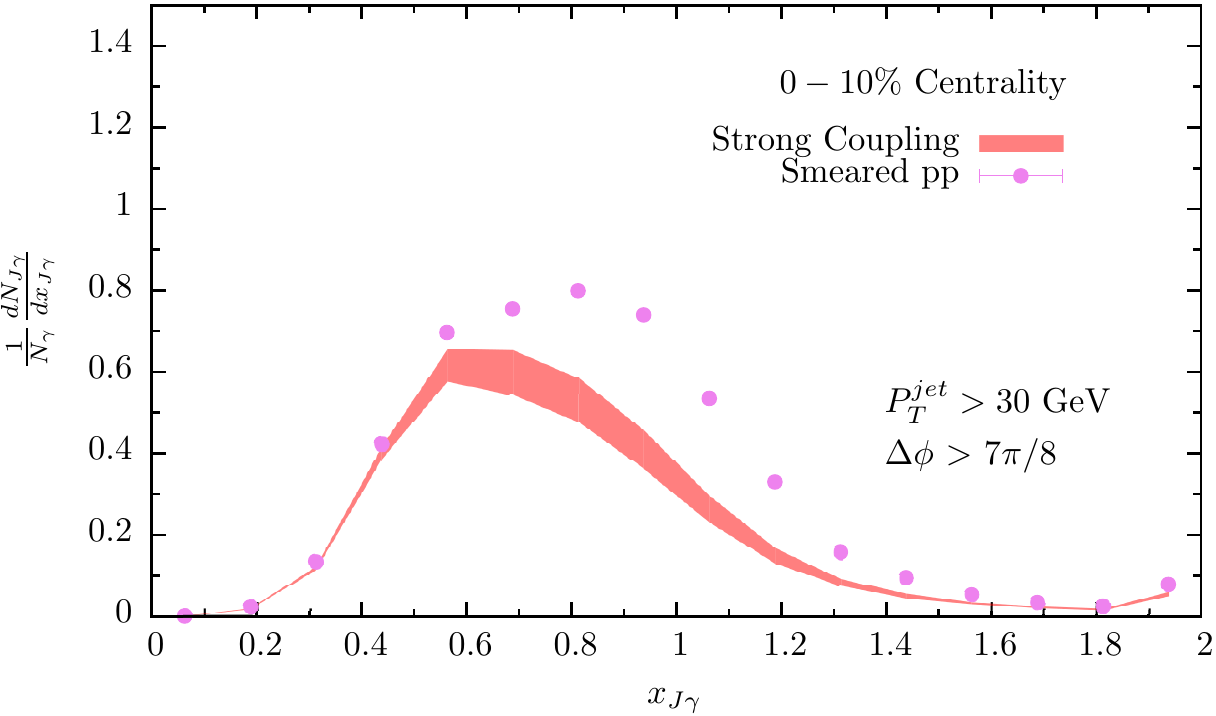}
\put(-180,110){\tiny{$\sqrt{s}=5.02~\rm{ATeV}$}}
&
\includegraphics[width=.5\textwidth]{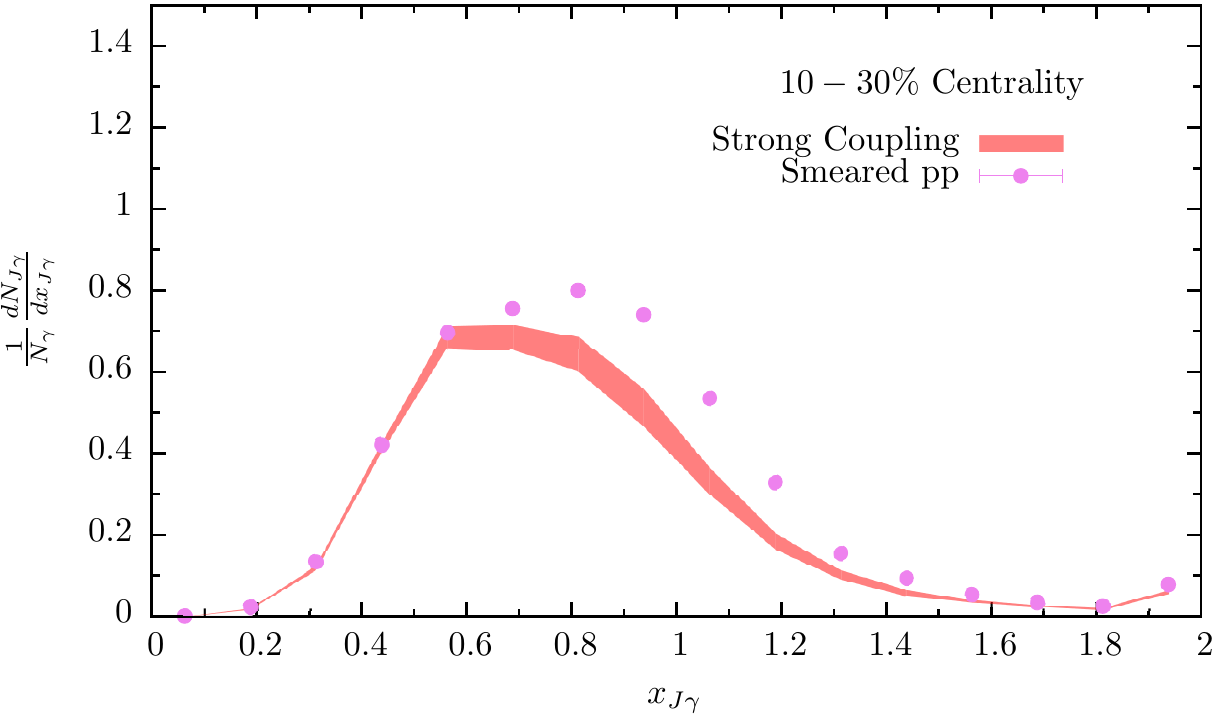}
\put(-180,110){\tiny{$\sqrt{s}=5.02~\rm{ATeV}$}}
\end{tabular}
\caption{\label{Fig:PJInv}  
Distribution of the transverse momentum imbalance of photon-jet pairs,  $x_{J\gamma}\equiv \pt^{\rm jet}/\pt^{\gamma}$, for Pb-Pb collisions. 
The left and right panels show the 
0-10\% and 10-30\% most central events, respectively. The upper panels show our results for collisions with 
$\sqrt{s}=2.76$~ATeV, as well as data from Ref.~\cite{CMS-HIN-13-006}. The lower panels show our predictions for heavy
ion collisions with $\sqrt{s}=5.02$~ATeV, anticipated for late this year. 
}
\end{figure}

In Fig.~\ref{Fig:PJInv}, we show the distribution of 
the imbalance in the transverse momentum of the associated jet relative to that 
of the photon, $x_{J\gamma}$, for events with two different centralities and for events with
two different collision energies.
Following the conventions established in the experimental analyses in Refs.~\cite{Chatrchyan:2012gt,CMS-HIN-13-006},
these distributions are normalized to the total number of photons, rather than to the total
number of photon-jet pairs; integrating each of the curves in Fig.~\ref{Fig:PJInv} therefore yields
a number below one.  Also as in the experimental analysis, we only
consider 
photons with  transverse momentum $\pt^{\gamma}>60$~GeV.
 The associated jet is reconstructed with the anti-$k_T$ algorithm with $R=0.3$;  
 we only count events in which the associated 
 jet has a transverse momentum $\pt^{\rm jet}>30$~GeV. 
 The widths of the colored  bands that illustrate our results in this Figure --- and in many Figures
 that follow --- incorporate both the uncertainty that comes from varying $T_c$ between 145 and 170 MeV
 and the uncertainty that comes from varying our model parameter $\kappa_{\rm sc}$ over
 its allowed range $0.32<\kappa_{\rm sc}<0.41$, determined in Section 2.   That is, we obtain four curves by repeating
 our calculation of the observable in question, here $x_{J\gamma}$, with
 $T_c$ and $\kappa_{\rm sc}$ each set to its lowest and its highest value, and plot the band
 that extends from whichever one of the four curves is lowest to whichever curve is highest at
 each point in the Figure.

 As a reference, in 
 Fig.~\ref{Fig:PJInv} we display the $x_{J\gamma}$ distribution in p-p collisions at the same nucleon-nucleon energy 
 as predicted by \pythia, {\it i.e.}~with no medium-induced parton energy loss. 
 In order to mimic the effects of jet-energy resolution on the transverse momentum
 of the associated jet obtained in the analysis of the Pb-Pb data after the subtraction of the background,
 in our p-p results 
 we have smeared the momenta of the associated jets obtained from our Monte Carlo
 calculation with a centrality-dependent Gaussian broadening.
 The parameters of this smearing, reported in Ref.~\cite{YYthesis}, 
 were tuned to reproduce the measured distributions 
 after embedding \pythia-generated photon-jet samples into real lead-lead events.
For this reason, the smeared proton-proton distribution is not identical in the left and right panels of Fig.~\ref{Fig:PJInv}, 
which display our results for Pb-Pb collisions with two different centralities. 
For the present, these smeared
p-p results are the correct reference to which 
both
the experimental results as measured by the CMS collaboration~\cite{Chatrchyan:2012nia} and 
our results for quenched jets in Pb-Pb collisions should be compared.  (See Ref.~\cite{ATLAS:CONF2012051} for 
fully unfolded experimental results to come.)
We apply the same smearing procedure to our simulated quenched jets, represented
by the colored band.
The smearing parameters are at present known only for $\sqrt{s}=2.76$~ATeV, not for $\sqrt{s}=5.02$~ATeV. For the present,
we have decided to employ the same smearing parameters at the higher collision energy. Our predictions therefore 
assume that the effects of background subtraction on these observables are similar at these two collision energies.

The differences between the (smeared) proton-proton Monte-Carlo data and the results of our
in-medium calculations displayed in Fig.~\ref{Fig:PJInv} are due to energy loss.  
The sample of isolated photons we have used to construct these distributions is 
dominated by prompt photons, which do not lose momentum when traversing the plasma.\footnote{The small fraction of fragmentation photons which fulfill the isolation requirement do suffer energy loss via the quenching of the partons from which they originate. This effect is small for this observable, but it can have consequences for more differential observables, as we will discuss below} 
The partons in the showers that become the 
associated jets, however, interact strongly with the medium and lose energy
according to (\ref{CR_rate}) as they propagate through it.  In some cases,
this pushes the transverse energy of the associated jet below 30~GeV, meaning
that the event does not get counted as a photon-jet in our Pb-Pb analysis although
it was counted in our p-p analysis.  This is why the integrals of  the curves
illustrating our  Pb-Pb results in Fig.~\ref{Fig:PJInv} are smaller than the
integrals of our p-p results. More on this below.
In other cases, when the transverse energy of the associated jet remains above 30 GeV,
the effect of the energy loss is to reduce $x_{J\gamma}$, displacing
the photon-jet imbalance distribution toward smaller values of $x_{J\gamma}$.

Keeping in mind that we fixed the single parameter in our hybrid model by comparing
it to the single-jet suppression $R_{AA}$, see Section 2, it is remarkable how well the
photon-jet imbalance distribution that we obtain from our model agrees with CMS
measurements in $\sqrt{s}=2.76$~ATeV collisions in both centrality bins in Fig.~\ref{Fig:PJInv}.
The fact that only one side of photon-jet pair loses energy makes the interpretation of
this observable cleaner than in the case of the dijet imbalance.  Of course, at present
the large statistical uncertainties in the photon-jet measurements is a limitation on their
use to differentiate between different model assumptions for the rate of energy loss $dE/dx$,
which is to say a limitation on their use as diagnostics of the mechanism of energy loss.
We illustrate this point in Appendix~\ref{Model_dep}, where we present the results that
we obtain by repeating our photon-jet and Z-jet calculations using the control models
for $dE/dx$ in (\ref{Eloss_equations}).  We find that these control models make predictions
that {\it are} distinct from those of our hybrid strong/weak coupling model, with $dE/dx$
given by (\ref{CR_rate}), but the distinctions are too small to be resolved by the present
data, with their statistical uncertainties.

Given these considerations, perhaps the most important results that we can use
our model to provide are our predictions for the upcoming runs of the LHC, at 
$\sqrt{s}=5.02$~ATeV, where the photon-jet data sets are expected to be 
larger by about an order of magnitude
and the statistical uncertainties are expected to be substantially smaller than at present.
For this reason, in the lower panels of Fig.~\ref{Fig:PJInv}  we show the predictions of our hybrid model, with its strongly
coupled form for the rate of energy loss, for the photon-jet imbalance distribution in 
Pb-Pb collisions at $\sqrt{s}=5.02$~ATeV. 
We have applied the same kinematical cuts used in current measurements (and our calculations)
at $\sqrt{s}=2.76$~ATeV.
The imbalance distribution shows little dependence on the collision energy.

\begin{figure}[tbp]
\centering 
\includegraphics[width=.49\textwidth]{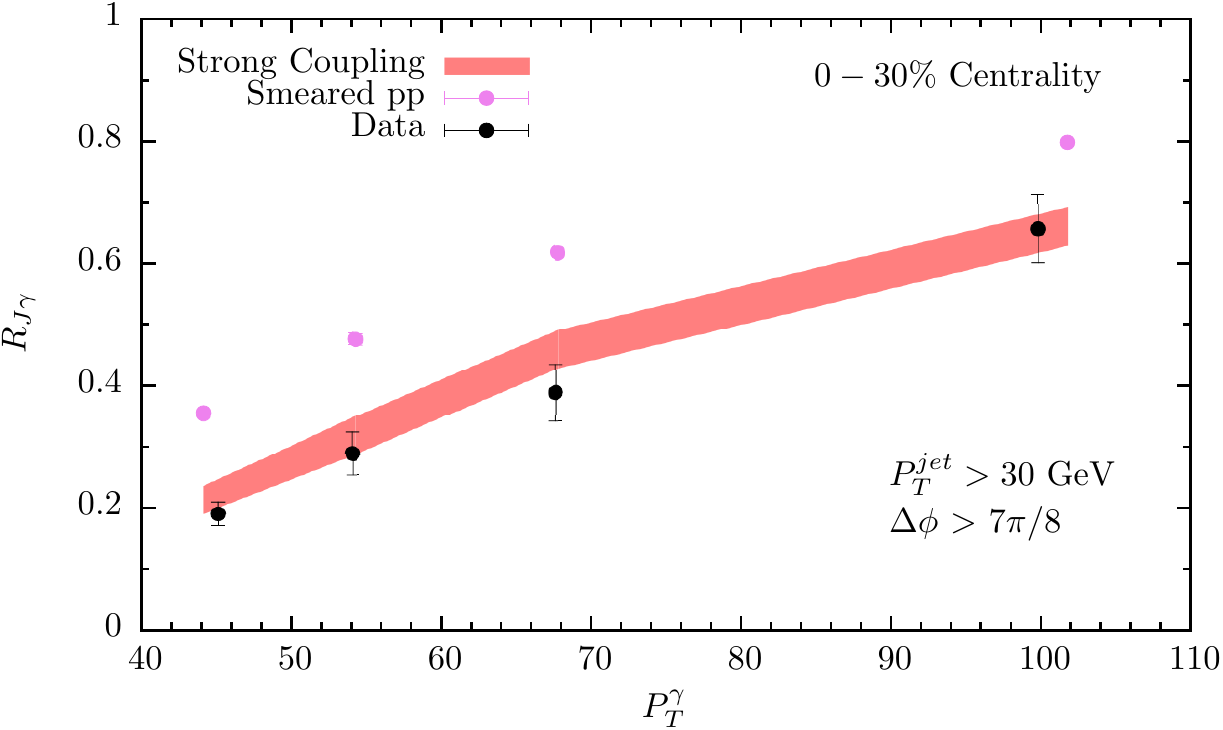}
\put(-176,90){\tiny{$\sqrt{s}=2.76~\rm{ATeV}$}}
\includegraphics[width=.49\textwidth]{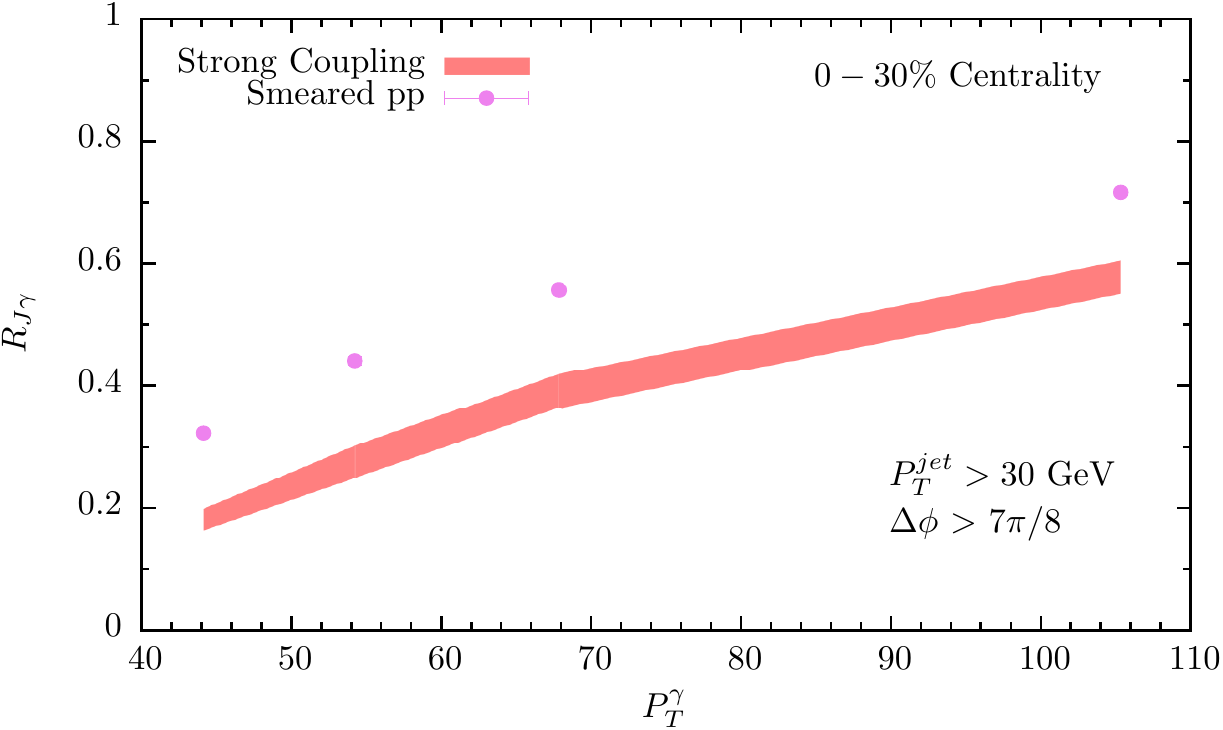}
\put(-176,90){\tiny{$\sqrt{s}=5.02~\rm{ATeV}$}}
\caption{\label{Fig:PJRJG}  
Fraction of events with an isolated photon in which we find a photon-jet pair, which is to say
in which we find an associated jet with $\pt^{\rm jet}>30$~GeV at an azimuthal angle
more than $7\pi/8$ away from that of the isolated photon.
We plot this fraction as 
a function of the photon transverse momentum, in collisions with $\sqrt{s}=2.76$~ATeV (left) and $\sqrt{s}=5.02$~ATeV (right). 
The colored band shows 
the results from our hybrid model, with its strongly coupled form for the rate of energy loss.
For comparison, the violet dots show the smeared p-p calculations (see text for details). 
Our results for $\sqrt{s}=2.76$~ATeV are compared to CMS data \cite{CMS-HIN-13-006}.
}
\end{figure}

To further test the success of our hybrid approach, we now turn to exploring other photon-jet observables. In Fig.~\ref{Fig:PJRJG} we show the fraction of isolated photons 
that come with
an associated jet, as reconstructed with the anti-$k_T$ algorithm
with $R=0.3$, that has $\pt^{\rm jet}$> 30 GeV and $\Delta \phi> 7 \pi /8$.
We plot this quantity, which we denote by  $R_{J \gamma}$, 
as a function of the transverse momentum of the isolated photon
for heavy ion collisions with 
$\sqrt{s}= 2.76$~ATeV (left) and $\sqrt{s}= 5.02$~ATeV (right).  
In the plots in Fig.~\ref{Fig:PJRJG}, the smeared proton-proton \pythia simulations are represented by the violet dots,
the results of our hybrid strong/weak coupling model are represented by the colored band, 
and the experimental results from Ref.~\cite{CMS-HIN-13-006}  for $\sqrt{s}= 2.76$~ATeV  are the black data points.\footnote{Note that 
the ratio of the integral of the photon-jet imbalance in Pb-Pb
collisions to that in p-p collisions, see Fig.~\ref{Fig:PJInv},
is the ratio of
$R_{J\gamma}$ 
for all photons with $\pt^\gamma>60$~GeV
in Pb-Pb collisions to that in p-p collisions. The fact
that the colored band lies below the violet dots in Fig.~\ref{Fig:PJRJG} was therefore foreshadowed
in Fig.~\ref{Fig:PJInv}.}
The broad $x_{J\gamma}$ distribution seen in Fig.~\ref{Fig:PJInv} implies that this ratio must be an increasing function of $\pt^\gamma$ for both collision energies, since the 30 GeV cut on the associated jet energy is more and more
easily satisfied as the momentum of the photon is made larger and larger. 
Quenching reduces $R_{J\gamma}$ since it pushes the energy of some of the
associated jets below 30 GeV.
As for the photon-jet imbalance distribution, 
we find good agreement between
the $R_{J \gamma}$ obtained from our hybrid model and that measured in present experiments with $\sqrt{s}=2.76$~ATeV,  in this case 
for all values of the photon momenta.\footnote{Note that in order to extend our calculations
of $R_{J\gamma}$ down to 40~GeV$<\pt^\gamma<$50~GeV we used a sample of Monte Carlo
events with the \pythia parameter $\pt^{\rm min}$ set to 20~GeV.} 
Our predictions for heavy ion collisions with $\sqrt{s}=5.02$~ATeV are shown in the right panel of Fig.~\ref{Fig:PJRJG}. 
As before, this observable shows little sensitivity to the collision energy, at least within these kinematical cuts. 

\begin{figure}[tbp]
\centering 
\begin{tabular}{cc}
\includegraphics[width=.5\textwidth]{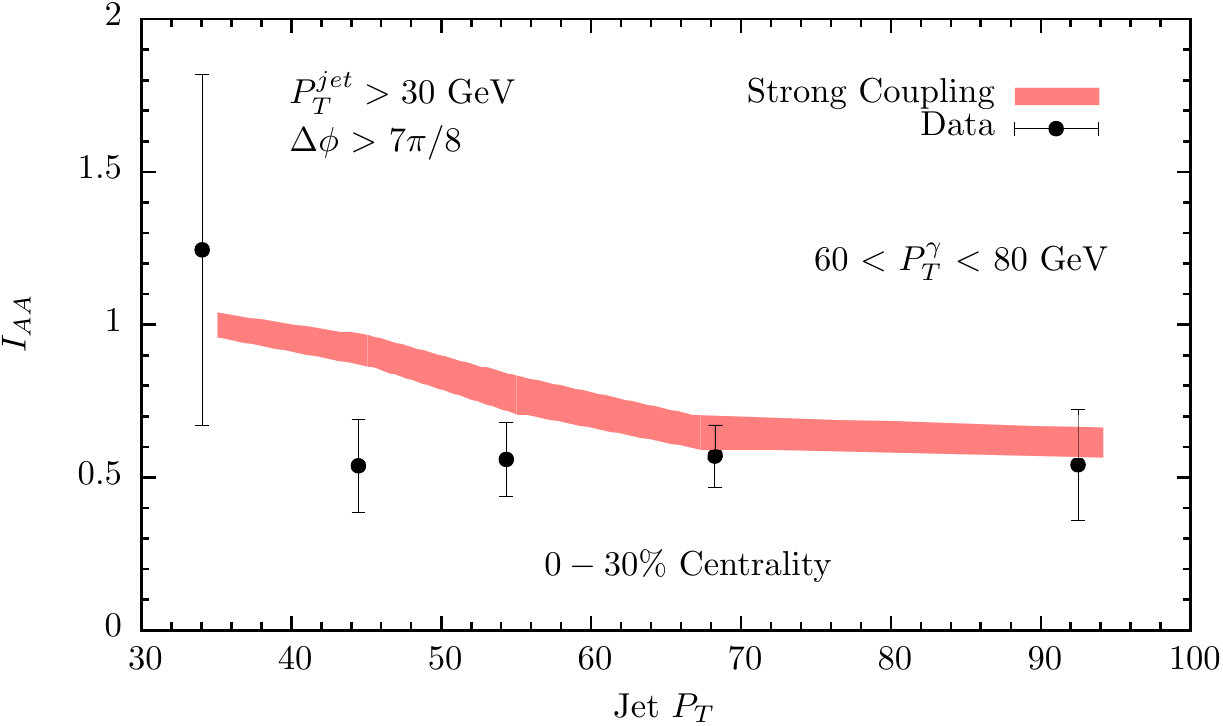}
\put(-165,90){\tiny{$\sqrt{s}=2.76~\rm{ATeV}$}}
&
\includegraphics[width=.5\textwidth]{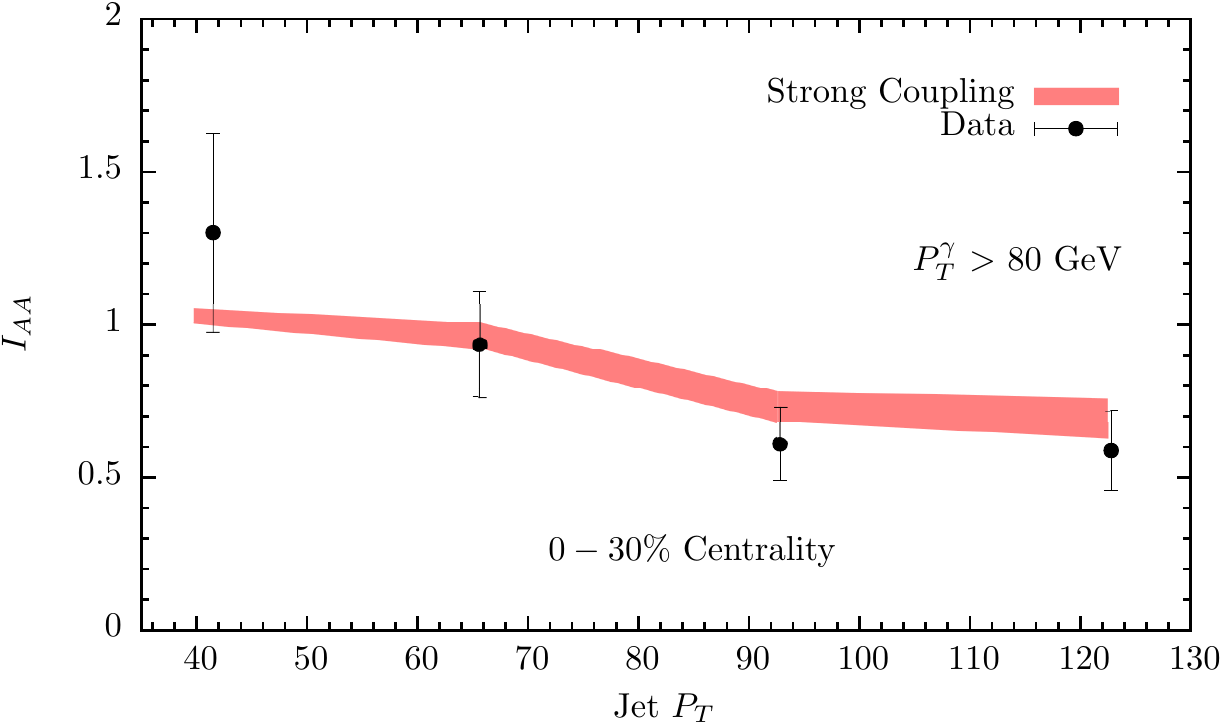}
\put(-165,90){\tiny{$\sqrt{s}=2.76~\rm{ATeV}$}}
\\
\includegraphics[width=.5\textwidth]{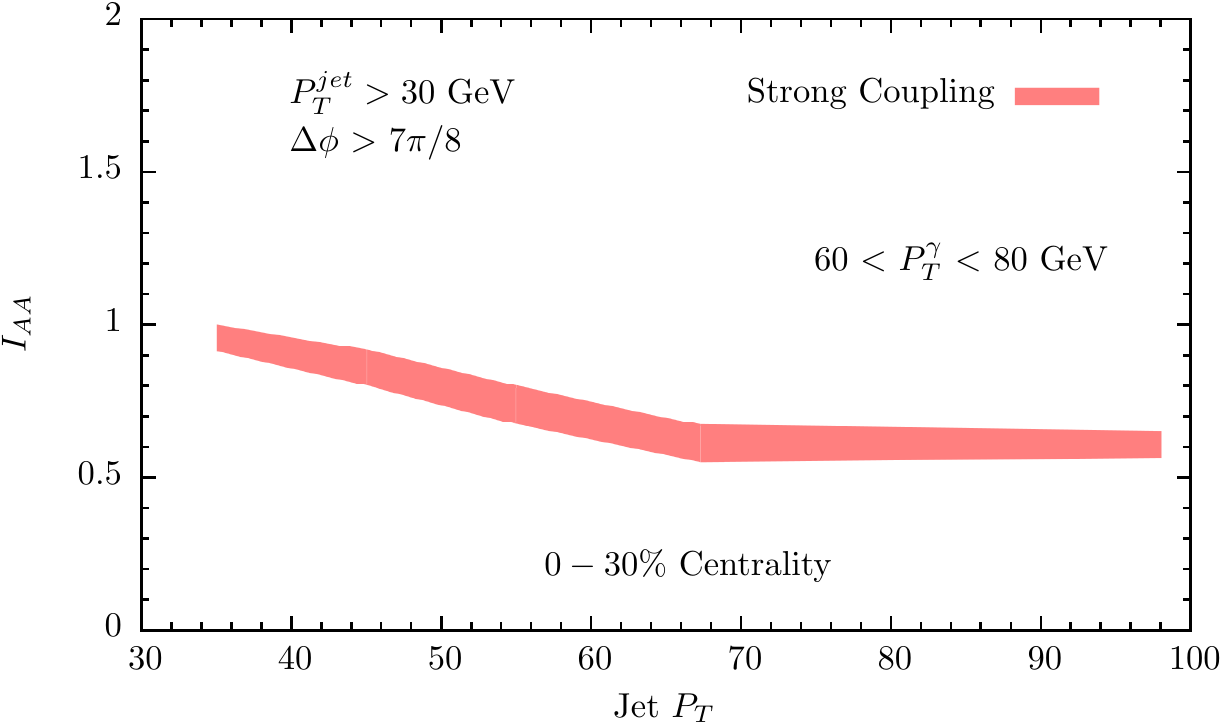}
\put(-165,90){\tiny{$\sqrt{s}=5.02~\rm{ATeV}$}}
&
\includegraphics[width=.5\textwidth]{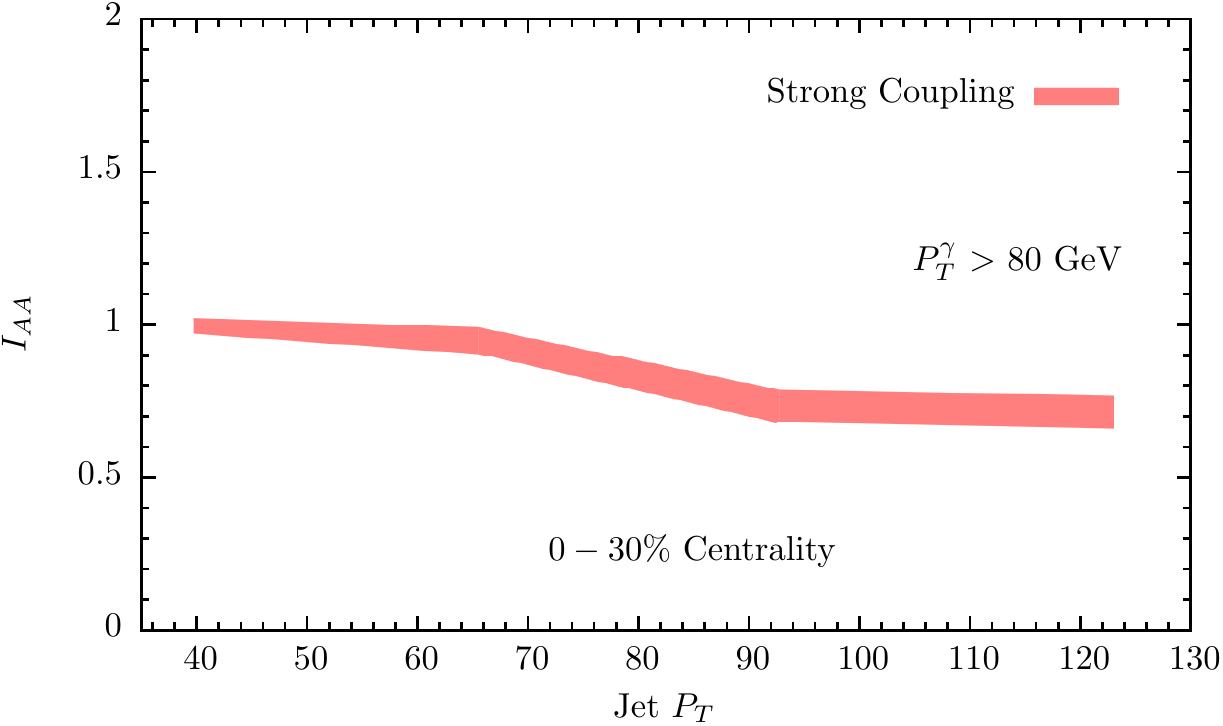}
\put(-165,90){\tiny{$\sqrt{s}=5.02~\rm{ATeV}$}}
\end{tabular}
\caption{\label{Fig:PJIAA}  
Ratio of the  transverse momentum spectra of jets associated with an isolated photon 
in Pb-Pb collisions to that in p-p collisions,
as a function of the 
jet transverse momentum, for two different ranges of the photon transverse momentum and for $\sqrt{s}=2.76$~ATeV (upper panels) and $\sqrt{s}=5.02$~ATeV (lower panels).  
 The hybrid model with its strongly coupled form for the rate of energy loss 
describes the available CMS data at $\sqrt{s}=2.76$~ATeV well.
}
\end{figure}

We have also analyzed the spectrum of jets produced in association with an isolated photon. 
In Fig.~\ref{Fig:PJIAA}, we show the ratio of the spectrum of associated jets in Pb-Pb collisions to that in proton-proton collisions, $I_{AA}$, 
for two different ranges of photon transverse momenta, 60~GeV$< \pt^\gamma<$80~GeV and $\pt^\gamma>80$~GeV. 
The observable $I_{AA}$, and in particular its suppression below 1, 
can be thought of as the photon-jet analogue of the single-jet suppression $R_{AA}$, but
instead of being constructed for inclusive jets as $R_{AA}$ is, $I_{AA}$ is constructed 
using only the associated jets back-to-back with an isolated photon.
This implies that the distribution of the jet energies as well as the fragmentation pattern and hence
the jet masses, of the p-p jets and of the Pb-Pb jets that enter into the calculations of 
$I_{AA}$ and $R_{AA}$ are different.  Furthermore, the distribution of the point in the
transverse plane at which the hard scattering event that produces a jet selected in
an $I_{AA}$ analysis is quite different from that for the jets selected in an $R_{AA}$ analysis.
It is therefore striking that even though we fitted the single parameter in our hybrid model
to a single measured value of $R_{AA}$, when we compare the results for the photon-jet
$I_{AA}$ obtained from the model
with data at $\sqrt{s}=2.76$~ATeV, displayed in the upper panels of In Fig.~\ref{Fig:PJIAA}, 
we see such good agreement, for both ranges of the photon energy and over the whole
range of jet $\pt$.
In the lower panels of Fig.~\ref{Fig:PJIAA}, we show the predictions from 
our hybrid model with its strongly coupled rate of energy loss 
for heavy ion collisions with $\sqrt{s}=5.02$~ATeV. As for the previous photon-jet observables, the spectrum of the associated jets hardly changes between these two collision energies.

The agreement between the predictions of our hybrid strong/weak coupling model, with
its \pythia branching and its strongly coupled form for the rate of energy loss, for all of these
photon-jet observables and the data available today is very encouraging.
Having fixed the single parameter in our model using a single measurement of $R_{AA}$ for 
inclusive jets, without introducing any new parameters 
we have obtained a good description of the experimental data for a total of 5 different observables, one involving inclusive jets, one
involving dijets, and three involving photon-jets, all with their centrality and energy dependence.
These observables sample different in-medium path length distributions of the quenched jets, 
different shapes of the original jet spectrum,
different fragmentation patterns and jet mass distributions, and different quark vs.~gluon compositions
of the observed jets.
Despite all these differences, our model is able to describe the systematics observed in all the data correctly. 
To avoid over-reaching in drawing conclusions, however, it is important to explore 
the predictions of the  different control models described in Section~\ref{sec:hybridmodel} for these observables. 
The results of this analysis can be found in Appendix \ref{Model_dep}. 
There are distinctions between the predictions of the control models and our hybrid model, but
these distinctions are small compared to the statistical uncertainties in present data.  
It is therefore not possible at present to use the agreement between our hybrid model and photon-jet data
to argue that the data favors a strongly coupled form for the rate of energy loss.
We therefore await the higher statistics data expected later this year in collisions with $\sqrt{s}=5.02$~ATeV, and
have provided the predictions of our hybrid model for all three photon-jet observables for collisions at this higher energy.
We have also provided such predictions for Z-jet observables below, in Section~\ref{sec:zjet_predictions}.
In addition, it is important to investigate the predictions of our hybrid model for intra-jet observables
like fragmentation functions, as we shall do in Section~\ref{Sec:FF}.   These are all paths toward
using jet data in the service of understanding the dynamics of the interaction between energetic
partons and strongly coupled plasma.

\subsection{\label{sec:zjet_predictions}Z-Jet Observables: Predictions for $\sqrt{s}=5.02$~ATeV}

In this section, we turn to the Z$^0$-jet observables that are complementary
to the photon-jet observables of the previous Section.
The large Z-mass ensures that 
Z-bosons seen in a heavy ion collision were almost
without exception produced promptly, in hard processes
dominated by short-distance physics: it is extremely unlikely
for a Z-boson to be produced during the fragmentation
of a parton in a jet, unless the jet energy is much higher than
is relevant to us.
The Z-boson production mechanism is therefore under good theoretical control.\footnote{Note, of course, that modifications of the nuclear parton distribution function in the 
relevant intermediate $x$-region can alter the production rate of Z-bosons, by modifying the composition of the initial flux of partons.}
Furthermore, because of their large width and short lifetime, Z-bosons decay during the very early stage of a
heavy ion collision, even prior to plasma formation. The Z-bosons that are identified in heavy ion collisions 
are those that decayed leptonically, in particular via
$Z^0\rightarrow \mu^+ \mu^-$. This means that their decay products do not interact
strongly with the pre-equilibrium partonic matter or with the strongly coupled plasma, once it forms.
Z-bosons are in this respect similar to prompt photons, making Z-jet events similar in
their utility to photon-jet events.
The Z-jet events come with
the added advantage 
almost all Z's are prompt Z's. 
As consequence, 
the experimental identification of Z-bosons via their $\mu^+\mu^-$ decays do not require isolation cuts, which leads to an arguably  cleaner determination of the associated jet energy.
The only disadvantage of Z-jet events is  that Z's are less numerous than photons.

We construct the same class of observables for Z-jet events that we constructed for photon-jet events in the previous Section.
Because of the low statistics of Z-boson production in $\sqrt{s}=2.76$~ATeV collisions, no constraining
measurements exist at present for the Z-jet observables that we shall construct.\footnote{For 
preliminary low-statistics measurements of Z-jet correlations at $\sqrt{s}=2.76$~ATeV, see Ref.~\cite{ATLAS_Zimbalance_conf}.}
We therefore present predictions for these observables in the $\sqrt{s}=5.02$~ATeV
heavy ion collisions coming soon in LHC heavy ion Run 2.  We have chosen the same kinematical
cuts for the Z-jet observables that we (and the CMS collaboration) have used in photon-jet observables, to
facilitate comparison between our results for the two cases.  Of course, once the experimentalists
decide on the cuts that they will use for their Z-jet analyses of the data-to-come, we can re-run
our analyses with their cuts.

 \begin{figure}[tbp]
\centering 
\begin{tabular}{cc}
\includegraphics[width=.5\textwidth]{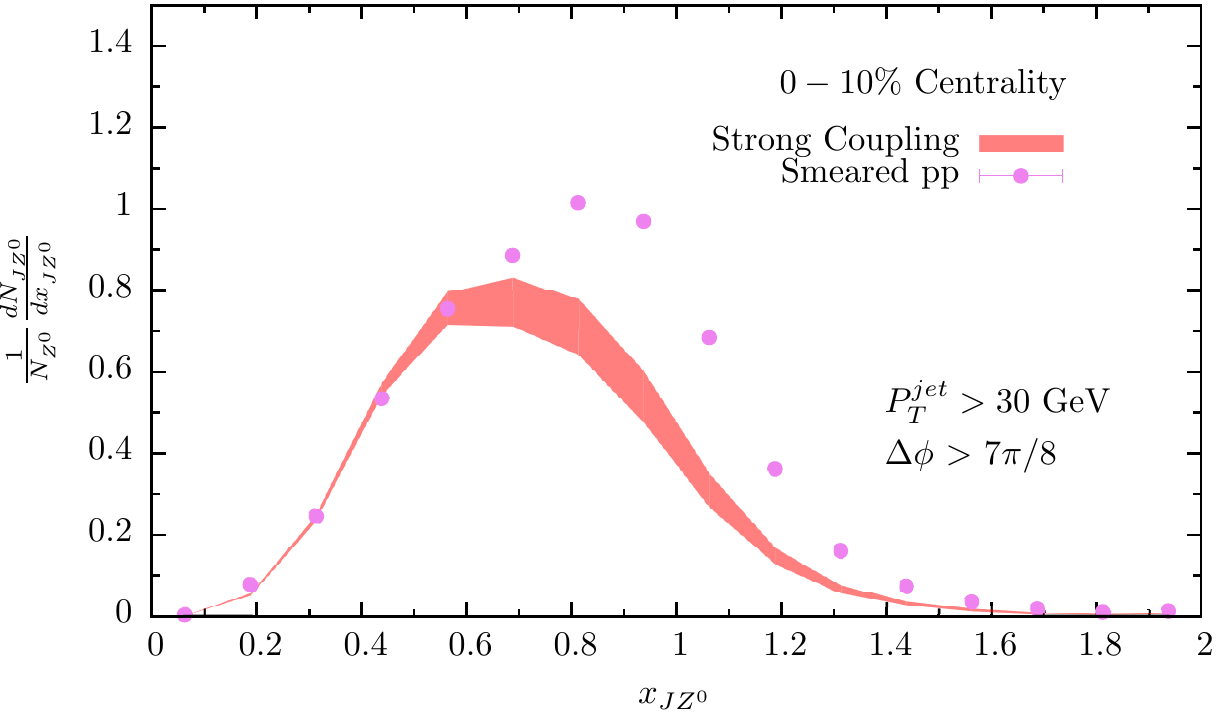}
\put(-172,99){\tiny{$\sqrt{s}=5.02~\rm{ATeV}$}}
&
\includegraphics[width=.5\textwidth]{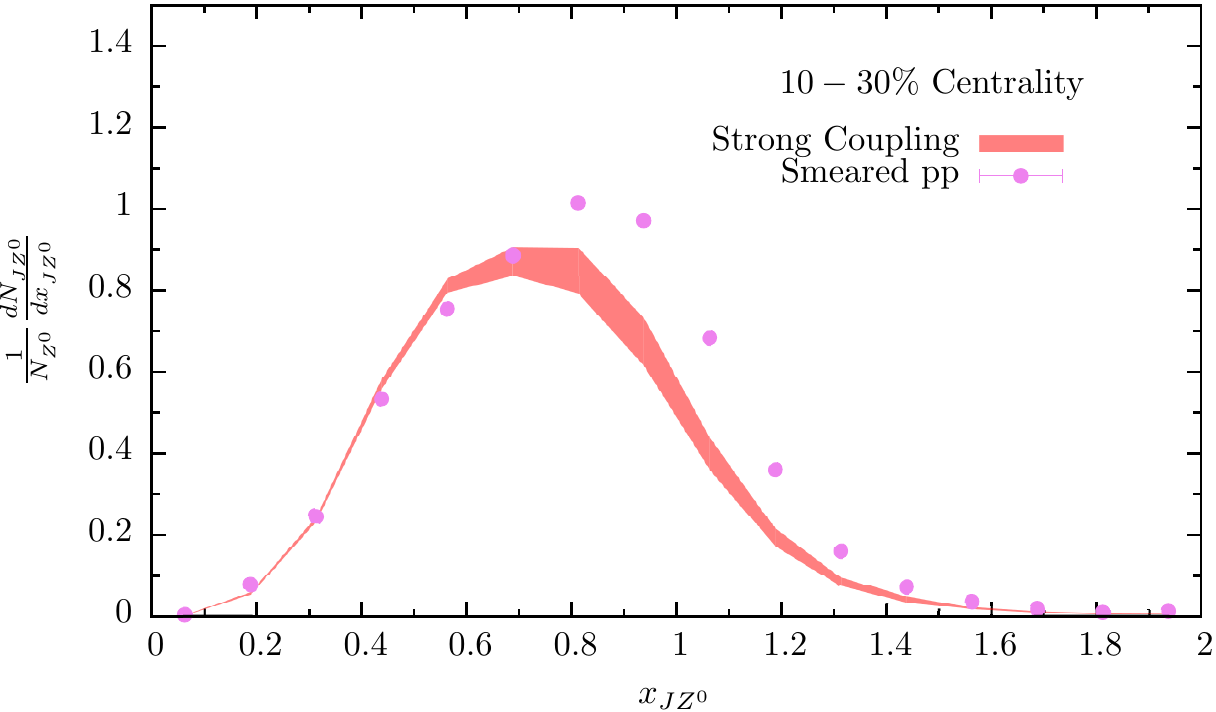}
\put(-172,99){\tiny{$\sqrt{s}=5.02~\rm{ATeV}$}}
\end{tabular}
\caption{\label{Fig:ZJInv}  
Distribution of the transverse momentum imbalance of Z-jet pairs,
$x_{J Z}\equiv \pt^{\rm jet}/\pt^{\rm Z}$, for Pb-Pb collisions with $\sqrt{s}=5.02$~ATeV.
The left and right panels show the 0-10\% and 10-30\% most central events, respectively.
Here and below, the colored
bands show the results from our hybrid model, with its strongly coupled form
for the rate of energy loss and the violet dots show the smeared p-p
calculations for comparison.
}
\end{figure}

In Fig. \ref{Fig:ZJInv}, we show the distribution of the Z-jet imbalance observable
$x_{JZ}\equiv \pt^{\rm jet}/\pt^{\rm Z}$, in heavy ion collisions with two different
ranges of centrality with  $\sqrt{s}=5.02$~ATeV.
In both cases the colored band shows the predictions of our hybrid model, with its
strongly coupled form (\ref{CR_rate})  for the rate of energy loss, for Pb-Pb collisions and the violet
dots show the distribution of $x_{JZ}$ for p-p collisions as predicted by \pythia.
As we did in our analysis of photon-jet observables, 
we have smeared the momenta of the associated jets
in both our Pb-Pb and p-p calculations.
For the present, before better guidance becomes available once
experimentalists have begun the analysis of LHC Run 2 data, we have
used the same smearing functions here as we (and the CMS collaboration)
used for the photon-jet observables that we discussed in the previous Section.
We obtained the $x_{JZ}$ distributions in Fig.~\ref{Fig:ZJInv} from a sample of events in which we required a Z-boson 
with $\pt^Z>60$~GeV and an associated jet reconstructed using the anti-$k_T$ algorithm
with $R=0.3$ 
that has $\pt^{\rm jet}>30$~GeV and is separated in azimuthal angle from the Z-boson by $\Delta \phi> 7 \pi/8$. 
We required that both the Z and the associated jet have $\left| \eta\right  | <1.6$. 
As in the case of the $x_{J\gamma}$ distribution in photon-jet events, 
the $x_{JZ}$ distribution is broad in p-p collisions, 
indicating the importance
of 
events with a Z and two jets, in particular those arising from initial state radiation.  This means
that at present Z-bosons are not substantially better as taggers of the associated jet energy
than photons are, which motivates the future development of methods to suppress events with
more than one jet in the final state.
Because the transverse momentum of the $Z$-boson and its $\mu^+\mu^-$ decay products
do not change in the medium, the difference between the Pb-Pb distribution and the p-p
distribution in our calculation is entirely due to the energy lost by the partons in the
associated jet in the Pb-Pb collisions due to their passage through the strongly coupled
plasma.
As for photon-jet events, we see a reduction in the integral of $x_{JZ}$, analyzed further below,
and a displacement of the distribution toward smaller $x_{JZ}$.  The
magnitude of this displacement is comparable to the corresponding 
shift in the $x_{J\gamma}$ distribution in photon-jet events, see Fig.~\ref{Fig:PJInv}.
As in that case, there are clear but small distinctions between the results
we obtain with our hybrid model, shown in Fig. \ref{Fig:ZJInv}, and those
we obtain when we use our control models for the rate of energy loss (\ref{Eloss_equations}) instead.
We present these in Appendix~\ref{Model_dep}.

\begin{figure}[tbp]
\centering 
\includegraphics[width=.5\textwidth]{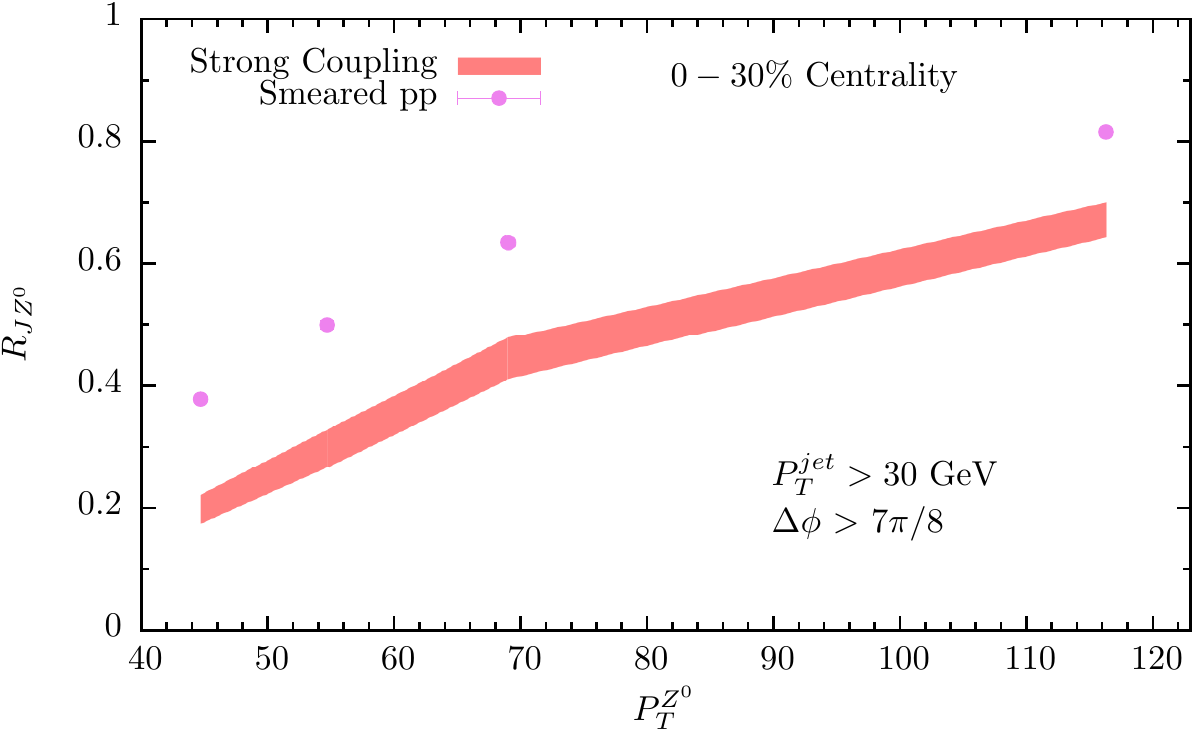}
\put(-174,100){\tiny{$\sqrt{s}=5.02~\rm{ATeV}$}}
\caption{\label{Fig:ZJRJG}  
Fraction of events with a Z-boson in which we find a Z-jet pair, which is to say in which
we find an associated jet with $\pt^{\rm jet}>30$~GeV at an azimuthal angle more than
$7\pi/8$ away from that of the Z-boson, 
in collisions with $\sqrt{s}=5.02$~ATeV.  
}
\end{figure}

In Fig.~\ref{Fig:ZJRJG}, we compute the 
fraction of Z-bosons in our sample
that come with an associated jet, as reconstructed with the anti-$k_T$ algorithm
with $R=0.3$, that has $\pt^{\rm jet}>30$~GeV and $\Delta\phi>7\pi/8$. We plot 
this quantity, which we denote by $R_{JZ}$, as a function of the transverse momentum
of the Z-boson.  We show the results obtained from our smeared proton-proton
\pythia simulations as the violet dots and the predictions for Pb-Pb collisions
from our hybrid strong/weak coupling model as the colored band.
As in the photon-jet case, 
as the partons in the associated jet shower lose energy 
the total energy of the associated jet can drop below our $\pt^{\rm jet}=30$~GeV cut, meaning that
energy loss leads to a reduction of the Z-jet yield as a fraction
of the number of Z-bosons.  This makes the integral under the
colored bands in Fig.~\ref{Fig:ZJInv} less than that under the violet
dots there, and it pushes the colored band in Fig.~\ref{Fig:ZJRJG} below the violet dots. 
The qualitative behavior and magnitude of this reduction is comparable to the reduction in  
the analogous photon-jet observable, see Fig.~\ref{Fig:PJRJG}.

\begin{figure}[tbp]
\centering 
\begin{tabular}{cc}
\includegraphics[width=.5\textwidth]{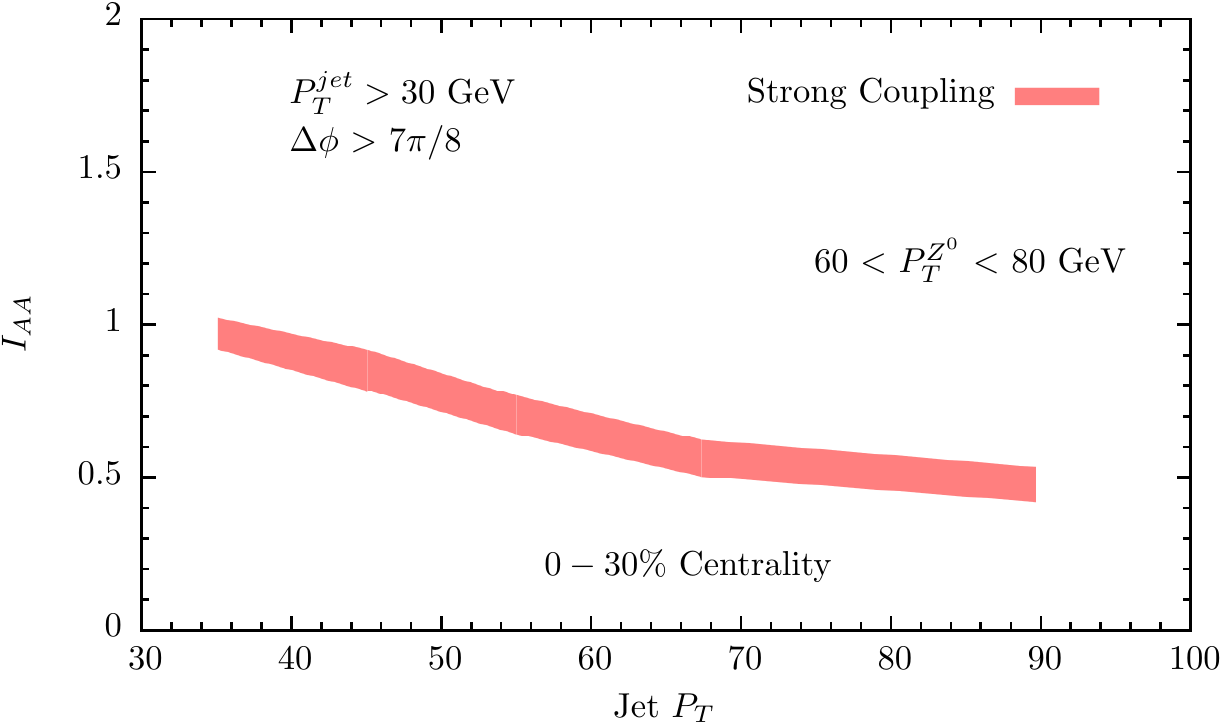}
\put(-165,90){\tiny{$\sqrt{s}=5.02~\rm{ATeV}$}}
&
\includegraphics[width=.5\textwidth]{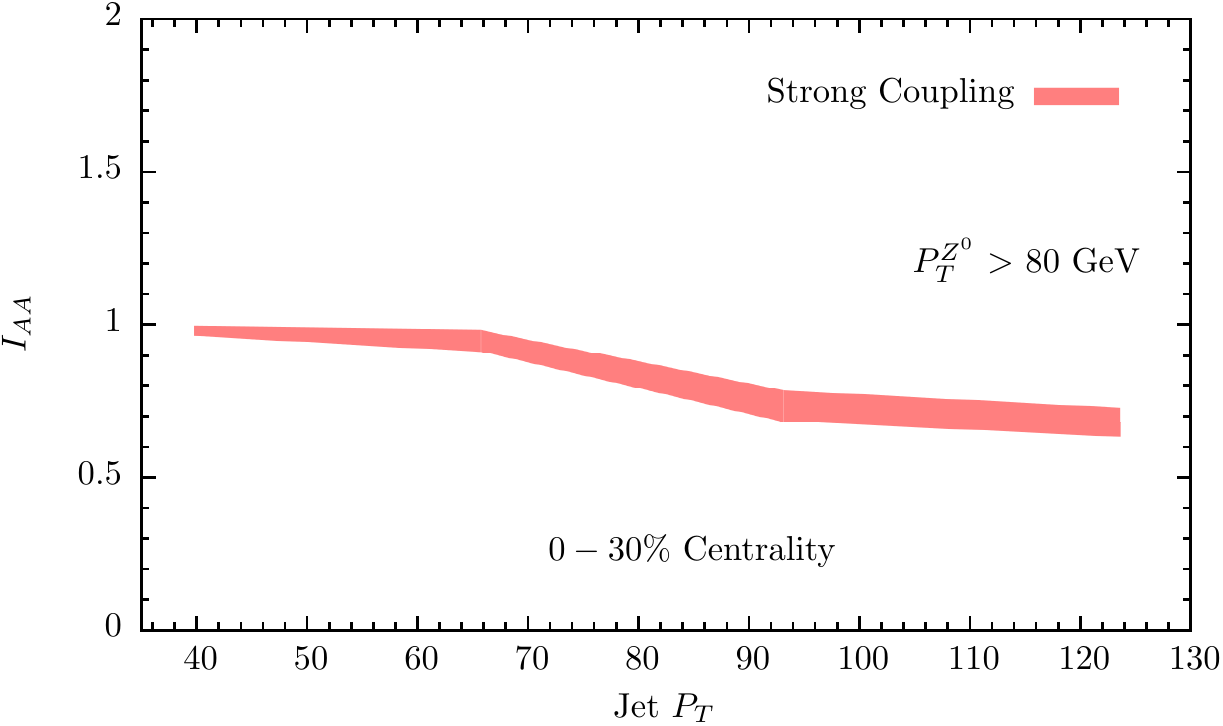}
\put(-165,90){\tiny{$\sqrt{s}=5.02~\rm{ATeV}$}}
\end{tabular}
\caption{\label{Fig:ZJIAA}  
Ratio of the transverse momentum spectra of jets associated with
a Z-boson in Pb-Pb collisions with $\sqrt{s}=5.02$~ATeV to that in p-p collisions as a function 
of the jet transverse momentum, for two different ranges of the Z-boson
transverse momentum.
}
\end{figure}

Finally, in Fig.~\ref{Fig:ZJIAA} we analyze the spectrum of jets produced in association with a Z-boson.
We show the ratio of the spectrum of associated jets in Pb-Pb collisions to that in p-p
collisions, $I_{AA}$, for two different ranges of Z-boson transverse momenta, 
60~GeV$<\pt^{\rm Z}<$80~GeV and $\pt^{\rm Z}>$80~GeV.
The effect of energy loss on this observable is again comparable to its effects on
the analogous photon-jet observable, see Fig.~\ref{Fig:PJIAA}.

In Appendix~\ref{Model_dep}, we repeat the analysis of $R_{JZ}$ and $I_{AA}$
for jets produced in association with a Z-boson in our control models, where
we use the expressions (\ref{Eloss_equations}) for the rate of energy loss.
We find that these observables exhibit little sensitivity to the form of $dE/dx$,
meaning little sensitivity to the microscopic dynamics via which the partons
in the jet shower interact with the strongly coupled plasma.

In Sections \ref{sec:photonjet_results} and \ref{sec:zjet_predictions}, we have provided
predictions for three photon-jet observables and three Z-jet observables 
in heavy ion collisions with $\sqrt{s}=5.02$~ATeV
obtained
from our hybrid strong/weak coupling model, with fragmentation taken from \pythia
and a rate of energy loss (\ref{CR_rate}) as at strong coupling.  We await the
data that will come from LHC heavy ion Run 2 with considerable anticipation.

\section{Fragmentation Functions}
\label{Sec:FF}

We now turn to the analysis of a more differential class of jet observables, namely fragmentation functions. 
We saw in Ref.~\cite{Casalderrey-Solana:2014bpa} and have confirmed in Fig.~\ref{Fig:FFdijet276}
of Appendix~\ref{update} that the predictions of our hybrid model and of our two control
models for the ratio of the partonic fragmentation function of inclusive jets in PbPb
collisions to that in p-p collisions are distinct for the three models.  This motivates the
hope that the higher statistics measurements expected from the coming LHC run may
serve to distinguish between models.  And, it motivates us to compute the
predictions of all three models for the partonic fragmentation functions of jets produced in
association with an isolated photon or a Z-boson.  We shall present
the results of these calculations in Section~\ref{sec:FFBJ}.
We were initially surprised to see that the partonic fragmentation
function ratio for jets produced in association with bosons turn out
to be more similar for the three models we are considering than was
the case for inclusive jets. Understanding this effect, which we shall do
by the end of Section~\ref{sec:FFBJ}, leads us, in Section~\ref{sec:ff_dijet},
to introduce a new observable constructed from the fragmentation functions
of jets in dijet pairs in Pb-Pb collisions, without the need for any p-p reference.
This turns out to be the observable that is most effective at differentiating
among our three models of any observable that we have considered to date.

Unlike the more inclusive observables described in the 
previous Section, the hadronic fragmentation functions that
experimentalists measure are quite sensitive to hadronization
effects.  
However,  the dynamics of hadronization, even in vacuum,  are not under full
theoretical control. In fact, the predictions of different Monte Carlo event generators
for fragmentation functions can differ among themselves, and in comparison
with p-p data, by as much as
20\% 
as shown in Refs.~\cite{atlas_ff_comp, d'Enterria:2013vba}. 
Since the modification of the fragmentation functions in Pb-Pb collisions with respect to those in p-p collisions
is itself on the order of several tens of percent at most (see the measurements reported
in Ref.~\cite{Chatrchyan:2012gw,Chatrchyan:2014ava,FFATLAS:HIN}) 
and since there are differences in hadronization dynamics (in particular, differences in the patterns of color recombination)
for jets in Pb-Pb and p-p collisions~\cite{Beraudo:2011bh,Beraudo:2012bq,Aurenche:2011rd}, it will be challenging to compare the ratios of {\it partonic} fragmentation
functions in Pb-Pb collisions to those in p-p collisions --- ratios which we calculate in
our models in Section~\ref{sec:FFBJ} ---  to data.  The in-medium effects of interest are
comparable in magnitude to the known uncertainties coming from our lack of understanding
of hadronization dynamics.  This is strong further motivation for the importance of the
observable that we introduce in Section~\ref{sec:ff_dijet}: the ratio of the fragmentation
function of inclusive jets in Pb-Pb collisions to that of the associated jets in dijet
pairs.  Since this is the ratio of the fragmentation functions of two different classes
of jets in Pb-Pb collisions, with no need for a p-p reference, many hadronization
uncertainties will cancel.

We shall restrict our calculations to partonic fragmentations
throughout this Section, meaning that our calculations are not
sensitive to hadronization and so are not affected by its challenges.
The observable that we introduce in Section~\ref{sec:ff_dijet} is the
one in this Section for which this will to the greatest degree possible
also be true in experimental data.  For this observable, as for
those we calculate in Section~\ref{sec:FFBJ}, however, the fragmentation
functions describing fragments with the lowest $\pt$'s cannot
be described reliably by our hybrid model or by the two control
models because none of these models include the contribution
to the low-$\pt$ component of a jet arising from the wake 
in the recoiling plasma that the jet plowing through it 
produces~\cite{Wang:2013cia,CasalderreySolana:2004qm,Chesler:2007an,Tachibana:2014lja,He:2015pra,Iancu:2015uja}.

\subsection{\label{sec:FFBJ}Fragmentation Functions of the Associated Jets in Photon-Jet and Z-jet Pairs}

Fragmentation functions are defined as the distribution of hadrons within a jet with a given fraction $z$ of the total 
longitudinal momentum $p_\parallel$ of the jet.  Longitudinal, here, means in the direction of the jet axis.
In p-p collisions, where most of the 
activity in events with hard jets comes from hadrons produced via the fragmentation and subsequent
hadronization of the virtual partons produced in an initial elementary partonic collision,
fragmentation functions
provide us with information about how the showering process via which the large virtuality of the
initial partons relaxes takes place. 
Furthermore, since final state effects are negligible, $p_\parallel$ obtained from the energy of a jet reconstructed with a sufficiently large reconstruction radius 
provides a good proxy of the initial energy of the hard parton that fragmented to form the jet.
This means that
the $z$-fraction of the final fragments are directly related to one of the QCD evolution variables. 
This is not the case in Pb-Pb collisions. 

We have already mentioned one of the complications in Pb-Pb collisions: at low $\pt$, some of the particles
in a reconstructed jet did not originate from the initial hard parton that was produced in the initial
elementary partonic collision.  Some soft particles reconstructed in the jet come, instead,
from the hadronization of moving quark-gluon plasma, set in motion by the momentum that
the jet passing through it transfers to the medium through which it is passing.  Operationally, this
enters the analysis because in analyzing jets in Pb-Pb collisions it is necessary and standard
to do a background subtraction to remove the hadrons formed from the quark-gluon plasma,
and this background subtraction procedure is based upon the assumption that the momenta
of these hadrons is uncorrelated with the direction of the jet.  To the extent that this is the case,
the background subtraction removes, on a statistical basis, particles in the jet cone that
are not part of the jet itself.  However, since the interaction of the jet with the plasma
transfers momentum to the plasma, this back reaction (or recoil) effect means that the
background subtraction procedure cannot remove all the particles from the medium:
there is no way to disentangle all of them from the products of
the jet shower; some of them must end up incorporated into jet observables. 
This means that, in Pb-Pb collisions, fragmentation functions at low $\pt$ are
sensitive to physical processes other than jet fragmentation.
Addressing these additional physical processes requires a dynamical treatment
of the response of the medium to the passage of the jet, which is beyond our
current model implementation.  For this reason, the results of our calculations
become less reliable at small $z$.  In this section, we will look at jets with  
$\pt>30$~GeV, meaning that $p_\parallel > 30$~GeV.  So, $\ln(1/z)=2.7$ or $\ln(1/z)=3.5$ corresponds
to fragments with $\pt>2$~GeV or $\pt>1$~GeV.  We will plot our
results out to larger values of $\ln(1/z)$, smaller values of $z$, 
but the effects of medium recoil that we are not including become
more and more important for $\ln(1/z)\gtrsim 3$. In
Section~\ref{sec:ff_dijet} where we consider inclusive jets and dijets,
rather than jets produced in association with a boson, we will look
at jets with $\pt > 80$~GeV, meaning that our results there will
be reliable out to somewhat larger values of $\ln(1/z)$.

There is a second complication in interpreting jet fragmentation
functions in Pb-Pb collisions, and this is that the $p_\parallel$ of
a jet, as reconstructed from the final state, is less than the
energy of the initial hard parton that fragmented into the jet
because the partons in the jet have lost energy as they propagate
through the strongly coupled medium produced in the  Pb-Pb collision.
At least some of the ``lost'' energy (according to the data~\cite{Chatrchyan:2011sx,CMS:2014uca}, a
significant amount of it) ends up as soft particles moving at large, random,
angles relative to the jet axis, and is not included when the
jet is reconstructed.  
Therefore, even when they are reconstructed with a large reconstruction radius, the total energy of the quenched in-medium jets 
is smaller than the total energy of the hard partons originating from an elementary partonic collision. 
This means that when a jet is reconstructed in a Pb-Pb collision, it is impossible to
make an experimental determination of the energy of the initial hard parton.  Consequently,
if one constructs a fragmentation function using the standard definition of $z$, namely the ratio between the momentum
of the momentum of an individual hadron to the $p_\parallel$ of the whole jet as reconstructed, this $z$
is not directly related to the evolution variable in a DGLAP shower and, more generally, is simply hard to
interpret.
We will nevertheless report the predictions of our hybrid model and the two control models
for fragmentation functions computed in this standard way, for comparison to future data.

Because we are looking at jets produced in association with a photon or Z-boson, however,
there is an obvious alternative.  We can define $z_\gamma$ or $z_Z$ (we shall denote
these variables  generically as $z_B$) as the ratio
of the momentum of an individual hadron in the jet to the momentum of the $\gamma$ or Z-boson, using
the momentum of the electroweak boson in the event (which cannot have lost any
energy since it does not interact with the quark-gluon plasma)
as a proxy for the momentum of the initial hard parton
that later fragmented and lost energy, forming the jet that the experimentalists reconstruct\footnote{
See Ref.~\cite{Wang:2013cia} for a similar definition of the scaling variable.}.
This is an improvement but it is not a panacea: we have seen in the previous Section that 
even in p-p collisions there is a broad distribution of the momentum imbalance between
a boson (photon or Z) and the associated jet in a boson-jet event.  One significant contributor to this imbalance
is the fact that in many events the elementary hard scattering process produces one photon or Z plus
more than one hard parton, not just one.  Regardless, this imbalance is 
entirely due to perturbative vacuum QCD physics, not to any in-medium effects. 
This means that even when we construct fragmentation functions using $z_\gamma$ or $z_Z$, 
we are not reliably dividing by the actual momentum of the initial hard parton that
fragmented into the jet we are looking at.
Still, by using $z_\gamma$ or $z_Z$ in Pb-Pb collisions we are using a variable that is as good a proxy
for what we want as is the case in the standard fragmentation function  in p-p collisions.

\begin{figure}[tbp]
\centering 
\begin{tabular}{cc}
\includegraphics[width=.5\textwidth]{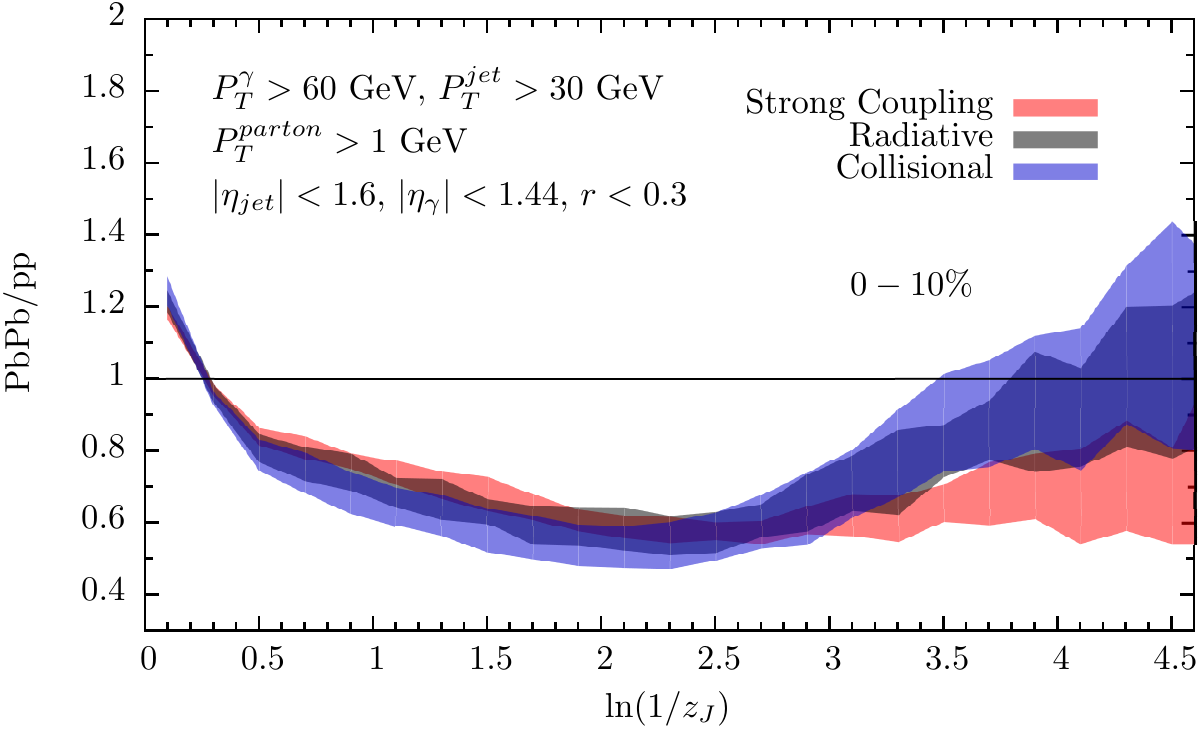}
\put(-180,80){\tiny{$\sqrt{s}=5.02~\rm{ATeV}$}}
&
\includegraphics[width=.5\textwidth]{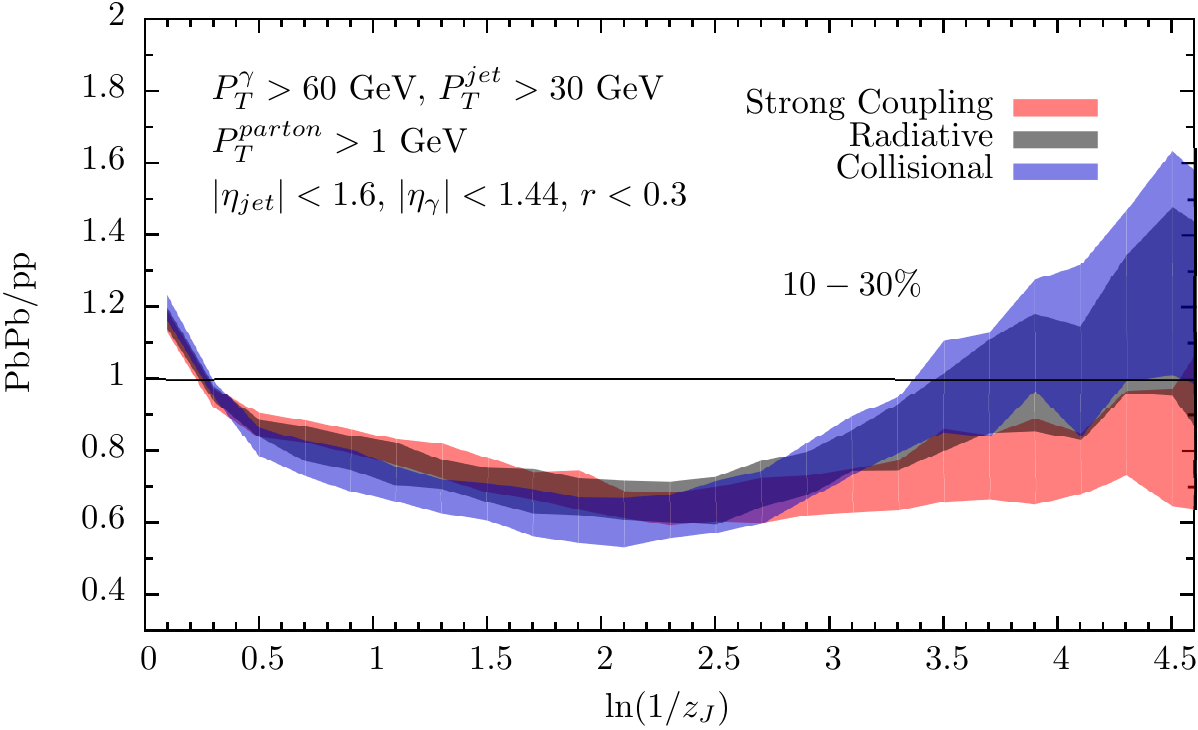}
\put(-180,80){\tiny{$\sqrt{s}=5.02~\rm{ATeV}$}}
\\
\includegraphics[width=.5\textwidth]{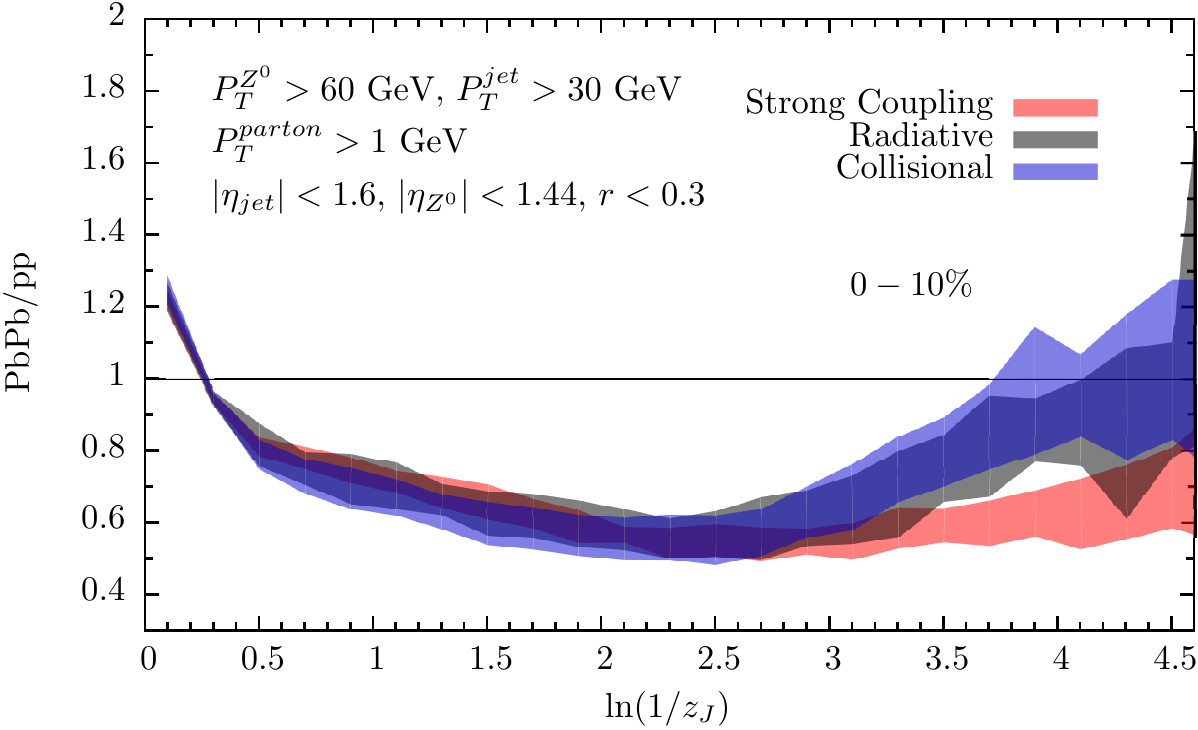}
\put(-180,80){\tiny{$\sqrt{s}=5.02~\rm{ATeV}$}}
&
\includegraphics[width=.5\textwidth]{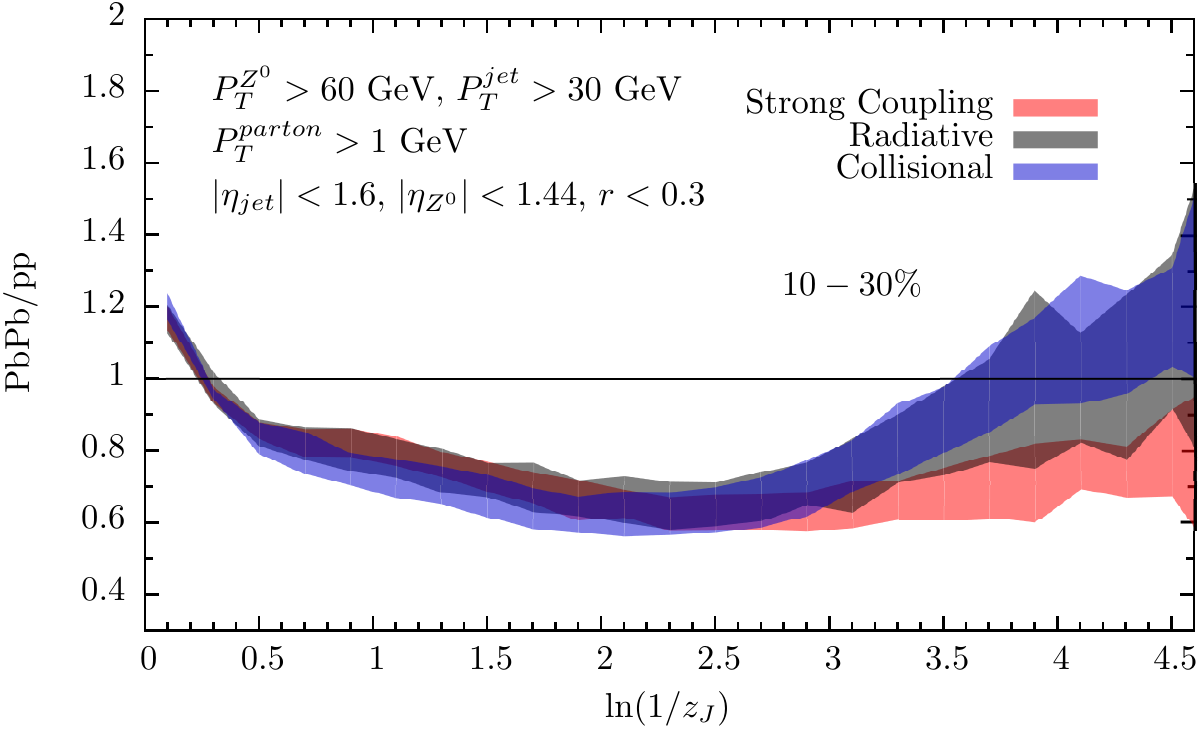}
\put(-180,80){\tiny{$\sqrt{s}=5.02~\rm{ATeV}$}}
\end{tabular}
\caption{\label{Fig:FFJ}  
Predictions of our hybrid model, with strongly coupled energy loss for
the partons in a \pythia shower, as well as our two control models 
for the partonic fragmentation function ratios (fragmentation function for jets in Pb-Pb collisions 
over that for jets in p-p collisions) 
for jets produced in association with an isolated photon (upper panels) 
or a Z-boson (lower panels) in Pb-Pb collisions at $\sqrt{s}=5.02$~ATeV
at two centralities (left and right panels).
The fragmentation functions are constructed   with respect to the
variable $z_J =p_\parallel^{\rm parton}/p_\parallel^{\rm jet}$.} 
\end{figure}

We first compute the  fragmentation functions constructed with respect to the reconstructed energy of the jet produced in association
with a photon or Z-boson. (We use the standard variable $z$, but here we denote it $z_J$ to emphasize
that its denominator is the momentum of the reconstructed jet.  It is defined by $z_J\equiv p_\parallel^{\rm parton}/p_\parallel^{\rm jet}$.) 
In Fig.~\ref{Fig:FFJ} we plot the ratio of the fragmentation functions in Pb-Pb to p-p as obtained in our framework for jets of $\pt>30$ GeV produced in association with isolated photons (upper panels) and Z-bosons (lower panels) with $\pt>60$ GeV.  
We have performed these simulations for two different Pb-Pb centralities, $0-10$~\%, shown on the left panels 
and $10-30$~\%, shown on the right.
Both photons and Z-bosons are required to have $\left| \eta \right| <1.4$ while the jets are constrained to $\left| \eta_{\rm jet} \right| <1.6$. 
The energy and direction of the jet axis are reconstructed with the anti-$k_T$ algorithm with radius $R=0.3$. 
Following the experimental analyses of inclusive fragmentation functions in Ref.~\cite{Chatrchyan:2014ava}, 
these fragmentation functions are constructed by including all particles surrounding the jet axis 
within an angular distance in $\eta$-$\phi$ space of $r=0.3$. 
Since our simulations do not include the underlying event, these correspond to particles from the hard scattering
process that remain correlated with the jet direction.  (As noted above, in the analysis of experimental data
a background subtraction is done, but soft particles from the plasma that is in motion following the jet
will not be subtracted. This means that our calculations should not reproduce the data-to-come at
large values of $\ln(1/z_J)$.)

The general features of the fragmentation function ratios plotted in Fig.~\ref{Fig:FFJ} 
are very similar to those that we found for inclusive jets in Ref.~\cite{Casalderrey-Solana:2014bpa} and have confirmed in Fig.~\ref{Fig:FFdijet276}
of Appendix~\ref{update}. 
All the models display an enhancement of the hardest part of the fragmentation function in Pb-Pb collisions relative
to p-p collisions. 
This is a generic behavior of any mechanism that removes soft particles
from the jet, either via energy loss as here or via deflecting them into a direction
far from that of the jet~\cite{CasalderreySolana:2010eh}.
Removing soft particles
increases the fraction of jets with a 
few hard fragments, which leads to the increase in  the hard part of the fragmentation function.  
For all models there is also a depletion in the Pb-Pb fragmentation function
at intermediate $z_J$. This is the expected result from quenching, which tends to reduce the energy of the fragments that propagate in plasma. 
Remarkably, for the energy range of bosons and jets explored in those figures, and with our current uncertainties, the pattern of fragmentation 
at large and intermediate $z_J$ is indistinguishable among the three models we explore, despite their very different path length and energy dependences. 
We comment further on this below.
At smaller values of $z_J$, the hybrid model with its strongly coupled energy loss suppresses soft fragments more than 
the control models. However, this separation between models occurs in a regime where the fragments have momenta smaller than 2 GeV, meaning that
our calculations of fragmentation functions are not reliable there.  Adding in the contributions from a medium that has picked
up momentum from the jet passing through it, meaning that it is not completely removed by the background subtraction,
would push the Pb-Pb fragmentation functions up in this soft region by an amount that our model does not permit us to estimate
at present.

\begin{figure}[tbp]
\centering 
\begin{tabular}{cc}
\includegraphics[width=.5\textwidth]{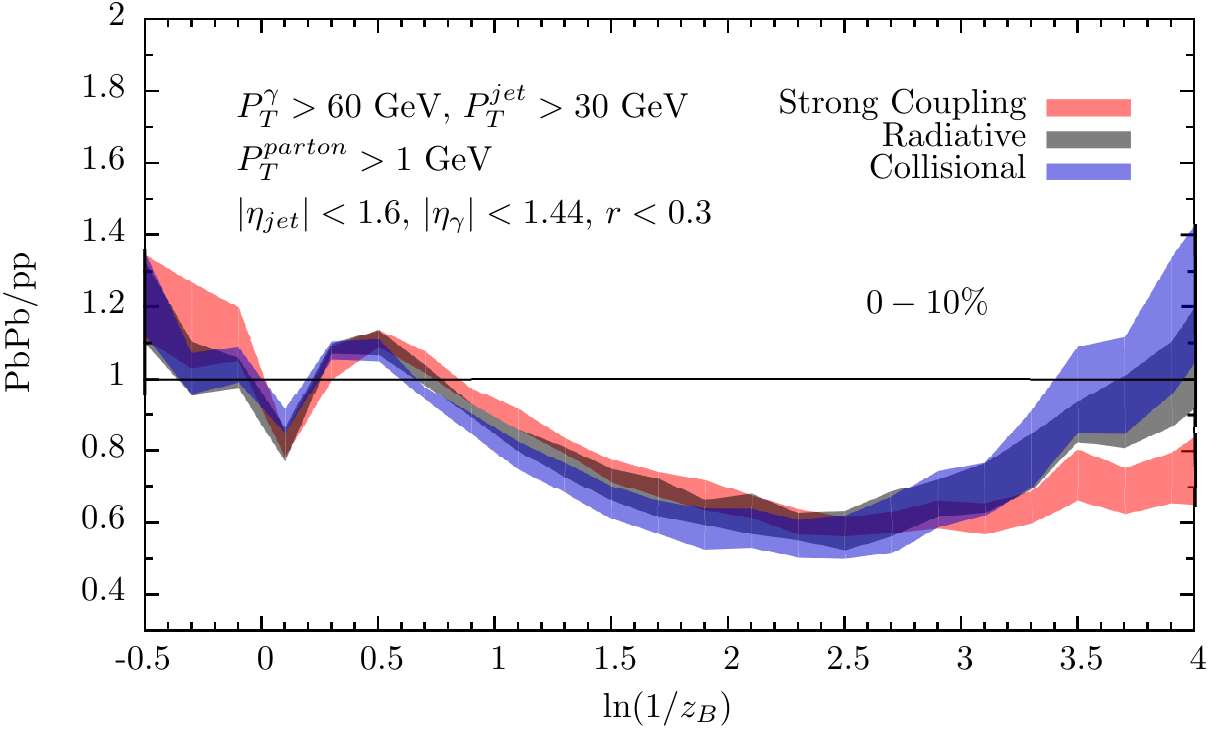}
\put(-180,30){\tiny{$\sqrt{s}=5.02~\rm{ATeV}$}}
&
\includegraphics[width=.5\textwidth]{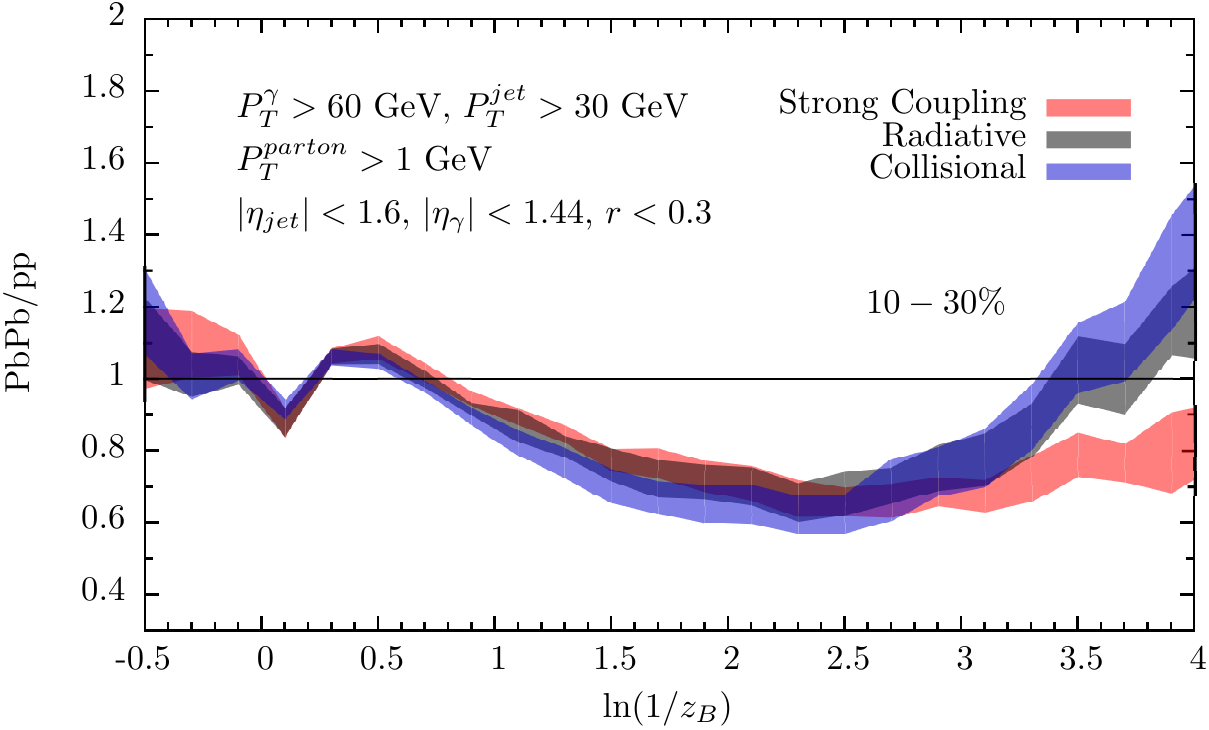}
\put(-180,30){\tiny{$\sqrt{s}=5.02~\rm{ATeV}$}}
\\
\includegraphics[width=.5\textwidth]{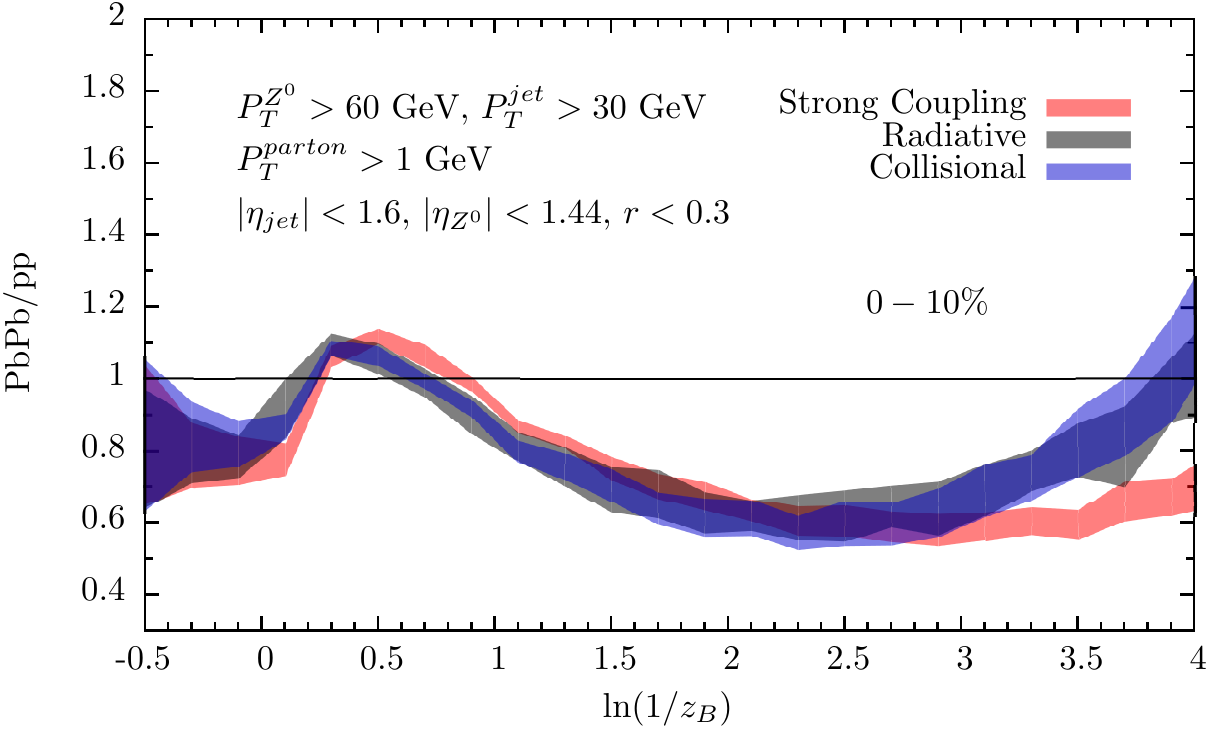}
\put(-180,30){\tiny{$\sqrt{s}=5.02~\rm{ATeV}$}}
&
\includegraphics[width=.5\textwidth]{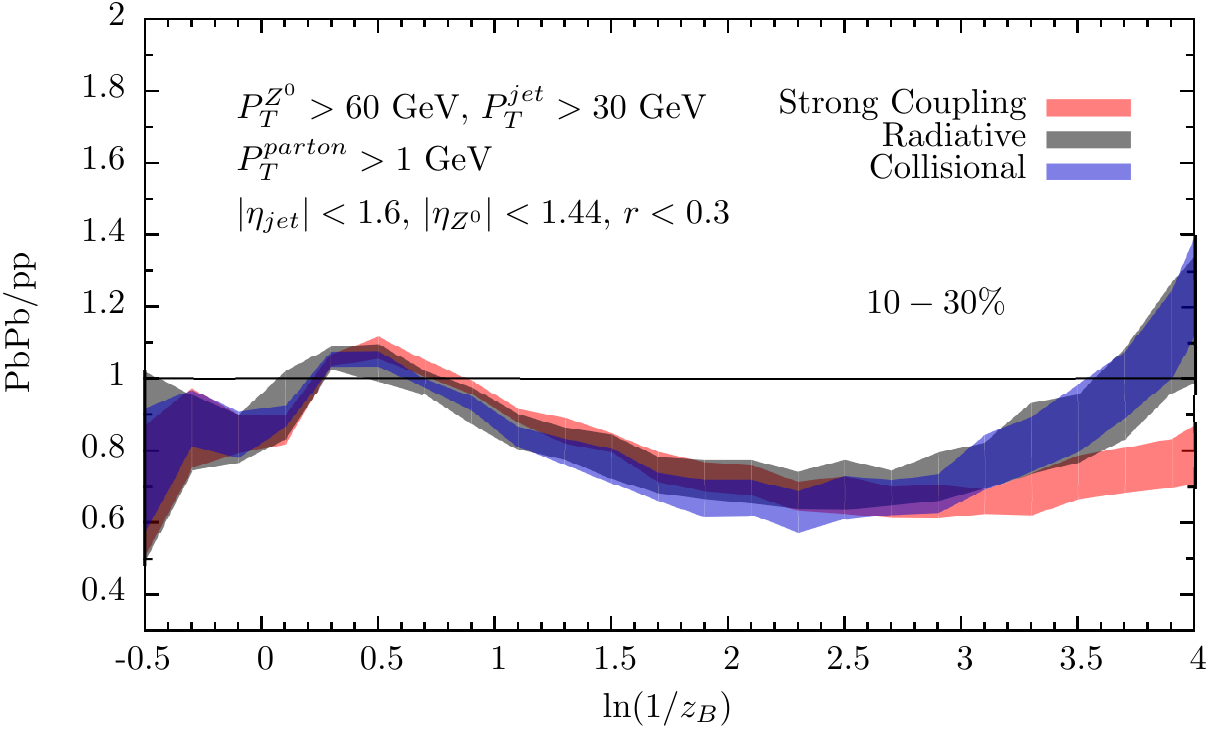}
\put(-180,30){\tiny{$\sqrt{s}=5.02~\rm{ATeV}$}}
\end{tabular}
\caption{\label{Fig:FFB}  
Predictions of our hybrid model and our two control models for the
partonic fragmentation function ratios (Pb-Pb over p-p)  for
jets produced in association with a boson
as a function of $z_B=-{\bf p}_T^{\rm parton}{\bf p}_T^{B}/(\pt^B)^2$  for two different centralities for photon-jet (upper) and Z-jet (lower) events in 
Pb-Pb collisions at $\sqrt{s}=5.02$~ATeV.
}
\end{figure}

In Fig.~\ref{Fig:FFB} we reanalyze the fragmentation function ratios, this time using
the boson momentum $z_B$ to define the scaling variable according to
$z_B\equiv -{\bf p}_T^{\rm parton}{\bf p}_T^{B}/(\pt^B)^2$,
with ${\bf p}^B_T$ the transverse momentum of the isolated photon or Z-boson.\footnote{Because of the fluctuations of the rapidity of the centre of mass of the elementary partonic collision that leads to boson-jet events, the rapidity of the boson and the jets do not need to be correlated, unlike the transverse momentum. For this reason, to construct $z_B$ we have chosen to use only transverse momenta and chosen not to project the momenta of fragments along the boson direction.}
As already mentioned, our main motivation for redefining the scaling variable is to have a better proxy for the jet energy prior to quenching.
If all bosons were prompt, their momentum  would be insensitive to in-medium effects and energy loss would only affect the numerator of $z_B$. For those prompt bosons, the mismatch between the boson and the initiator parton of the jet originates entirely in vacuum processes.
This is the case for $Z$-jet correlations, where the possibility of producing a $Z$ in a jet shower is highly suppressed.  In contrast, in the isolated photon sample we use, there is a small fraction of fragmentation photons even after we make our isolation cut, and for these photons 
energy loss effects are present via the quenching of their parent parton.  Therefore, for photons, there is also a small dependence of the denominator of $z_B$ on quenching; the effect of this small dependence on the fragmentation function ratio depends on the $z_B$ range.

Note that as a consequence of the broad distribution
of the boson-jet imbalance even in p-p collisions, $z_B$ can be larger than one: there are jets with $\pt$ larger than the momentum of the
boson they are associated with. 
Those events populate the region of negative $\log 1/z_B$ in Fig.~\ref{Fig:FFB}. 
In this region, 
all models lead to an enhancement of the in-medium fragmentation function of jets associated to photons. 
At first this seems puzzling,
but there are in fact two reasons for this effect.
The first 
is the fact that  jet quenching does reduce the energy
of fragmentation photons, as their parent partons lose energy.
The second reason comes from imposing an isolation
cut on photons in the events in our calculations, which do
not include the particles corresponding to the medium.
Since quenching affects all partons in our events, prompt photons in events with more than one jet 
are more isolated in our Pb-Pb simulations than in vacuum. 
This leads to an enhancement of the fragmentation function ratio for jets produced in association with 
isolated photons in the region of $x_J>1$. However, it is not clear to us whether this effect will persist in a full simulation of  photon-jet 
events in which particles from the Pb-Pb background are incorporated in the sample. We leave the study of the fate of this enhancement for future work. For the 
$Z$-jet correlation neither of these two effects are present, and the negative $\log z_B$ region is
slightly suppressed, as expected.

The ratio of fragmentation functions also exhibits a non-trivial structure in the vicinity of $\log 1/z_B \sim 0.5$ ($z_B \sim 0.6$). This structure is correlated with the position of the maximum of the in-medium boson-jet imbalance in Figs. \ref{Fig:PJInv}
and \ref{Fig:ZJInv}, $x_{J \gamma} \sim 0.6$ and $x_{J Z} \sim 0.6$.
Since the maximum of the in-medium imbalance distribution is shifted towards smaller $x$ values, the non-monotonic behavior of the fragmentation function ratio in this region reflects the behavior of the imbalance distribution. Indeed, the relative abundance of associated jets with $x_{J\gamma}\sim 0.6$ or $x_{JZ}\sim 0.6$ is enhanced in Pb-Pb collisions with respect to $p$-$p$. The fact that all three models exhibit the same behavior is a consequence of the coincidence of the imbalance distribution in the three models.

As for fragmentation functions constructed with the variable $z_J$, 
here too the strongly coupled model only separates from the control models for soft particles 
where the physics of how the medium responds to the passage of the jet, physics that none
of our models includes, becomes important.

In summary, the analysis of the fragmentation functions of jets produced in
association with photons and Z-bosons 
indicates that mechanisms of energy loss that do not increase the number of hard fragments in jets, like the ones we have explored, lead to robust modifications to the fragmentation pattern of these jets. 
This also means that this type of observable is not very sensitive to the microscopic 
mechanism of parton energy loss.

We close this section by recalling our initial motivation --- the separation between the predictions
of our models for the fragmentation functions of inclusive jets seen in Ref.~\cite{Casalderrey-Solana:2014bpa} 
and in Fig.~\ref{Fig:FFdijet276} --- and asking why that separation between model predictions is less in
the fragmentation functions for jets produced in association with photons and Z-bosons
seen in Figs.~\ref{Fig:FFJ} and \ref{Fig:FFB}.  The answer comes in understanding
the selection effects in a sample of ``inclusive jets''.  Because the jet production spectrum is
a steeply falling function of $\pt$, and because usually two or more jets are produced in an event, 
most of the jets in a sample of inclusive jets are the most energetic jet in an event.  In selecting
a sample of inclusive jets, one is preferentially selecting jets that are the jet in their event that
has lost the least energy.  In contrast, when one selects jets by first identifying an isolated
photon or Z-boson and then reconstructing an associated jet there is no such selection effect.
This means that, on average, the inclusive jets whose fragmentation functions are shown 
in Fig.~\ref{Fig:FFdijet276} have travelled through the medium over a shorter path-length 
and, again on average, they are jets that fragmented less.  Fewer fragments, i.e. a jet
with a lower jet mass and a smaller opening angle and a harder fragmentation function,
means that within the jet there are fewer partons losing energy in the medium, and 
therefore means less energy loss.  Both these effects are likely small compared to
the event-by-event variation.  But, on average, inclusive jets contain somewhat 
fewer fragments\footnote{There are actually two reasons why they contain somewhat
fewer fragments.  First, as we shall see only in the next Section, even in p-p collisions the higher energy
jet in a dijet pair --- and this is what most jets in an inclusive jet sample are ---  tends to  
have a harder fragmentation function.  And, second, jets with fewer, harder, fragments
lose less energy \cite{CasalderreySolana:2012ef}
in the plasma produced in a Pb-Pb collision and so are even more
likely to end up being the higher energy jet in a dijet pair.}
and
traverse somewhat less plasma.  What are the consequences in our models
of the fact that in going from Fig.~\ref{Fig:FFdijet276} to Figs.~\ref{Fig:FFJ} and \ref{Fig:FFB}
there is an increase in path length?  The path-length dependence of $dE/dx$ in (\ref{CR_rate})
is stronger than in the control models (\ref{Eloss_equations}) meaning that
in the mid-range of $z$ where we see the effects of quenching (say $1<\log 1/z <3$) 
we expect the increase in path length to push the predictions of our
hybrid strongly coupled model down relative to the control models.
What are the consequences of the fact that in going from Fig.~\ref{Fig:FFdijet276} to 
Figs.~\ref{Fig:FFJ} and \ref{Fig:FFB} the jets become wider and there is some reduction
in the energy of the partons within them?  In (\ref{CR_rate}), a reduction in the energy
of the partons means a reduction in the stopping length $x_{\rm stop}$ and an
increase in $dE/dx$.  So, for this reason also we expect to see the predictions
of our hybrid strongly coupled model pushed down relative to those of 
the control models.  And, this is indeed what we see.  Unfortunately,
the predictions of the hybrid model get pushed down just to the extent
that in Figs.~\ref{Fig:FFJ} and \ref{Fig:FFB} they are essentially
on top of the predictions of the control models.

In the next Section, we show that it is possible to select a sample of jets in which
the predictions of the hybrid strongly coupled model for the fragmentation
functions in the mid-range $1<\log 1/z <3$ are pushed down even farther.

\subsection{\label{sec:ff_dijet}Fragmentation Functions of the Associated Jets in Dijet Pairs}

Motivated by the results and discussion above, we now turn our attention to the
fragmentation functions of the associated (less energetic) jets in dijet events.
Whereas in the inclusive jet sample of Fig.~\ref{Fig:FFdijet276} we have
selected jets that are likely to be those among the jets in their event that 
have been quenched the least, by selecting associated jets
in dijet events we will likely be selecting those that have been quenched the most.
That means we will be selecting those that have, on average, traversed
a longer path-length of medium and those that were produced, on average,
with a larger jet mass and jet opening angle and that, on average,
contained more, and therefore lower energy, fragments before quenching. 

\begin{figure}[tbp]
\centering 
\begin{tabular}{cc}
\includegraphics[width=.5\textwidth]{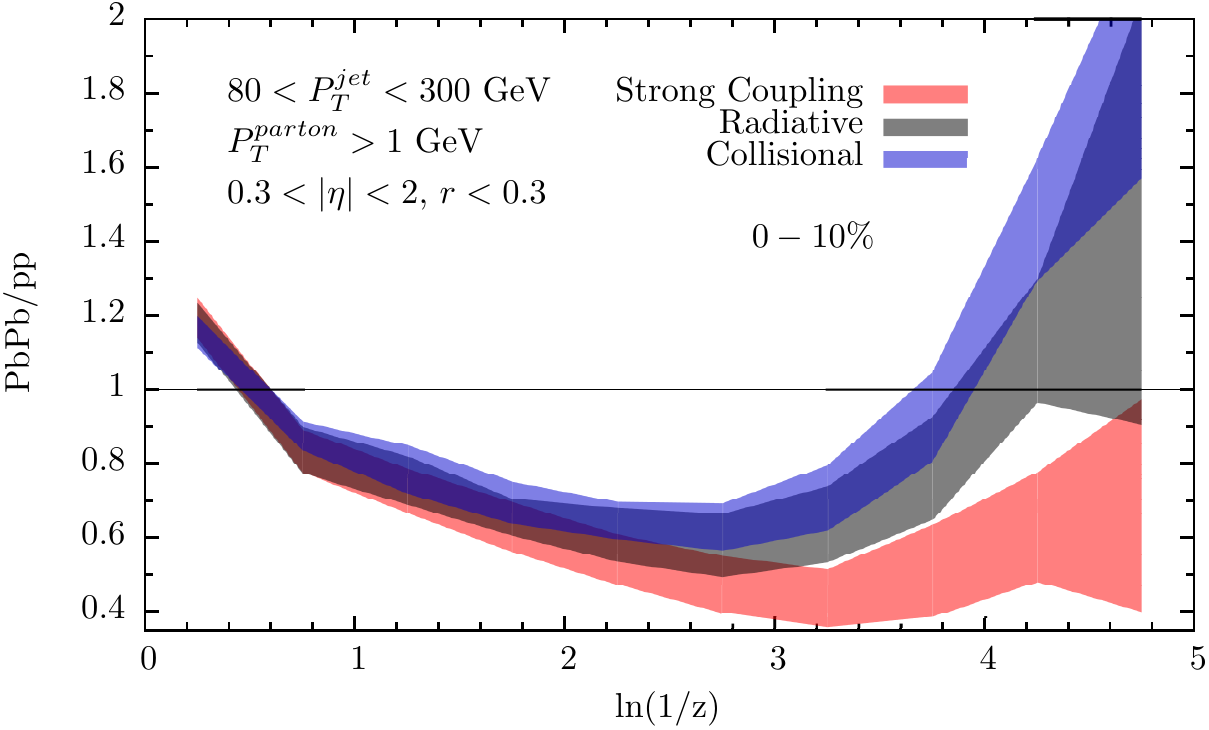}
\put(-175,83){\tiny{$\sqrt{s}=5.02~\rm{ATeV}$}}
&
\includegraphics[width=.5\textwidth]{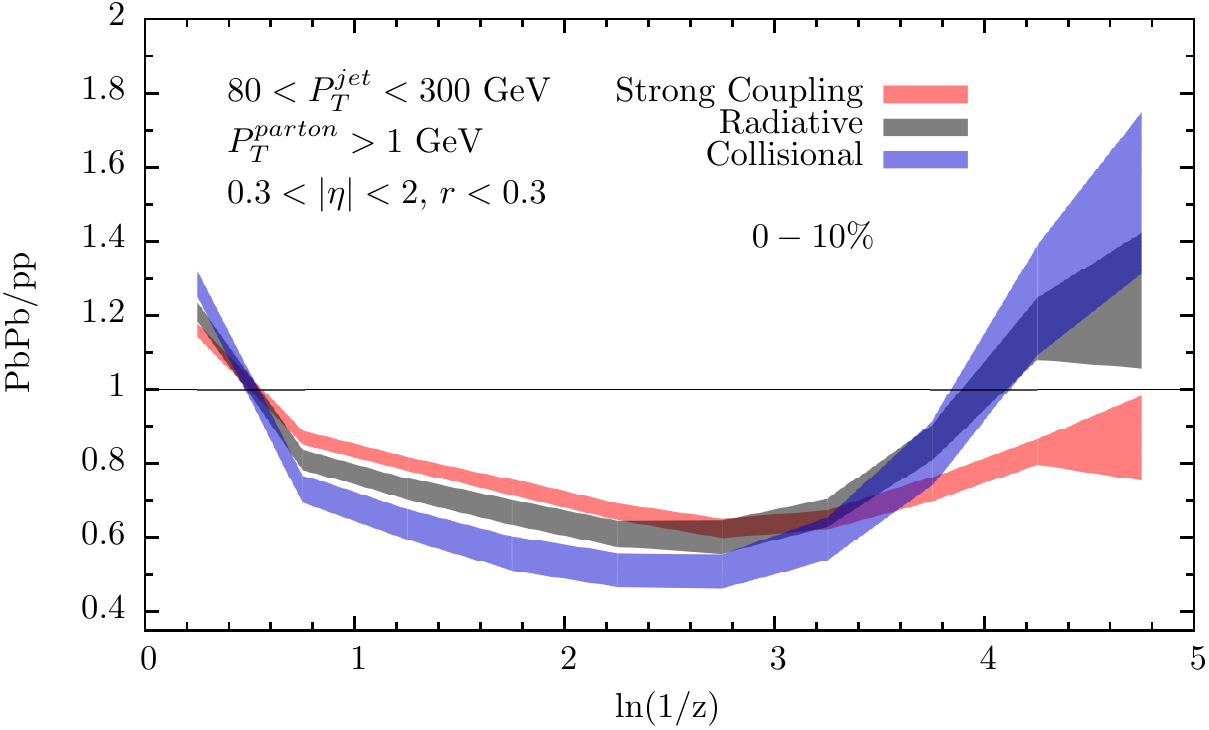}
\put(-175,83){\tiny{$\sqrt{s}=5.02~\rm{ATeV}$}}
\end{tabular}
\caption{\label{Fig:FF_I_A}  
Left: Model predictions for the Pb-Pb to p-p ratio of fragmentation functions
of associated jets (lower energy jets) in dijet events at $\sqrt{s}=5.02$~ATeV 
with a leading jet of $\pt>120$~GeV. Right: Same for inclusive jets with the
same range of $\pt$ and $\eta$ as the associated jets in the left panel.
The right panel is similar to Fig.~\ref{Fig:FFdijet276}, but is for a different $\pt$ range and is for the higher LHC heavy ion Run 2 collision energy.
}
\end{figure}

In the left panel of Fig.~\ref{Fig:FF_I_A}, we
show the ratio of fragmentation functions for jets of $80<\pt<300$ GeV 
with $0.3< \left| \eta \right| <2$
produced in association 
with a jet of $\pt>120$ GeV in Pb-Pb collisions to that in p-p collisions. 
By the associated jet in a dijet pair we will always mean the jet with the lower energy.
For comparison, in the right panel of Fig.~\ref{Fig:FF_I_A} we have analyzed inclusive jet fragmentation in a lower energy range, 
such that the momentum of the inclusive jets is in the same range as the momentum of the associated jets in the left panel of Fig.~\ref{Fig:FF_I_A}. 
Although by careful comparison to Fig.~\ref{Fig:FFdijet276} we see that the modification pattern 
seems to be slightly dependent on the jet energy, the basic features in the right panel of Fig.~\ref{Fig:FF_I_A}
and in particular the ordering of the predictions of the three models, is the
same as in Fig.~\ref{Fig:FFdijet276}.  The inclusive jets are selected in a way that makes them
likely to be the less quenched jets in their event with, on average, a shorter path length
and fewer fragments.  The jets in Section~\ref{sec:FFBJ}, selected via having been
produced in association with photons or Z-bosons, have no such selection effects.
And, the associated jets in  the left panel of Fig.~\ref{Fig:FF_I_A}
are selected in a way that makes them likely to be the more quenched
jets in their event with, on average, a longer path length and more fragments.
Sure enough, 
we see that the
predictions of our hybrid strongly coupled model --- with a $dE/dx$ that depends strongly on
path-length and that increases at lower energies as $x_{\rm stop}$ decreases --- are pushed
lower than those of the control models
in the left panel of Fig.~\ref{Fig:FF_I_A}.

\begin{figure}[tbp]
\centering 
\includegraphics[width=.5\textwidth]{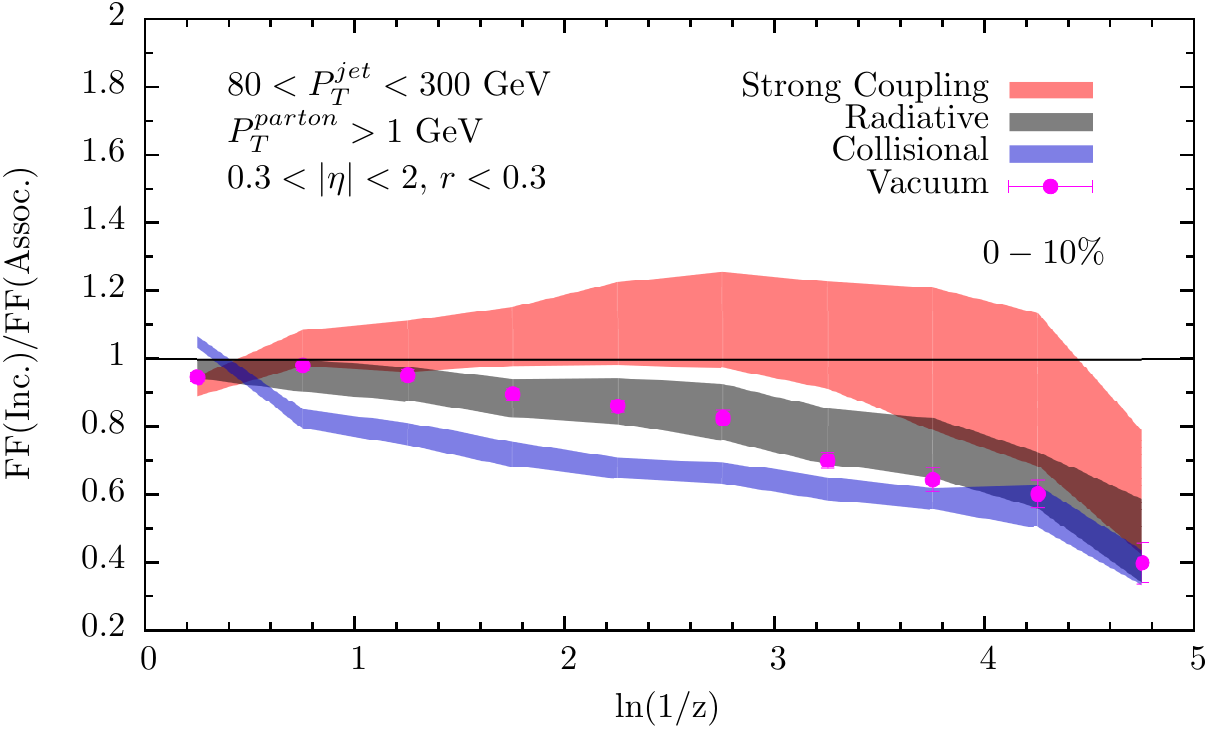}
\put(-175,85){\tiny{$\sqrt{s}=5.02~\rm{ATeV}$}}
\caption{\label{Fig:FIA_Ratio}  
Model predictions for the ratio of the fragmentation functions of inclusive jets to the fragmentation functions of 
associated jets in dijet events whose most energetic jet has $\pt>120$~GeV in Pb-Pb collisions at $\sqrt{s}=5.02$~ATeV. 
Both the inclusive jet and associated jet samples are constrained to the same $\eta$ and $\pt$ range, see the text.
We see a striking, and robust, separation between the predictions of our hybrid strongly coupled
model and our two control models.  We also show the same ratio, constructed with the same
kinematic cuts, for inclusive and associated jets in p-p collisions.  The ratio in p-p collisions should
not be compared to that in Pb-Pb collisions: each stands on its own and has its own implications.
(Comparisons between Pb-Pb and p-p are shown in Fig.~\ref{Fig:FF_I_A}.)
Here, the p-p results illustrate the differences between jets selected inclusively and jets
selected by association with a leading jet in vacuum, with no jet quenching. And, the comparison
between the Pb-Pb results from the different models shows how the interplay between these differences
and the path-length and energy dependence of different expressions for $dE/dx$ yields 
predictions for this ratio that depend sensitively on the underlying microscopic dynamics of jet quenching.
}
\end{figure}

The model-dependence of the fragmentation function ratios seen by comparing the
left and right panels of Fig.~\ref{Fig:FF_I_A} is striking:  the ordering of the predictions
of the three models is opposite in the two panels, with the hybrid strongly coupled
model predicting the least depression of the fragmentation function of the inclusive
jets and the most depression of the fragmentation function of the associated jets.
Seeing this motivates us to introduce a new observable in Fig.~\ref{Fig:FIA_Ratio} in which we, in
effect, take the ratio of the right panel of Fig.~\ref{Fig:FF_I_A} to the left panel of Fig.~\ref{Fig:FF_I_A}.  
That is, we propose
to compare the fragmentation function of inclusive (which is to say leading) jets
to the fragmentation function of associated (lower energy) jets in Pb-Pb dijet events.  
To avoid trivial kinematic differences between the energies of the inclusive and associated jets, we have constructed this ratio with 
jets in the same interval of $\pt$ and rapidity, $80<\pt<300$ GeV and $0.3<\left |\eta \right |<2$. 
As in Fig. \ref{Fig:FF_I_A}, the associated jets whose fragmentation functions constitute
the denominator in the ratio plotted in Fig.~\ref{Fig:FIA_Ratio} are produced in association with
leading jets of $\pt> 120$ GeV. 
(We have not investigated other choices of leading jet $\pt$.) 
The predictions
of our hybrid strongly coupled model and of our two control models for this
new observable are displayed in Fig.~\ref{Fig:FIA_Ratio}.   This observable
yields the largest separation between the predictions of our models of any
observable that we have investigated.  This means that it is particularly
sensitive to the underlying microscopic dynamics behind the modification
of jets in medium.

The new observable that we have introduced in Fig.~\ref{Fig:FIA_Ratio} has
the added virtue that its construction does not require fragmentation functions
from jets in p-p collisions.  It is the ratio of the fragmentation functions of
differently selected Pb-Pb jets with the same kinematics.  This means that
none of the uncertainties coming from the differences in the way hadronization
occurs in Pb-Pb collisions and p-p collisions come into play.  Our
calculations behind Fig.~\ref{Fig:FIA_Ratio} are, like those in Section~\ref{sec:FFBJ}
and Appendix~\ref{update}, calculations of partonic fragmentation functions. Here,
though, 
we expect that the ratio of fragmentation functions displayed in Fig.~\ref{Fig:FIA_Ratio}
should be a better  predictor
 for the same ratio of the hadronic fragmentation functions that
experimentalists can measure than in the case of any of the other fragmentation function
ratios that we have constructed
because much of the uncertainty in our understanding
of hadronization should cancel in this ratio.  
Of course, in the softest
region, at the smallest $z$'s, it continues to be the case that we are leaving
out contributions to the fragmentation functions from the
moving quark-gluon plasma, set in motion by the jet passing through it. 
This effect should at least partially cancel in the ratio plotted in Fig.~\ref{Fig:FIA_Ratio};
this is yet one more advantage of using the ratio of fragmentation functions
of differently selected jets with the same kinematics in the same Pb-Pb collisions, without introducing jets from p-p collisions.

In Fig.~\ref{Fig:FIA_Ratio} we have also plotted (as purple dots) the same ratio of the fragmentation
function of inclusive jets to that of associated jets for p-p collisions from \pythia, with the same kinematic cuts
as in our Pb-Pb analysis.
We show this solely to confirm that even in the absence of any medium there is a difference
between the inclusive jets and the associated jets.  Selecting inclusive jets selects jets
with fewer and therefore harder fragments.  Selecting associated jets selects jets
with more and therefore softer fragments.  In Pb-Pb collisions, then, in the presence
of a medium the differences between the rates of energy loss come into play as
we have discussed.  This effect, plus the difference between the path lengths
seen by inclusive and associate jets on average, are amplified by the different
path-length and energy dependence of the rates of energy loss in our hybrid
strongly coupled model and our two control models.  The result is the large
separation between the predictions of the models for the ratio plotted in Fig.~\ref{Fig:FIA_Ratio},
making this ratio so discriminating.  We look forward to seeing experimental measurements
of this ratio; they have great potential to teach us about the microscopic dynamics
that results in the observed modification of jets in heavy ion collisions.

\section{\label{conclusions}Conclusions and Outlook}

In this paper we have explored a broad range of jet observables in Pb-Pb collisions at LHC energies within the context of the 
hybrid strong/weak coupling model that we introduced in Ref.~\cite{Casalderrey-Solana:2014bpa}. 
This is a phenomenological approach to the physics of jet quenching in which we
aim to synthesize the very different types of physical 
processes involved in the production, branching,
and subsequent  in-medium dynamics of jets in heavy ion collisions. 
In particular, 
our model separates the short distance physics which controls the production and 
hard branching evolution of jets, 
which behaves as in vacuum, from the longer distance processes that control the 
interaction of the jet shower with the strongly coupled fluid produced in energetic heavy ion collisions. 
Our goal is to describe the short distance physics with standard tools for describing
the weakly coupled physics of jets in vacuum, in particular with \pythia, and to describe the longer distance processes
using insights from calculations done using strong coupling methods. In particular, to
model the long distance physics we adopt results obtained by analyzing 
the rate of energy loss $dE/dx$ of an energetic massless parton propagating through the hot liquid
plasma of strongly coupled
$\mathcal{N}=4$ SYM theory, as obtained via gauge/gravity duality.
We apply these results in QCD upon assuming that the differences between
the hot liquid plasma phases of the two theories can be absorbed in a single
parameter which controls the stopping distance of energetic excitations in 
plasma. 
We fitted this one model parameter  to jet $R_{AA}$ data 
in our previous publication. 
Our model, now fully specified, 
yields a very good description of the suite of inclusive jet observables and dijet observables 
discussed in our previous publication and, as our explicit analysis in this paper shows, of all currently available data
on photon-jet correlations.
All these observables in sum span a wide range of energies, originate from different hard production processes with
different primordial spectra, and correspond to jet samples with different selection biases, different ratios of quark jets to gluon jets, 
different fragmentation patterns and jet mass distributions,
and that traverse different distributions of path lengths.
Nevertheless, we have obtained a satisfactory description of all of these observables using our simple, hybrid, implementation of
strongly coupled dynamics --- namely by taking jets from \pythia and applying  the strongly coupled rate of
energy loss $dE/dx$ parton-by-parton to the partons in a parton shower.

Although at one level we are pleased by the now increasingly many successes of our hybrid
model approach, at another level they are frustrating. Our model is simple, gluing together two
rather different descriptions of physics at different scales and in so doing
incorporating much unknown physics into just one parameter.  Of course the first goal
in creating such a model is to capture some of the physics correctly, and it seems that
we have done that.  However, it could be even more interesting to see the model breaking down
and to use ways in which it fails to describe some feature of some experimental data 
to understand which of the aspects of the physics that the model leaves out are
important, and how, and why, and where.  The string of successes in the comparison
between the results of our model calculations and experimental data preclude investigations of this nature at present.

All that said, our explicit analysis of two control models with parametrically different expressions for $dE/dx$ 
cautions us that
we should not rush to conclude from the successes of our hybrid model 
that the experimental measurements of the various jet, dijet and photon-jet observables 
favor the strongly coupled form for $dE/dx$ over other possibilities.
Rather, we must acknowledge that these observables are not strongly sensitive
to the parametric form of the rate of energy loss $dE/dx$.  The experience
that we have gained from our analysis suggests that as long as the vacuum-like
branching processes that are at the core of jet dynamics are described well,
any mechanism that is able to quench particles, in particular the softer partons in 
a jet, can capture the bulk features of the measured distributions of jet observables, including
the various dijet and photon-jet asymmetries and correlations, as long as the parameter that
governs the overall magnitude
of $dE/dx$ is chosen appropriately.

Nevertheless, it is certainly heartening and perhaps even remarkable that the range of values
of the one parameter in our hybrid strong/weak coupling model, $\aSC$, that we find provides
a good description of so much jet data agrees so well with {\it a priori} expectations.  $\aSC$ should
be smaller than but of order one, exactly as we have obtained.  A stopping length for energetic partons
in the strongly coupled QGP of QCD that is three to four times longer than that in the strongly coupled
${\cal N}=4$ SYM plasma with the same temperature, as we find, is an eminently reasonable result.

These conclusions are all conclusions that we reached in our previous publication.  The analyses of photon-jet observables presented here
serves to reinforce them in many ways.

Looking ahead, we hope that among the many calculations from our hybrid model that we have presented the ones that
will be most important will be the many predictions that we are making for experimental measurements that are anticipated
in the near future.  These can be grouped into three categories, the first two of which are:
\vspace{-0.5pt}
\begin{itemize}
\setlength{\itemsep}{-0.5pt}
\item
We have provided the predictions of our hybrid model for inclusive jet observables, dijet observables,
and photon-jet observables in heavy ion collisions with collision energy $\sqrt{s}=5.02$~ATeV, in anticipation
of LHC heavy ion Run 2.  This is particularly important for the photon-jet observables.  The present data 
have low statistics, and correspondingly large error bars, making the fact that our hybrid model
describes them well less impressive than it could be.  In the run to come, the statistics will be 
greater by about an order of magnitude, meaning that the error bars should be significantly smaller, making the confrontation between
the predictions of our model and these measurements much more constraining.
\item
We have provided the predictions of our hybrid model for Z-jet observables, which we hope will be tested
in future LHC runs.
\end{itemize}
Confrontation between these predictions and the data to come should serve either to further strengthen
our confidence in the approach to jet quenching that we have introduced or to identify and quantify 
ways in which it fails, ideally pointing
toward which aspects of the physics that we have left out are most important and guiding the
improvement of the model.  It is also possible that as the experimental uncertainties shrink
measurements of these observables could serve to differentiate between the different
assumptions about the dynamics of parton energy loss, and consequent different
forms for $dE/dx$.  However, the distinctions between the predictions of our hybrid
model and our two control models for these observables are small, limiting the
discriminating power of these observables even as the experimental uncertainties shrink.

It seems clear that in order to find observables that provide more discrimination
among the different possible dynamical processes via which the partons in a shower
lose energy as they traverse the strongly coupled plasma we will need to investigate
intrajet observables.  Utilizing those observables that involve the angular shape of jets 
must wait, as it will require adding further physics to the model, including for example
the transverse momentum picked up by the shower partons as they interact with the medium,
and so must involve the addition of at least one new parameter to the model.  There is 
good motivation for such investigations, but we leave them to future work.    With
the one parameter model that we have constructed here, the class of intrajet observables
that we may be able to describe is those constructed from fragmentation functions:
\vspace{-0.5pt}
\begin{itemize}
\item
The third category of predictions that we have made are predictions for
various ratios of fragmentation functions.  We have taken advantage of 
having a tool with which we can compute multiple observables (here, fragmentation
functions for jets in p-p collisions, inclusive jets and jets
in dijet pairs in Pb-Pb collisions, and jets produced in association with photons
or Z-bosons in either Pb-Pb or p-p collisions) in varying kinematic regimes to search for
new discriminating observables which are particularly sensitive to differences between
mechanisms of energy loss.
We have provided the results of our hybrid model calculations for ratios between
partonic fragmentation functions in photon-jet and Z-jet events in Pb-Pb collisions to
those in p-p collisions, but for reasons that we have understood qualitatively these
turn out not to be particularly discriminating.  Also, these predictions are sensitive to
differences between hadronization dynamics in Pb-Pb and p-p collisions that
are not under good theoretical control at present, reducing the reliability of
our predictions for these ratios.  
The most discriminating observable that we have found is the ratio of the fragmentation
function of inclusive jets in Pb-Pb collisions to the fragmentation functions of jets
in the same kinematic regime, in the same collisions, that are the lower energy jets
in a dijet pair. (We refer to the latter as associated jets.)
Unlike most jet observables, this measurement does not require any p-p reference data.
And, since the ratio we propose is constructed from jets in the same kinematic regime in the same 
PbPb collisions, differing only in how they were selected, we expect that many of the theoretical uncertainties associated with the modification of hadronization in medium should cancel. 
This makes the ratio of partonic fragmentation functions that we compute a better proxy to the
ratio of fragmentation functions that experimentalists will measure.
Furthermore, we find that the predictions of our hybrid strong/weak coupling model
and our two control models for this ratio, see Fig.~\ref{Fig:FIA_Ratio}, are well separated
over a wide range of $z$, making the discriminating power of this observable robust
even after the softening of the fragmentation functions expected after hadronization.
And, perhaps best of all, we have a good qualitative understanding of why this
ratio is such a discriminating observable. First, in Pb-Pb collisions as in vacuum 
inclusive jets tend to contain fewer, harder, fragments than associated jets on average.  
Second, the distribution
of the path length of the medium through which a sample of inclusive jets has propagated
is, on average, shorter than that for a sample of associated jets.  
Both these effects mean that, on average, the inclusive jets have lost less energy than the
associated jets. And, we have shown that
both  these effects push the predictions of our hybrid strong/weak coupling
model and our control models apart.  Measurement of this ratio of fragmentation
functions should be a particularly effective way to gain information about the
dynamics via which energetic partons lose energy as they traverse strongly
coupled plasma.
The experimental determination of this ratio in the imminent LHC heavy ion Run 2 
can therefore shed light on the microscopic dynamics of jets in quark-gluon plasma. 
\end{itemize}

\acknowledgments

We gratefully acknowledge the assistance of Chun Shen and Ulrich Heinz, who
have generously provided us with hydrodynamic solutions for heavy ion collisions
at $\sqrt{s}=2.76$ and 5.02~ATeV.
We are grateful to Aaron Angerami, Paul Chesler, Yen-Jie Lee, Gunther Roland, Konrad Tywoniuk, Wilke van der Schee and Xin-Nian Wang 
for  helpful conversations over the course of this work.
KR is grateful to the CERN Theory Division for hospitality
at the time this research was completed.
The work of JCS was  supported by a Ram\'on~y~Cajal fellowship.  The work of JCS and DP was 
supported by  the  Marie Curie Career Integration Grant FP7-PEOPLE-2012-GIG-333786, by grants FPA2013-46570 and  FPA2013-40360-ERC
and MDM-2014-0369 of ICCUB (Unidad de Excelencia `Mar\'ia de Maeztu') 
 from the Spanish MINECO,  by grant 2014-SGR-104 from the 
Generalitat de Catalunya 
and by the Consolider CPAN project. 
The work of DCG and KR was supported by the U.S. Department of Energy
under Contract Numbers DE-SC0011088 and DE-SC0011090, respectively.
The work of JGM was supported by Funda\c{c}\~{a}o para a Ci\^{e}ncia e a Tecnologia (Portugal) under  project  CERN/FP/123596/2011 and contract `Investigador FCT -- Development Grant'.

\appendix

\section{\label{BI-Eloss}Energy Loss in a Boosted Fluid}

In this Appendix, we explicitly perform the Lorentz transformation
that, as we explained in Section~\ref{Sec:FlowEffects}, is needed in order to
determine the rate of energy loss $dE/dx$ in the collision center-of-mass
frame (the frame in which we do the overall computation of the modifications
to the energies of the partons in the shower)
from the rate of energy loss in the local fluid rest frame (the frame
in which the fluid at the spacetime location of a particular parton is at rest).
This transformation is particularly important for partons with significant rapidity, as
they propagate through fluid that is moving with a significant velocity, meaning that
the boost from the local fluid rest frame back to the collision center-of-mass frame is substantial.

In the local fluid rest frame, the change in the four-momentum of a parton which propagates for an infinitesimal time $d t_F$  is given by 
\be
d P^\mu_F &=& \mathcal{F}_F \left(x_F, E^F_{\rm in}\right)  \frac{P^\mu_F}{E_F} dt_F \,, 
\label{fig:diff_rate_fluid}
\ee
where 
and $P^\mu_F$ and
$E_F$ are the four momentum and energy of the parton in the local fluid rest frame and
$\mathcal{F}_F \left(x, E^F_{\rm in}\right) $ is the functional form of the rate of energy loss in that frame, in the notation that we  introduced in Eq. (\ref{dEdxFluidFrame}),
and is given by the right-hand side of (\ref{CR_rate}) in
our hybrid model or by the right-hand side of one of the expressions (\ref{Eloss_equations}) in our control models.
In writing the expression (\ref{fig:diff_rate_fluid}) we have used our assumption that the exchanges of momentum and energy
between the parton and the medium do not change  the direction of the parton significantly.

The Lorentz structure of the expression (\ref{fig:diff_rate_fluid}) simplifies the boost back to the collision center-of-mass frame.  First of all, it is easy to show that 
\be
\frac{dt_F}{E_F}=\frac{dt}{E} \,,
\ee
where $t$ and $E$ are the time and energy in the collision center-of-mass frame. 
 Second, after a Lorentz transformation, $P^\mu_F \rightarrow P^\mu$ and $dP^\mu_F \rightarrow dP^\mu$, with $P^\mu$ and $dP^\mu$ the four-momentum and the infinitesimal four-momentum loss in the collision center-of-mass frame. Therefore
\be
\frac{d P^\mu}{dt_F}&=& \mathcal{F}_F \left(x_F, E^F_{\rm in}\right) \frac{P^\mu}{E} \,,
\label{eq:AppAresult}
\ee
where the arguments of $\mathcal{F}_F$ are still expressed in terms of quantities in the local fluid rest frame. 

We next express the initial parton $E^F_{\rm in}$ in the local fluid rest frame
as a function of quantities in the collision center-of-mass frame via
the Lorentz transformation
\be
E^F_{\rm in}= E_{\rm in} \,\gamma_F \left(1-\w \v\right)\, ,
\label{eq:EinLT}
\ee
where ${\bf w}={\bf P}/E$ is the velocity of the parton in the collision center-of-mass frame and 
${\bf v}$ is the local velocity of the fluid in the same frame, which is to say it is the velocity vector
for the boost between the two relevant frames.
$\gamma_F$ is the gamma factor of the local fluid velocity ${\bf v}$.

The relation between $x_F$ and $x$, the distances travelled
in the two frames, requires further discussion. In the derivation of the rate of energy loss by one of
the partons in the shower, $x_F$ is the distance that that parton has travelled through the fluid.
However, 
as partons propagate through the hot plasma created in an heavy ion collision, 
the temperature and velocity of the fluid at their location in space and time changes.
We will assume that $x_F$ is the accumulated distance of the parton 
summed in such a way that each infinitesimal contribution $dx_F$ 
is evaluated in the local fluid rest frame.
This means that (if the rate of energy loss depends on $x_F$, as in (\ref{CR_rate}) and the first
expression in (\ref{Eloss_equations}))
the energy lost by a  parton traversing some $dx_F$ depends on the total $x_F$ accumulated
by that parton over its previous passage through the flowing plasma.
With this prescription, which neglects gradient effects, we have
\be
d\x_F= \w dt + \gamma_F \left(\w_L-\v\right) dt \,,
\ee
where $\w_{T\,, L}$ are the transverse and longitudinal components of the parton velocity 
in the collision center-of-mass frame. 
After some algebra, the increment in the accumulated distance is given by 
\be
\left( \frac{dx_F}{d t} \right)^2
&=& \w^2 + \gamma_F^2\left(\v^2 - 2 \v \w + (\v \w)^2\right)\,.
\ee 
Summing over the previous history of the parton, we  obtain
\be
\label{Eq:xF}
x_F(t)=\int_{t_0}^{t} d t \sqrt{\left[\w^2 + \gamma_F^2\left(\v^2 - 2 \v \w + (\v \w)^2\right) \right]} \,,
\ee
where $t_0$ is the creation time of the parton.  We have used (\ref{eq:AppAresult}), (\ref{eq:EinLT}) and (\ref{Eq:xF}) in 
(\ref{eq:biform}), (\ref{eq:EF}) and (\ref{eq:xF}).

\section{\label{update}Update on Single-jet and Dijet Observables at $\sqrt{s}=2.76$~ATeV}

\begin{figure}[tbp]
\centering 
\begin{tabular}{cc}
\includegraphics[width=.5\textwidth]{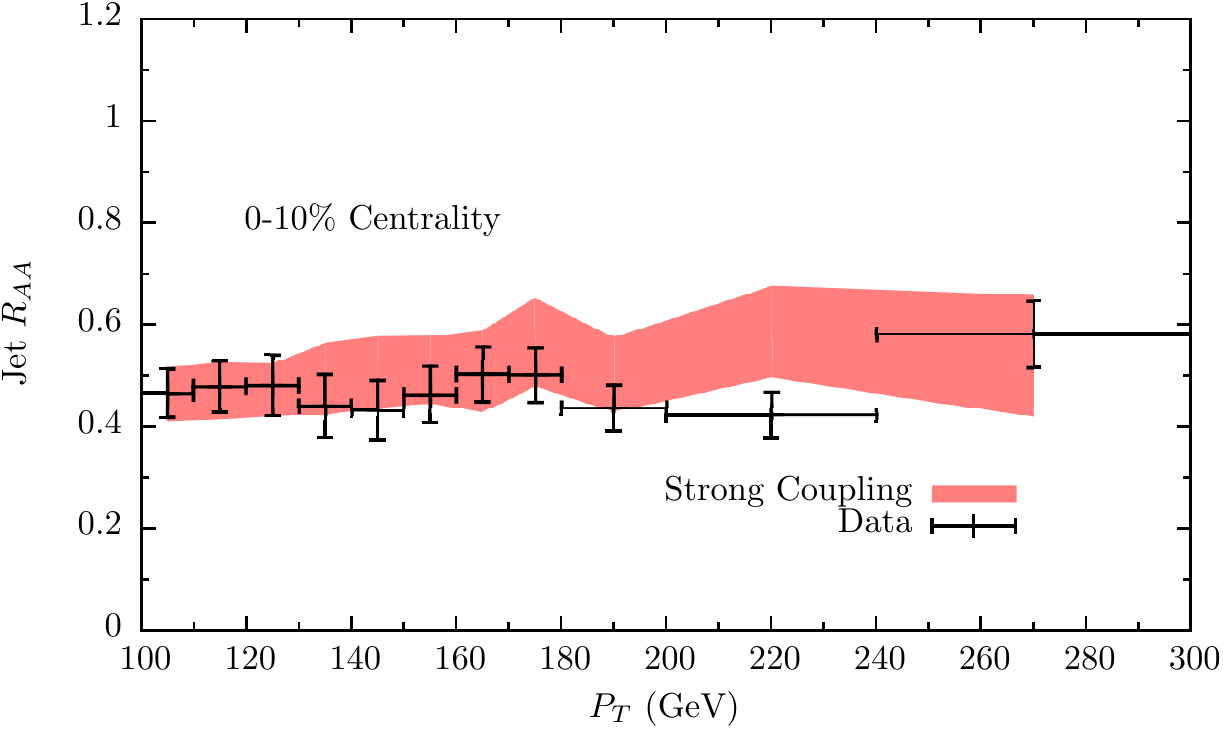}
\put(-175,34){\tiny{$\sqrt{s}=2.76~\rm{ATeV}$}}
&
\includegraphics[width=.5\textwidth]{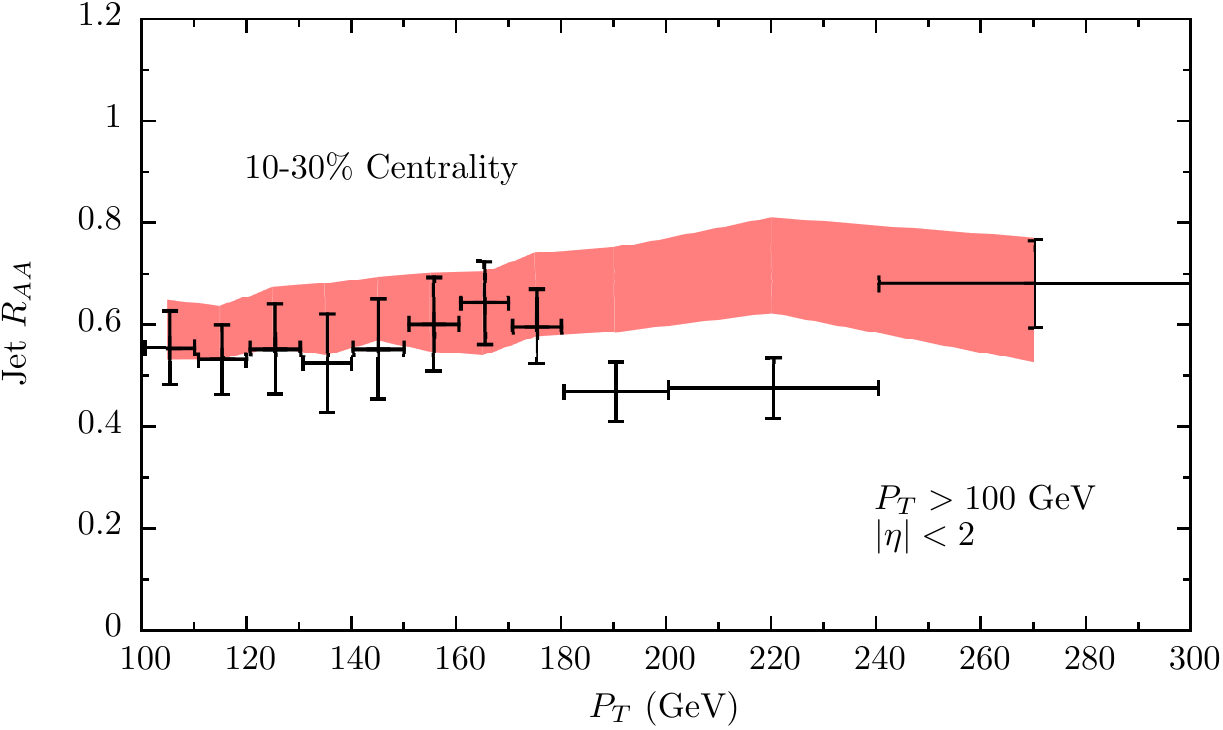}
\put(-175,34){\tiny{$\sqrt{s}=2.76~\rm{ATeV}$}}
\\
\includegraphics[width=.5\textwidth]{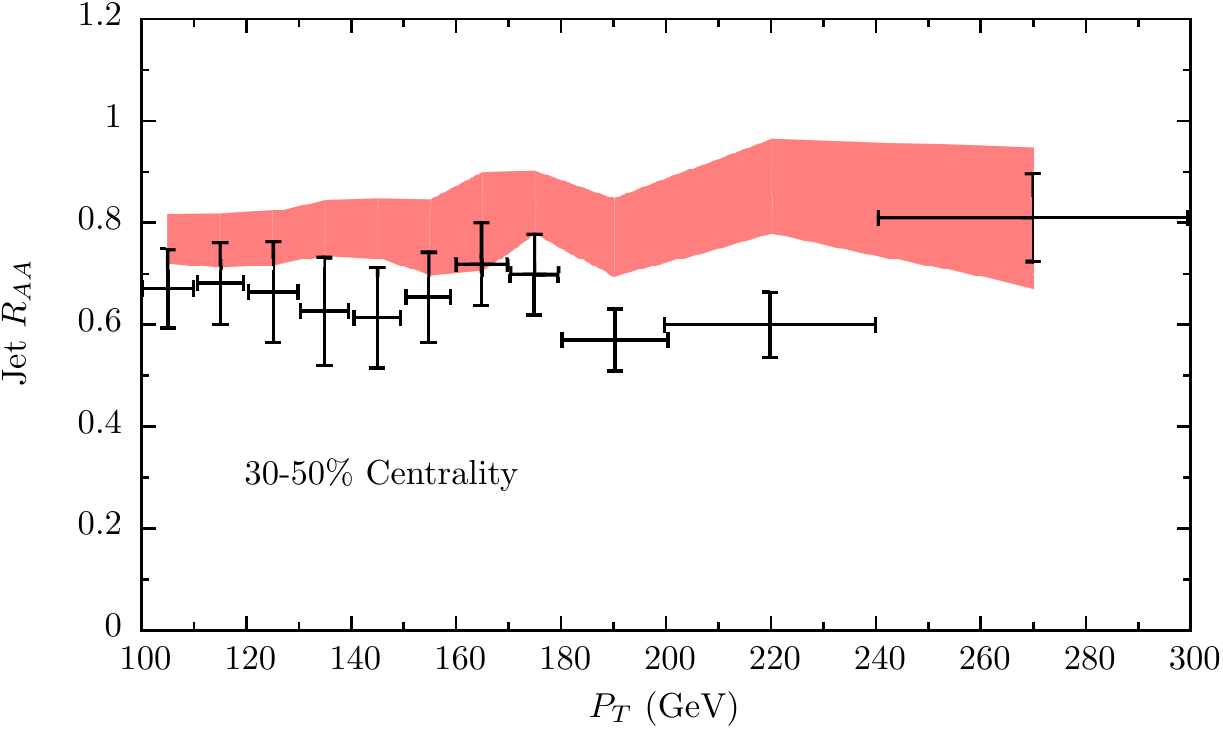}
\put(-175,34){\tiny{$\sqrt{s}=2.76~\rm{ATeV}$}}
&
\includegraphics[width=.5\textwidth]{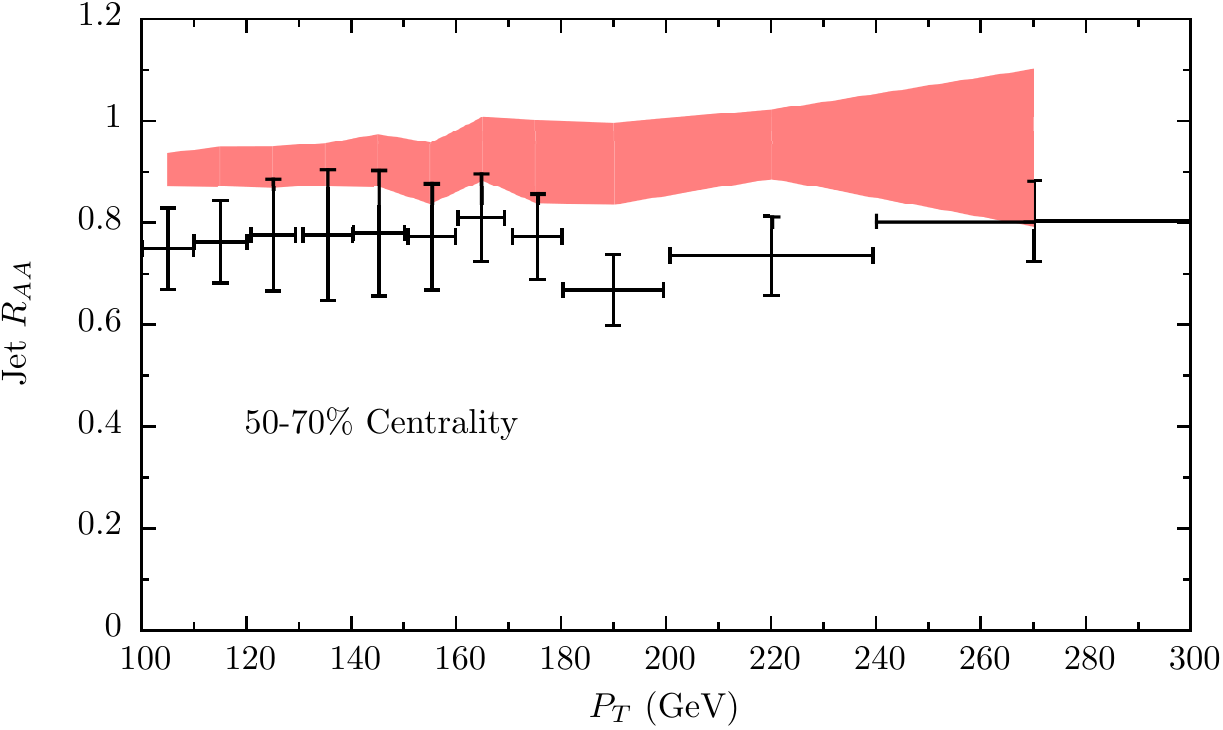}
\put(-175,34){\tiny{$\sqrt{s}=2.76~\rm{ATeV}$}}
\end{tabular}
\caption{\label{Fig:RAA}  Jet $\Raa$ as a function of $\pt$ for different centralities 
from our hybrid strong/weak coupling model (colored bands)
compared to preliminary CMS data from Ref.~\cite{Raajet:HIN}. 
The value of the single parameter in the model, $\kappa_{\rm sc}$,
is fitted to the  left-most data point in the top-left panel, namely the jets with $100~{\rm GeV}< \pt<110~{\rm GeV}$
in the most central collisions.  
All the rest of the features of the colored bands are results from our hybrid  model. 
In this Appendix and throughout
this paper,  single-jet, dijet, photon-jet, Z-jet and fragmentation function observables are all fully specified
once the single parameter in the model has been fixed.
}
\end{figure}

\begin{figure}[tbp]
\centering 
\vspace{-0.1in}
\begin{tabular}{cc}
\includegraphics[width=.5\textwidth]{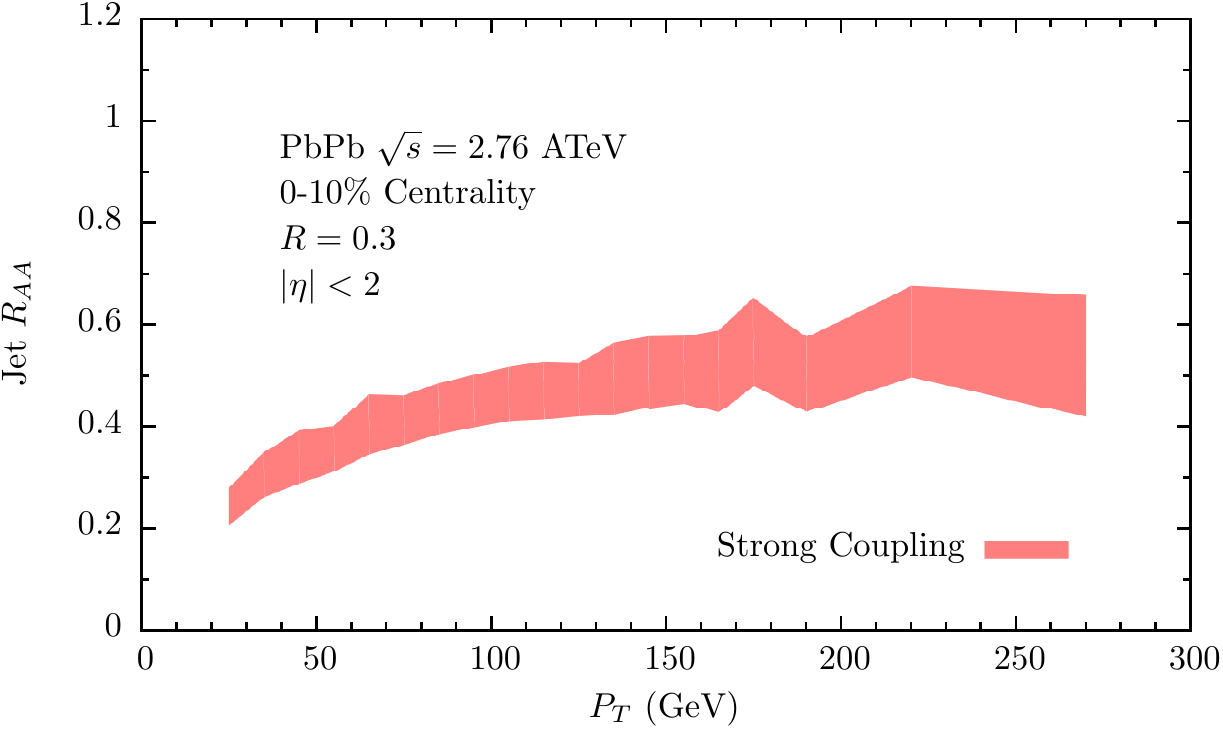}
&
\includegraphics[width=.5\textwidth]{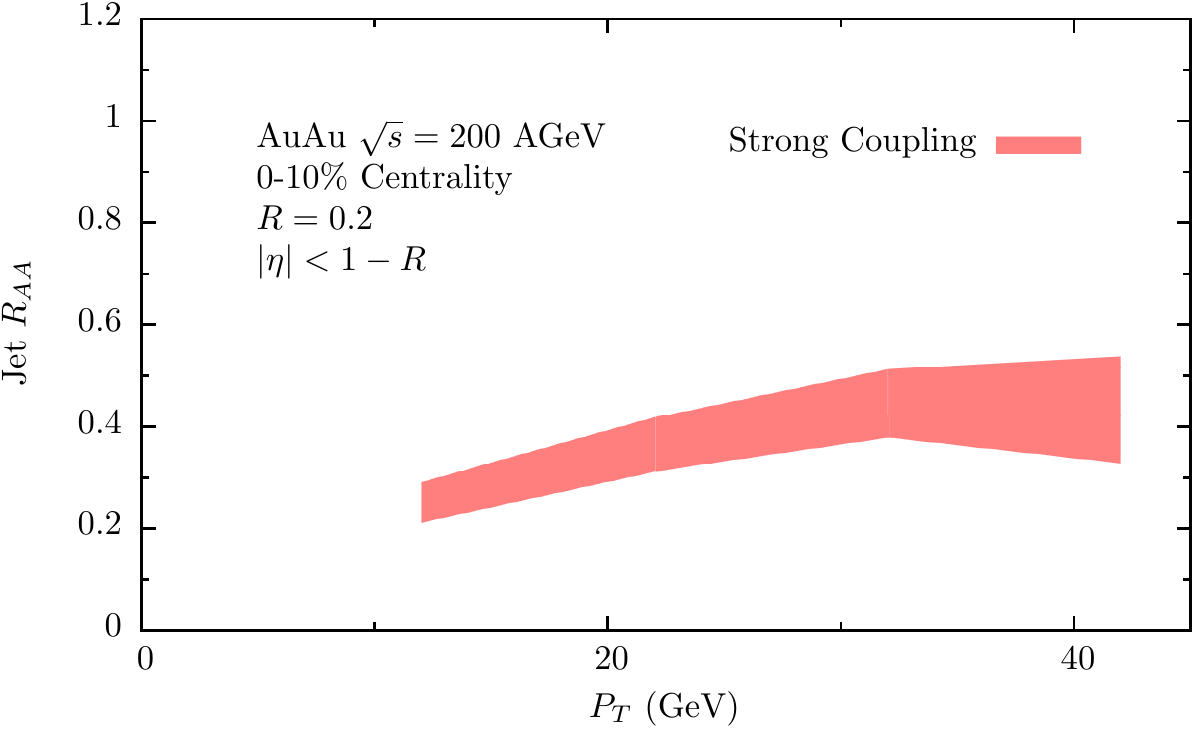}
\end{tabular}
\vspace{-0.1in}
\caption{\label{RHICANDLHC} Predictions of our hybrid strongly coupled
model for jet $\Raa$ for central Pb-Pb collisions at
the LHC with $\sqrt{s}=2.76$~ATeV as a function of $\pt$, extended down to $\pt=15$~GeV (left) and for central Au-Au collisions at
RHIC with $\sqrt{s}=200$~AGeV (right).  In both cases, we only show
our results for collisions in the 0-10\% centrality bin.
 }
\end{figure}

\begin{figure}[tbp]
\centering 
\begin{tabular}{cc}
\includegraphics[width=.5\textwidth]{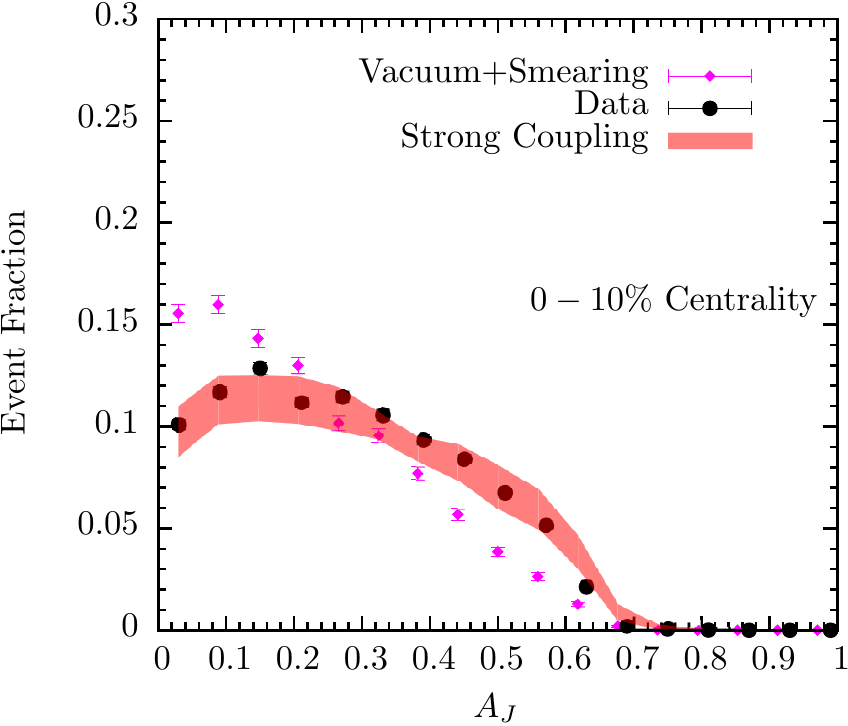}
\put(-85,88){\small{$\sqrt{s}=2.76~\rm{ATeV}$}}
&
\includegraphics[width=.5\textwidth]{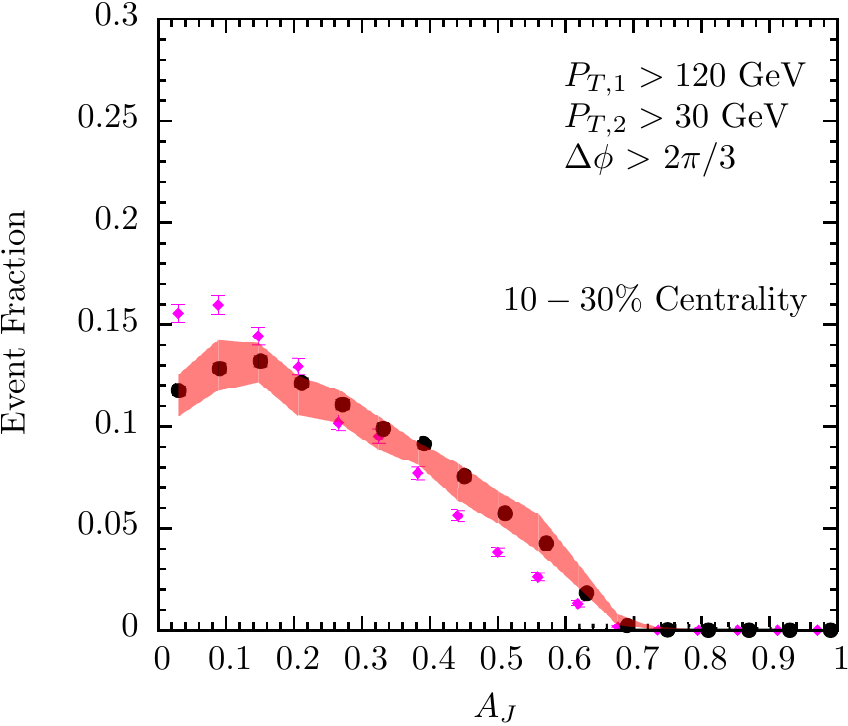}
\put(-88,88){\small{$\sqrt{s}=2.76~\rm{ATeV}$}}
\\
\includegraphics[width=.5\textwidth]{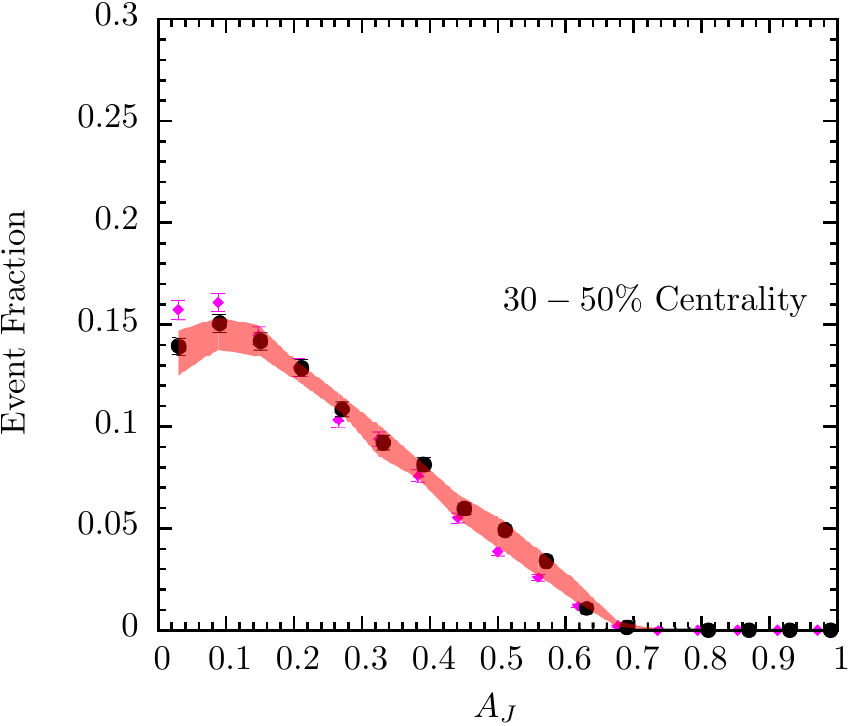}
\put(-88,88){\small{$\sqrt{s}=2.76~\rm{ATeV}$}}
&
\includegraphics[width=.5\textwidth]{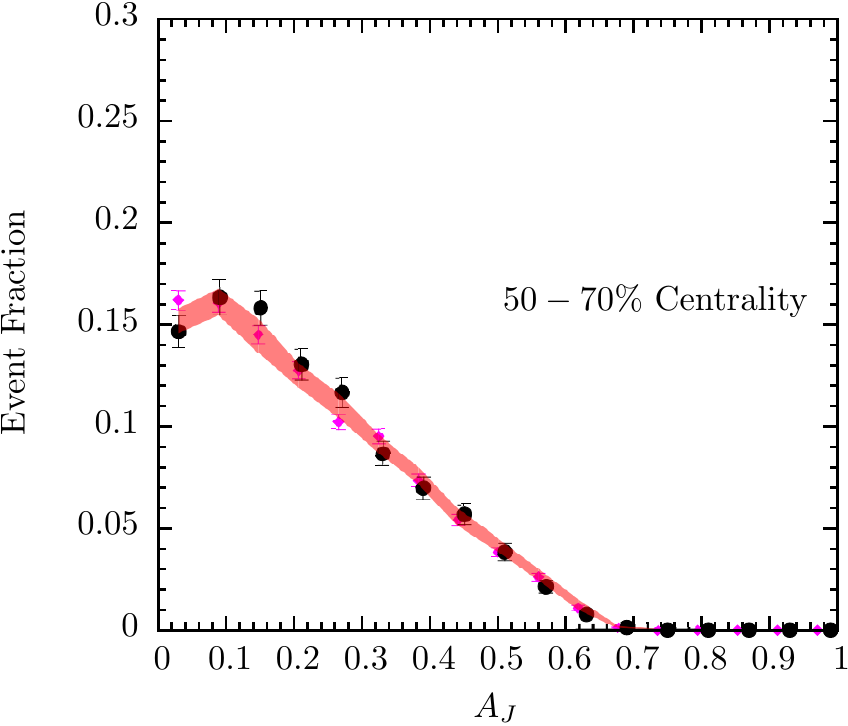}
\put(-88,88){\small{$\sqrt{s}=2.76~\rm{ATeV}$}}
\end{tabular}
\caption{\label{Fig:Asym}  Dijet imbalance $A_J$
in heavy ion collisions with $\sqrt{s}=2.76$~ATeV from our hybrid model (colored bands)
compared to CMS data from Ref.~\cite{Chatrchyan:2012nia} (black) and \pythia-generated
 proton-proton collisions (violet). Both the hybrid model calculations and the proton-proton reference are
 smeared according to the prescription in Ref.~\cite{YYthesis}.
 }
\end{figure}

\begin{figure}[tbp]
\centering 
\begin{tabular}{cc}
\includegraphics[width=.5\textwidth]{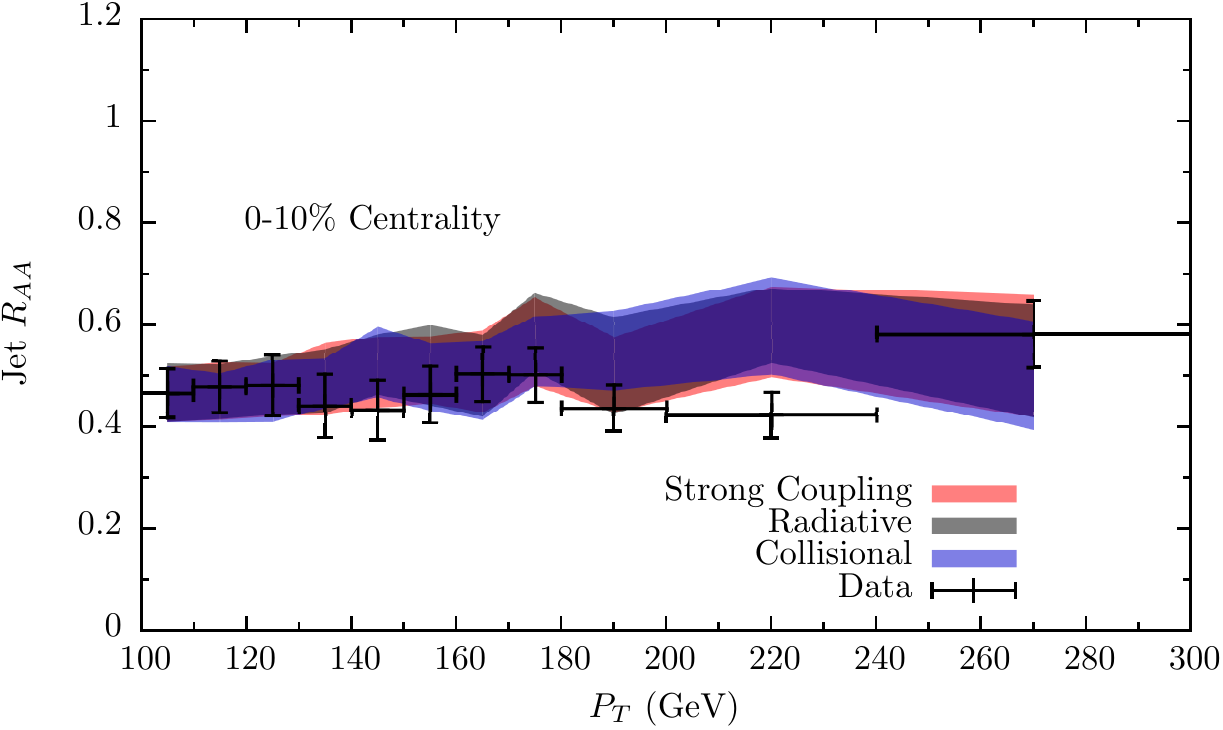}
\put(-175,34){\tiny{$\sqrt{s}=2.76~\rm{ATeV}$}}
&
\includegraphics[width=.5\textwidth]{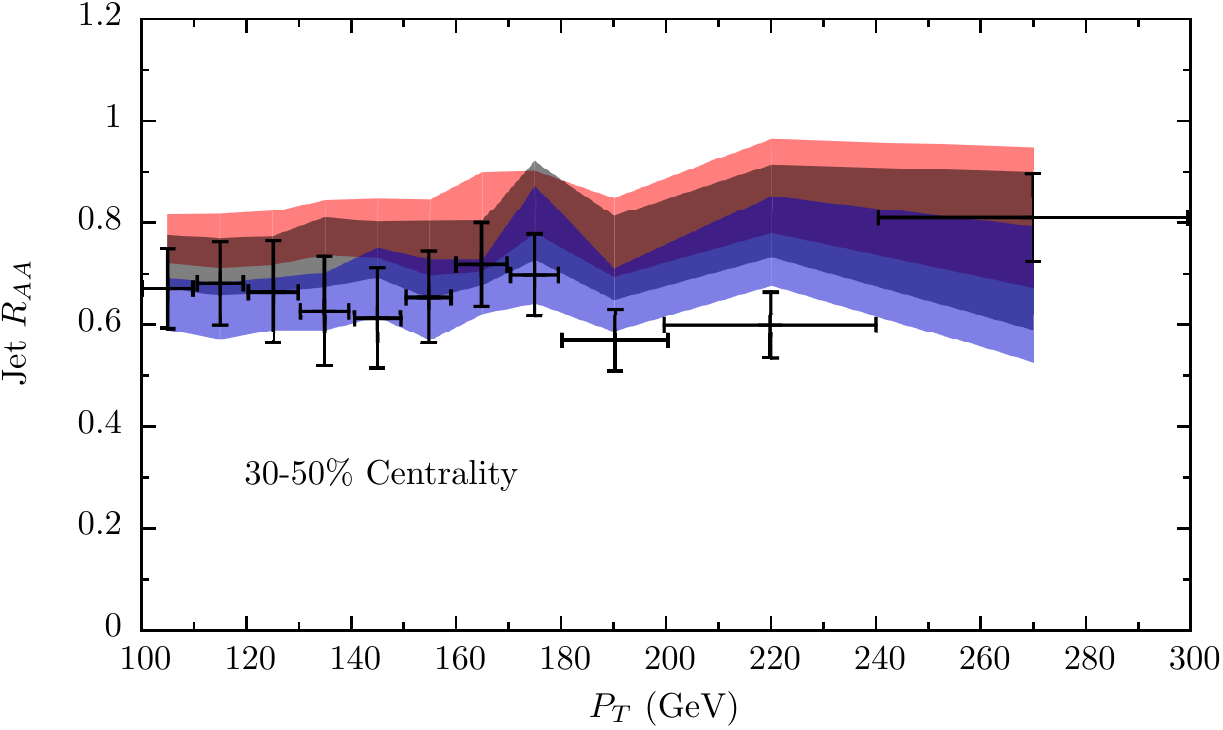}
\put(-175,34){\tiny{$\sqrt{s}=2.76~\rm{ATeV}$}}
\\
\includegraphics[width=.5\textwidth]{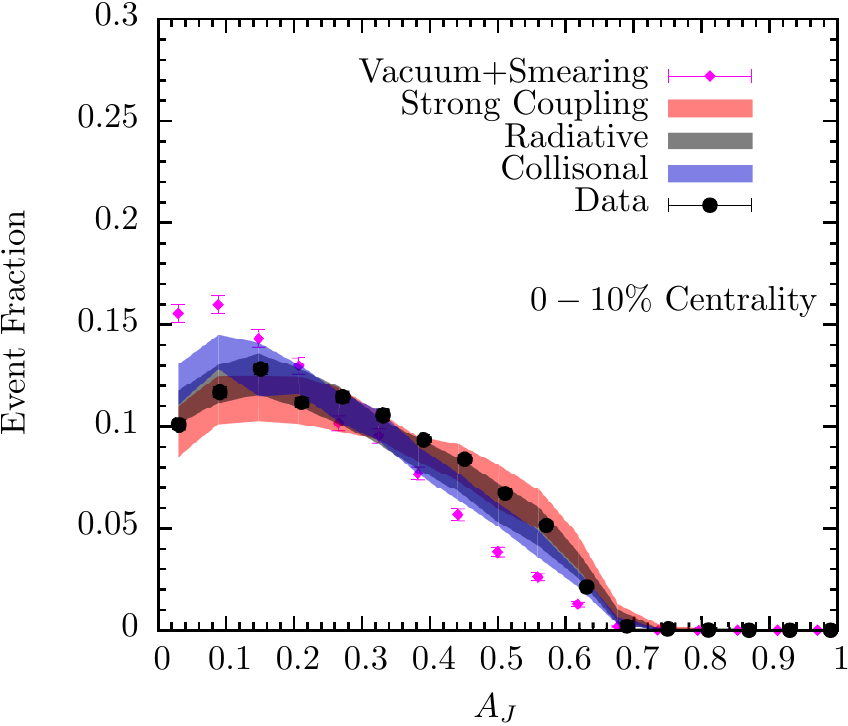}
\put(-85,88){\small{$\sqrt{s}=2.76~\rm{ATeV}$}}
&
\includegraphics[width=.5\textwidth]{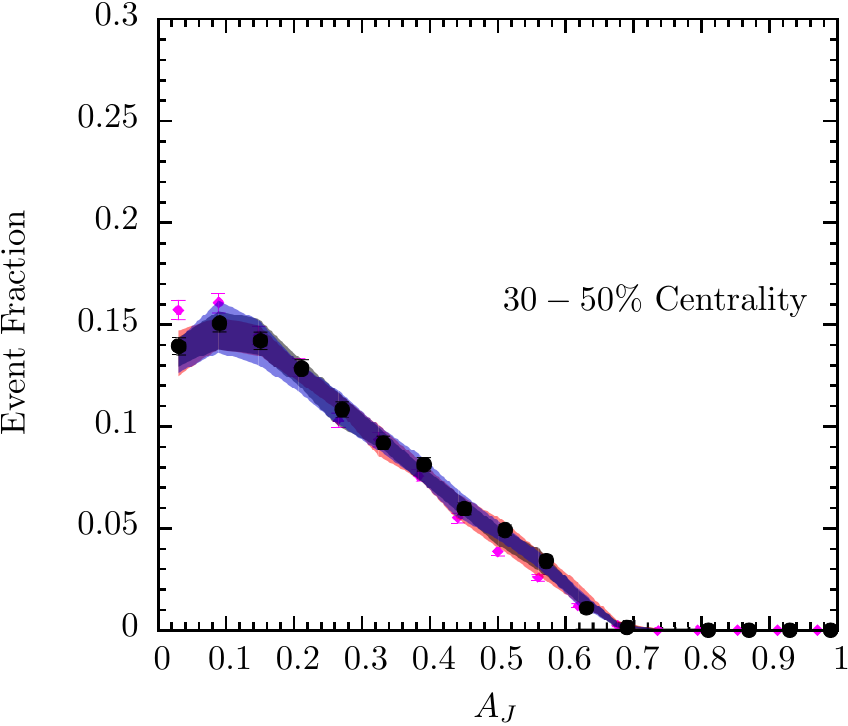}
\put(-88,88){\small{$\sqrt{s}=2.76~\rm{ATeV}$}}
\end{tabular}
\caption{\label{Fig:RAAcomp}  
Upper panel: Jet $\Raa$ as a function of $\pt$ for heavy ion collisions with $\sqrt{s}=2.76$~ATeV
in two different centrality bins for the hybrid model with its strongly coupled rate of energy
loss and the two control models from
Section~\ref{sec:hybridmodel}, as compared to preliminary CMS data \cite{Raajet:HIN}. 
Lower panel: Dijet imbalance distribution in two different centrality bins for those three energy loss models, as compared to CMS data
from Ref.~\cite{Chatrchyan:2012nia}. 
Each of the three models
for the rate of energy loss $dE/dx$ includes one free parameter, and in each case we have
fitted the value of this parameter to obtain agreement between the model and the data
for $100~{\rm GeV}<\pt<110~{\rm GeV}$ in the most central ($0-10$\%) collisions.
}
\end{figure}

\begin{figure}[tbp]
\centering 
\begin{tabular}{cc}
\includegraphics[width=.5\textwidth]{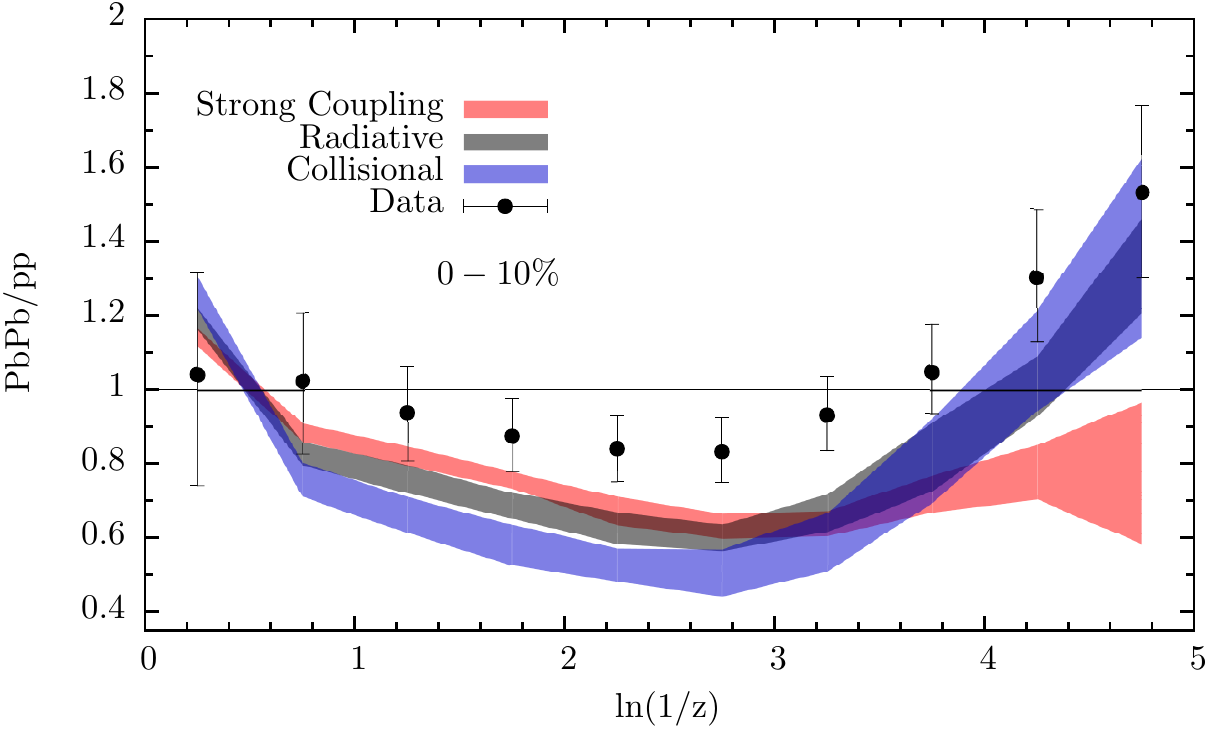}
\put(-90,108){\tiny{$\sqrt{s}=2.76~\rm{ATeV}$}}
&
\includegraphics[width=.5\textwidth]{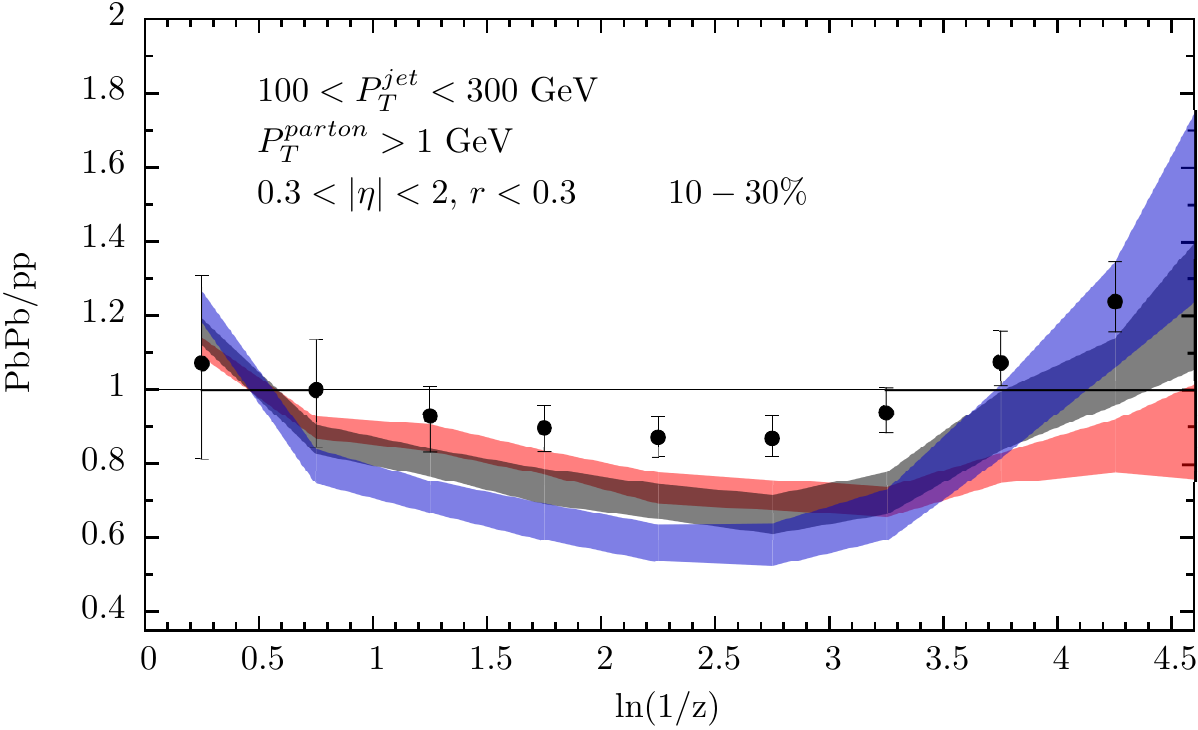}
\put(-97,108){\tiny{$\sqrt{s}=2.76~\rm{ATeV}$}}
\\
\includegraphics[width=.5\textwidth]{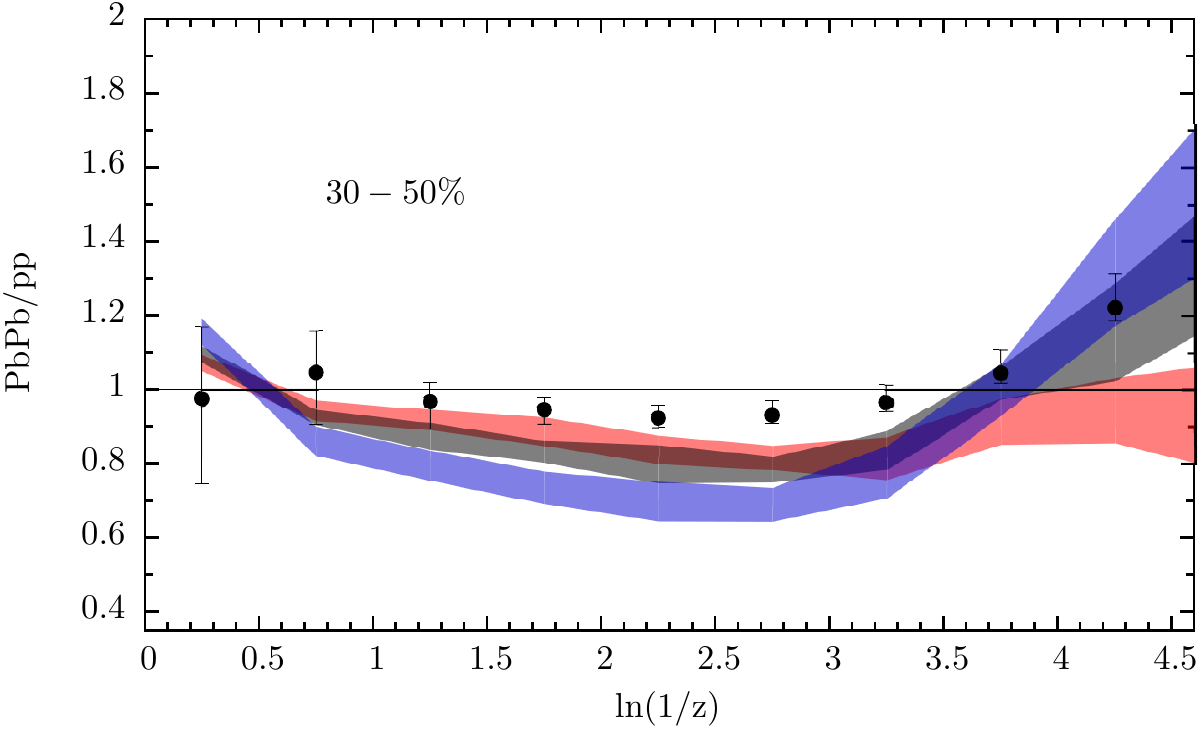}
\put(-90,108){\tiny{$\sqrt{s}=2.76~\rm{ATeV}$}}
&
\includegraphics[width=.5\textwidth]{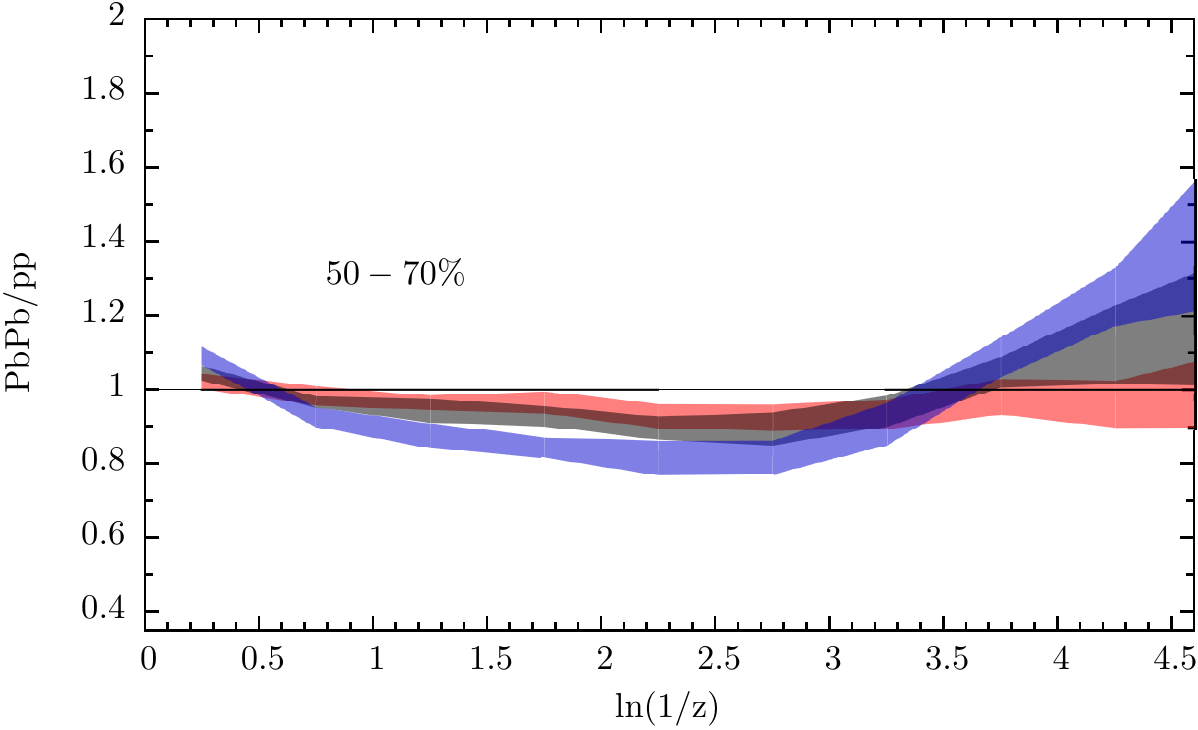}
\put(-90,108){\tiny{$\sqrt{s}=2.76~\rm{ATeV}$}}
\end{tabular}
\caption{\label{Fig:FFdijet276}  
  Partonic fragmentation function ratios (fragmentation function for jets in Pb-Pb collisions over that for
  jets in p-p collisions) for jets with $100<\pt^{\rm jet} <300$~GeV in
  heavy ion collisions with $\sqrt{s}=2.76$~ATeV with four different centralities 
  from the hybrid model and the two control models
  of the rate of energy loss. Jets are reconstructed with the anti-$k_T$ algorithm with $R=0.3$. 
  The jet fragments consist of final state partons within a cone of radius $r=0.3$ around the jet axes 
  determined by the reconstruction algorithm. These partons are classified with respect to the 
  longitudinal variable $z=p^\parallel/ p^{\rm jet}$ with $p^\parallel$ the component of the 
  momentum of the fragment along the jet axis.  Note that
in  the softest region of the fragmentation function, say $\ln (1/z)\gtrsim 3.5-4$,
there is an additional contribution to the fragmentation function that is not included
in the model: as the jet deposits energy and momentum into the medium, some part of the medium
ends up moving as a wake in the direction of the jet.  This will serve to push up
the softest region of the fragmentation function in all our models, by an amount that
should be model independent to a good approximation. This effect is not included in our models,
but does contribute in the experimental data.
The three models are compared to the data from Ref.~\cite{Chatrchyan:2014ava}. 
The separation between the models is cause for optimism that the higher statistics 
measurements  expected from LHC heavy ion Run 2 may serve to distinguish between models.  This optimism
must be tempered, however, given that the predictions of the models do not differ in 
a qualitative way and given that we have left out
both hadronization and the response of the medium
to the passage of the jet.
}
\end{figure}

In this Appendix, we update our analysis of the single-jet and dijet observables that
we presented in Ref.~\cite{Casalderrey-Solana:2014bpa}
to include the effect of fluid flow on the rate of energy loss, as described in Section~\ref{Sec:FlowEffects}. 
We also employ the viscous hydrodynamic simulations of Ref.~\cite{Shen:2014vra}, as described in Section~\ref{sec:MC} and
as in all results presented in this paper.   
As discussed in Section \ref{sec:effects_flow},  
we have followed the same fitting procedure as we used in Ref.~\cite{Casalderrey-Solana:2014bpa} 
to determine the value of the single parameter which controls each the rate of energy loss in our hybrid model and in our
two control models. The results of these fits are summarized in Table~\ref{alphatable}. 
Since the change in all the observables is minor, in this Appendix we will not describe each observable in detail.
We simply plot the results in Figs.~\ref{Fig:RAA}, \ref{RHICANDLHC}, \ref{Fig:Asym}, \ref{Fig:RAAcomp} and \ref{Fig:FFdijet276}, and refer the reader to the our previous publication for all discussion. 
  
\clearpage
\section{\label{DijetPredictions}Predictions for Single-jet and Dijet Observables at $\sqrt{s}=5.02$~ATeV}

In this Appendix, we provide predictions for the single jet and dijet observables that we presented in detail in 
 our previous publication \cite{Casalderrey-Solana:2014bpa} and updated in Appendix \ref{update}, now 
 for Pb-Pb collisions at $\sqrt{s}=5.02$~ATeV. 
We study $10^6$  dijet events in p-p collisions  
at $\sqrt{s}=5.02$~TeV 
generated by  \pythia 8.183~\cite{Sjostrand:2007gs} without any underlying event. 
For each of the centrality bins we consider, we embed these hard scattering processes into 
 the hydrodynamic simulations along the lines of those in Ref.~\cite{Shen:2014vra} as we described
 in Section~\ref{sec:effects_flow}, but simulations
 of heavy ion collisions with $\sqrt{s}=5.02$~ATeV. 
 We then follow the procedure for determining the energy
 loss outlined in Section \ref{sec:MC}. 
 As in our previous computations,  we smear our predictions to simulate resolution effects. However, 
 and similarly to the procedure we have followed in obtaining
 our photon-jet  and Z-jet predictions described in Sections~\ref{sec:photonjet_results}  and \ref{sec:zjet_predictions}, 
 we use the smearing functions determined at $\sqrt{s}=2.76$~ATeV,
 since they are yet unknown at $\sqrt{s}=5.02$~ATeV.

\begin{figure}[tbp]
\centering 
\begin{tabular}{cc}
\includegraphics[width=.5\textwidth]{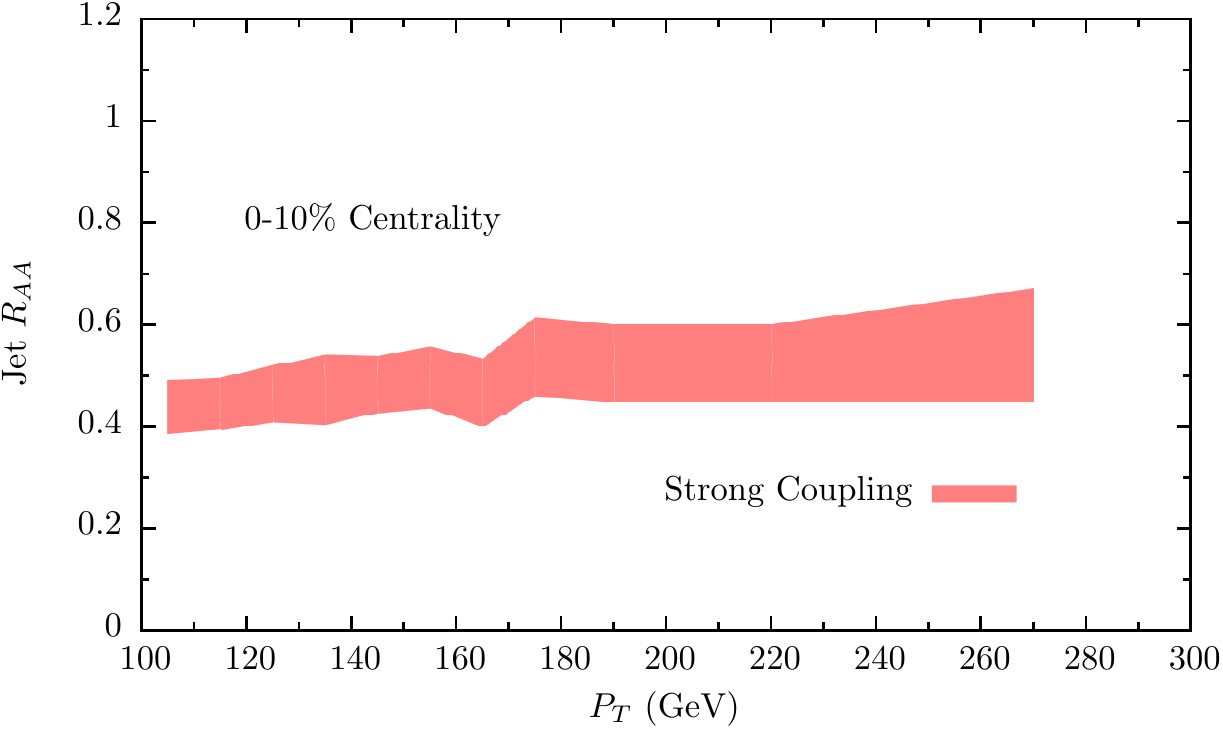}
\put(-175,31){\tiny{$\sqrt{s}=5.02~\rm{ATeV}$}}
&
\includegraphics[width=.5\textwidth]{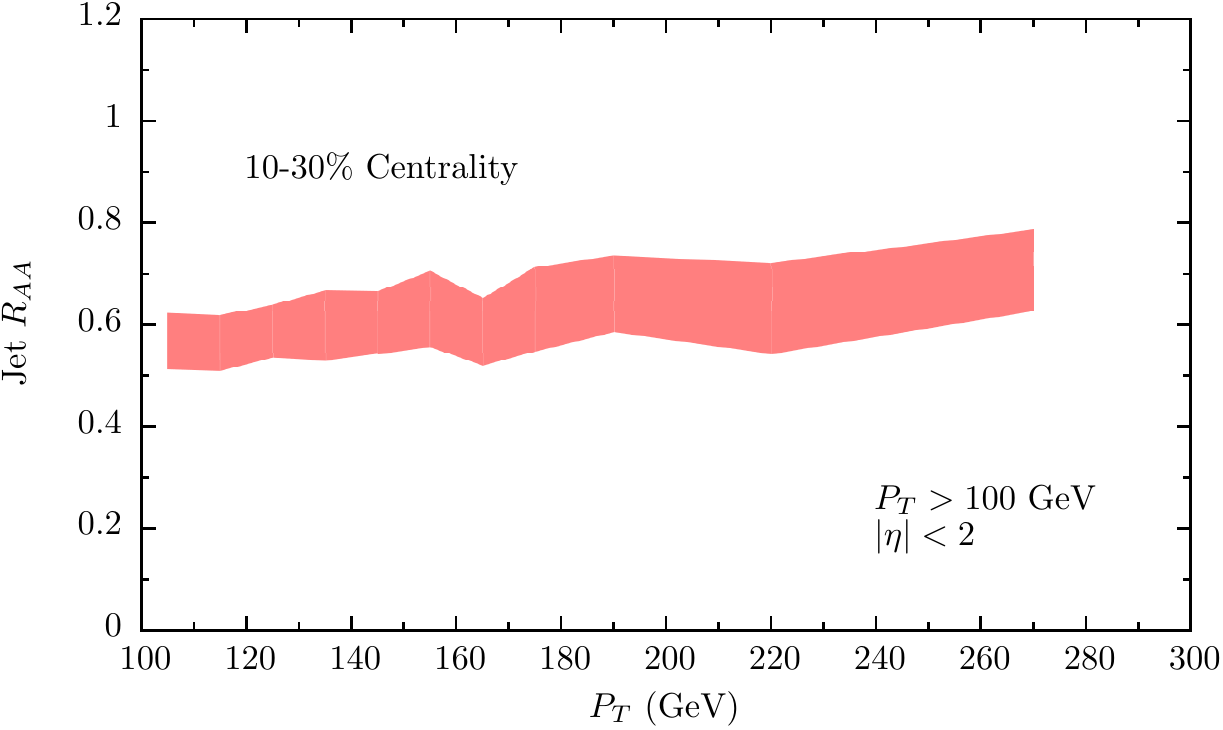}
\put(-175,31){\tiny{$\sqrt{s}=5.02~\rm{ATeV}$}}
\\
\includegraphics[width=.5\textwidth]{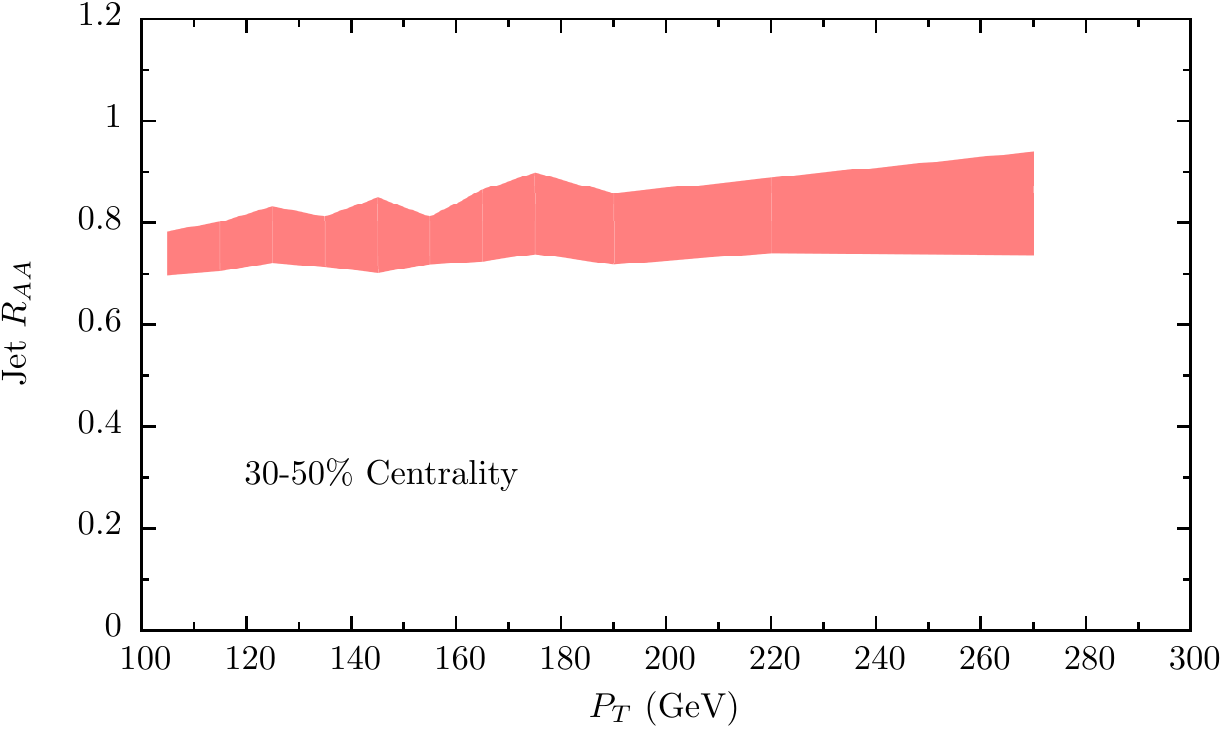}
\put(-175,31){\tiny{$\sqrt{s}=5.02~\rm{ATeV}$}}
&
\includegraphics[width=.5\textwidth]{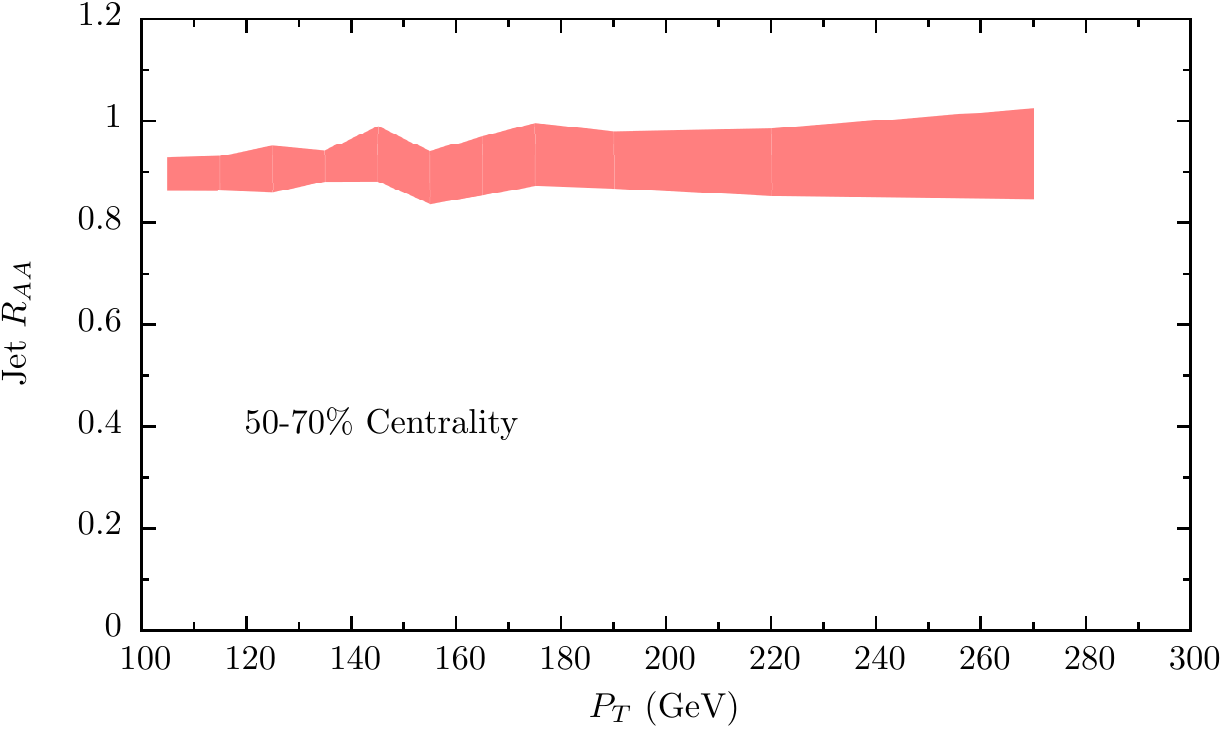}
\put(-175,31){\tiny{$\sqrt{s}=5.02~\rm{ATeV}$}}
\end{tabular}
\caption{\label{Fig:RAA_prediction} 
Hybrid model prediction for 
Jet $\Raa$ as a function of $\pt$ for different centralities at $\sqrt{s}=5.02$~ATeV. The single model parameter has been fitted to $\sqrt{s}=2.76$~ATeV data 
previously and no additional parameters have been introduced. 
}
\end{figure}

\begin{figure}[tbp]
\centering 
\begin{tabular}{cc}
\includegraphics[width=.5\textwidth]{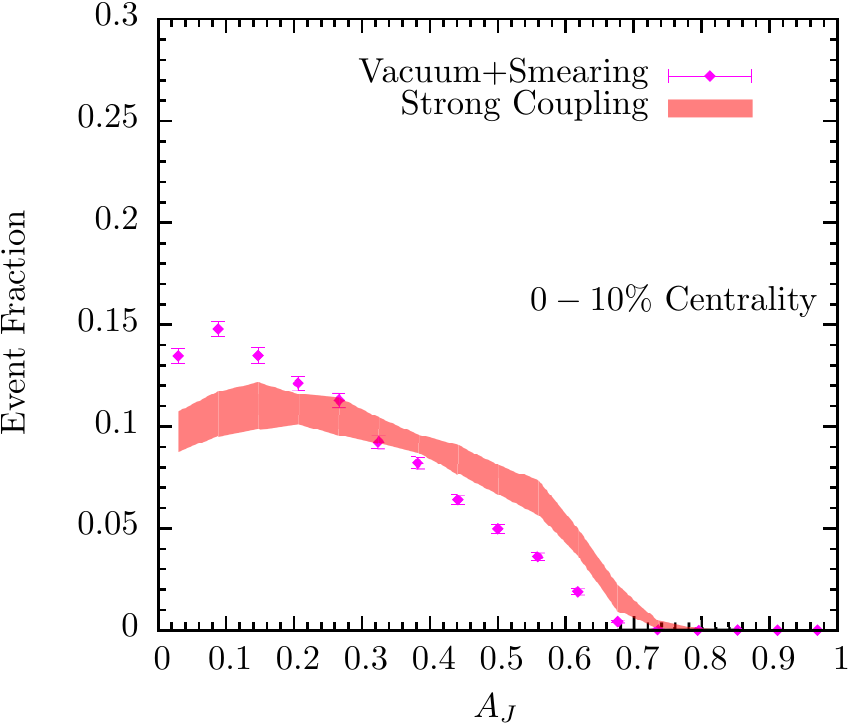}
\put(-85,88){\small{$\sqrt{s}=5.02~\rm{ATeV}$}}
&
\includegraphics[width=.5\textwidth]{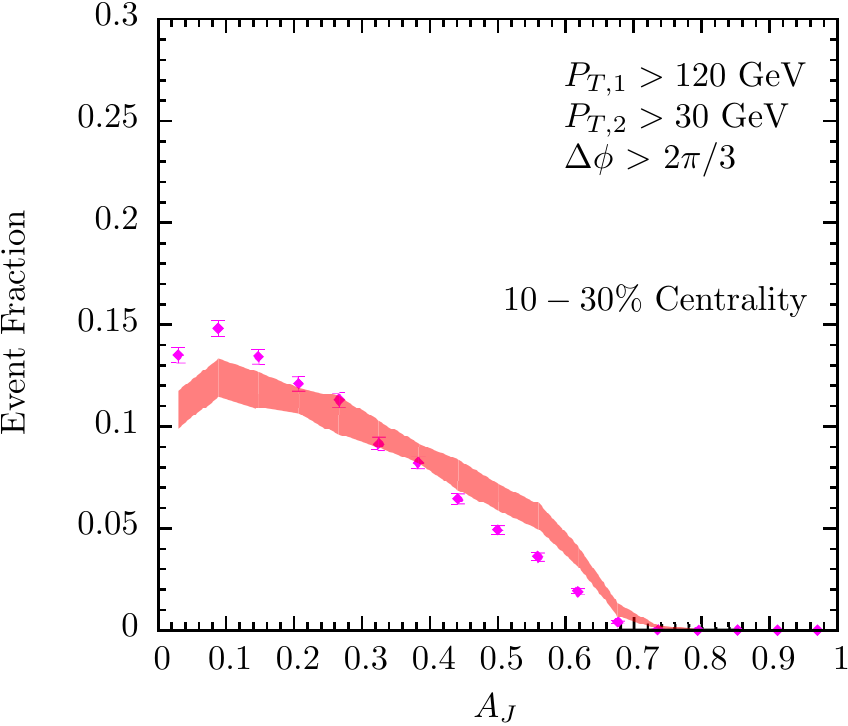}
\put(-88,88){\small{$\sqrt{s}=5.02~\rm{ATeV}$}}
\\
\includegraphics[width=.5\textwidth]{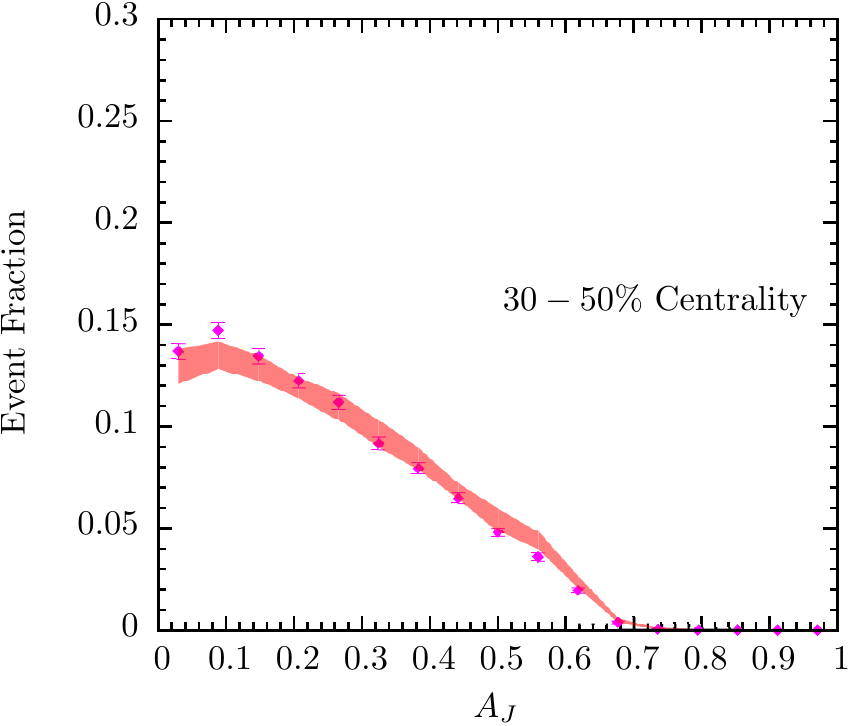}
\put(-88,88){\small{$\sqrt{s}=5.02~\rm{ATeV}$}}
&
\includegraphics[width=.5\textwidth]{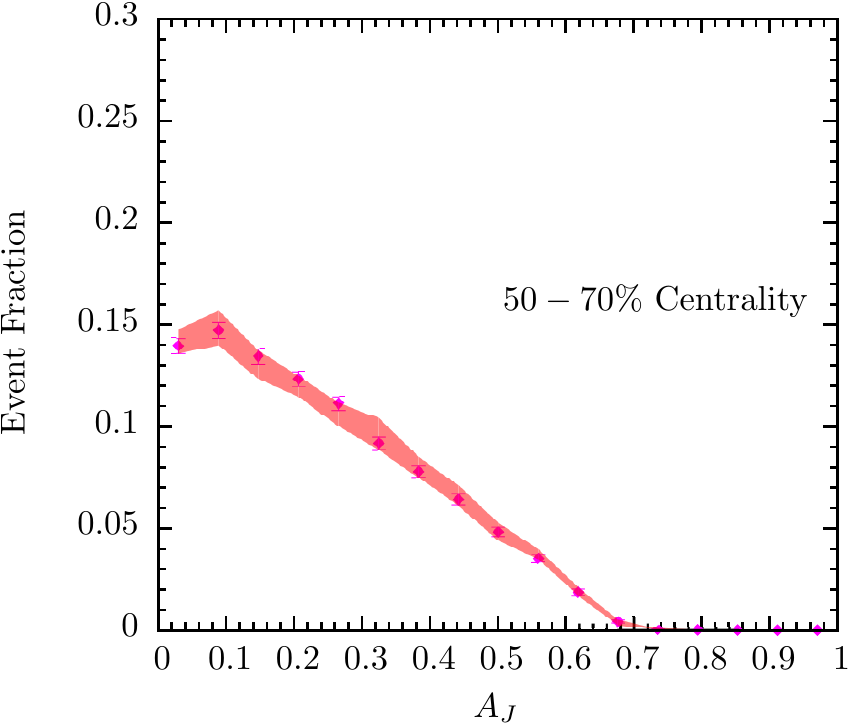}
\put(-88,88){\small{$\sqrt{s}=5.02~\rm{ATeV}$}}
\end{tabular}
\caption{\label{Fig:Asym_prediction}   
Hybrid model predictions for the  dijet imbalance $A_J$
in Pb-Pb collisions at $\sqrt{s}$=5.02~ATeV with different centralities. 
Both the theoretical calculations and the proton-proton reference are
 smeared according to the prescription in Ref.~\cite{YYthesis}.
}
\end{figure}

Fig.~\ref{Fig:RAA_prediction} shows the hybrid strong/weak coupling model 
predictions for the suppression factor $R^{\rm jet}_{AA}$ of jets in heavy ion
collisions with $\sqrt{s}=5.02$~ATeV reconstructed with the anti-$k_T$ algorithm with $R=0.3$. 
We do not fit even a single parameter here, or anywhere in this Appendix.  The model is fully constrained by
the data on collisions with $\sqrt{s}=2.76$~ATeV, with its one parameter having been fixed
as described in Section~\ref{sec:effects_flow} and Appendix~\ref{update}.
The width of the displayed bands is a combination of our theoretical uncertainties (estimated
by varying the temperature $T_c$ below which we stop quenching, see Section~\ref{sec:Section2})
together with the experimental uncertainties
in the data that we use to fix the one parameter in each model.
Both the centrality dependence and the transverse momentum dependence of 
$R^{\rm jet}_{AA}$ in collisions with  $\sqrt{s}=5.02$~ATeV are very similar to 
what we have seen previously at $\sqrt{s}=2.76$~ATeV, with a slight increase 
in the suppression of $R^{\rm jet}_{AA}$ at the higher collision energy. Similar 
conclusions can be drawn from the strong coupling predictions for the dijet imbalance in
collisions with  $\sqrt{s}=5.02$~ATeV, displayed in Fig.~\ref{Fig:Asym_prediction}. The centrality dependence and the momentum dependence of this observable are very similar at the two  collision energies as well.

\begin{figure}[tbp]
\centering 
\begin{tabular}{cc}
\includegraphics[width=.5\textwidth]{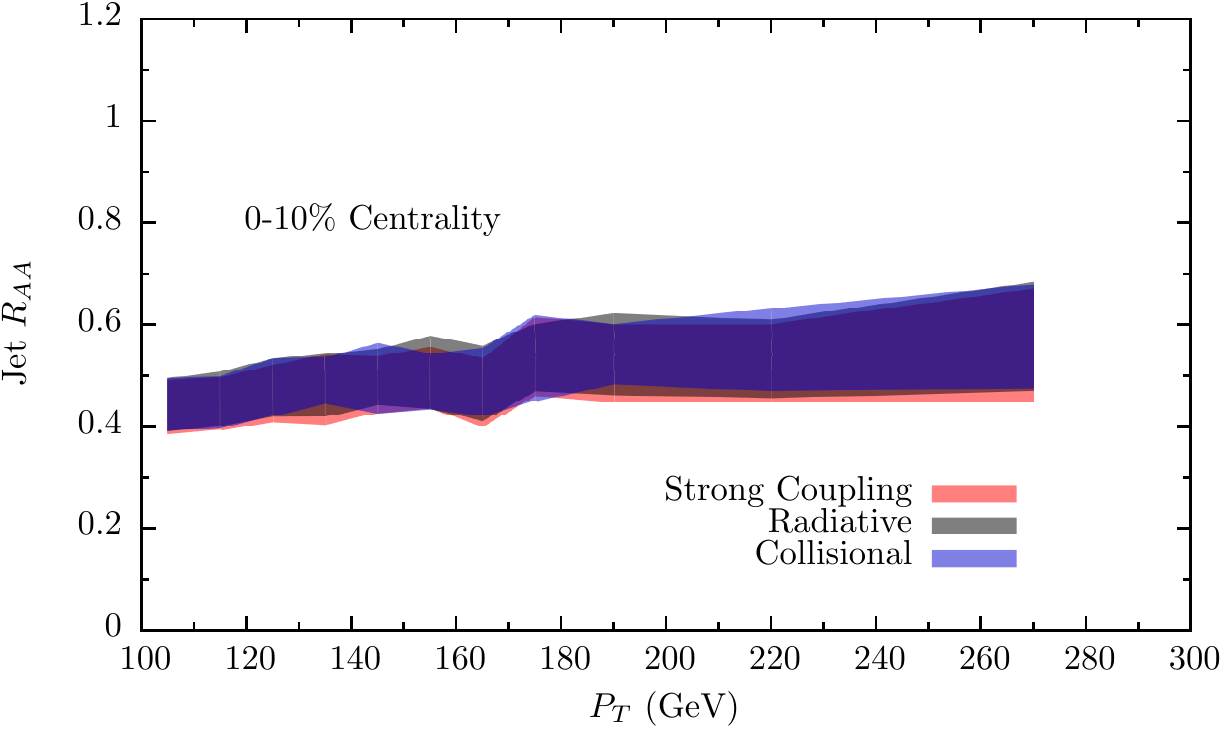}
\put(-175,34){\tiny{$\sqrt{s}=5.02~\rm{ATeV}$}}
&
\includegraphics[width=.5\textwidth]{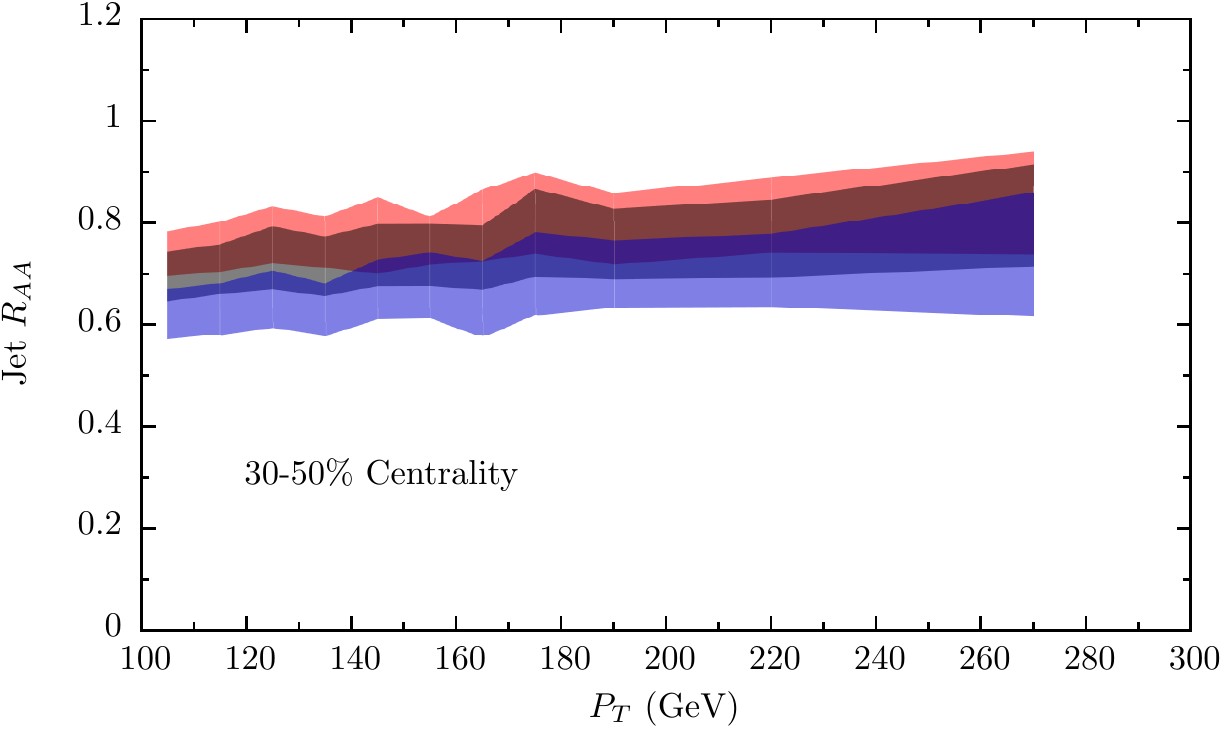}
\put(-175,34){\tiny{$\sqrt{s}=5.02~\rm{ATeV}$}}
\\
\includegraphics[width=.5\textwidth]{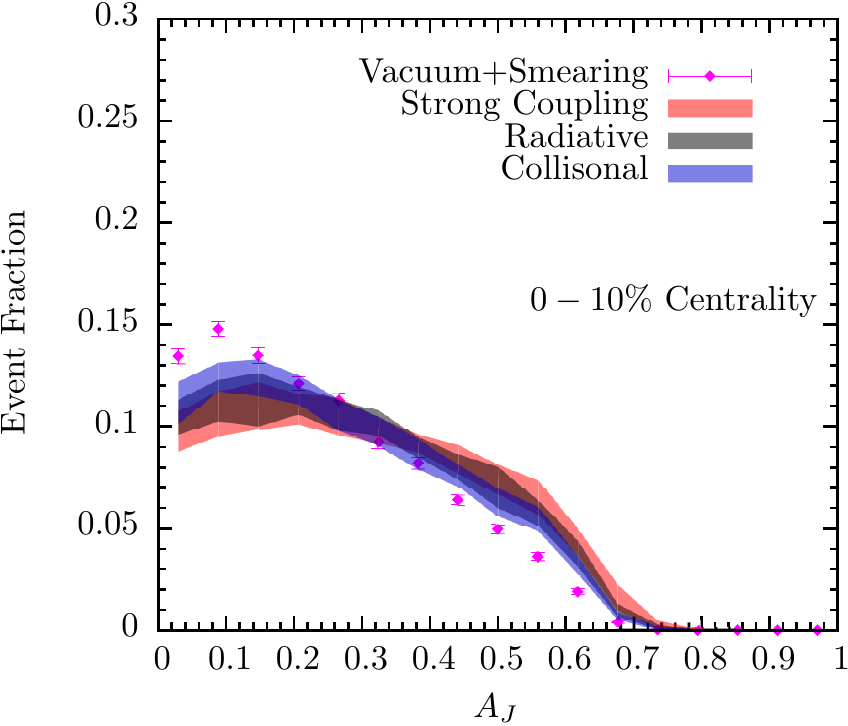}
\put(-85,88){\small{$\sqrt{s}=5.02~\rm{ATeV}$}}
&
\includegraphics[width=.5\textwidth]{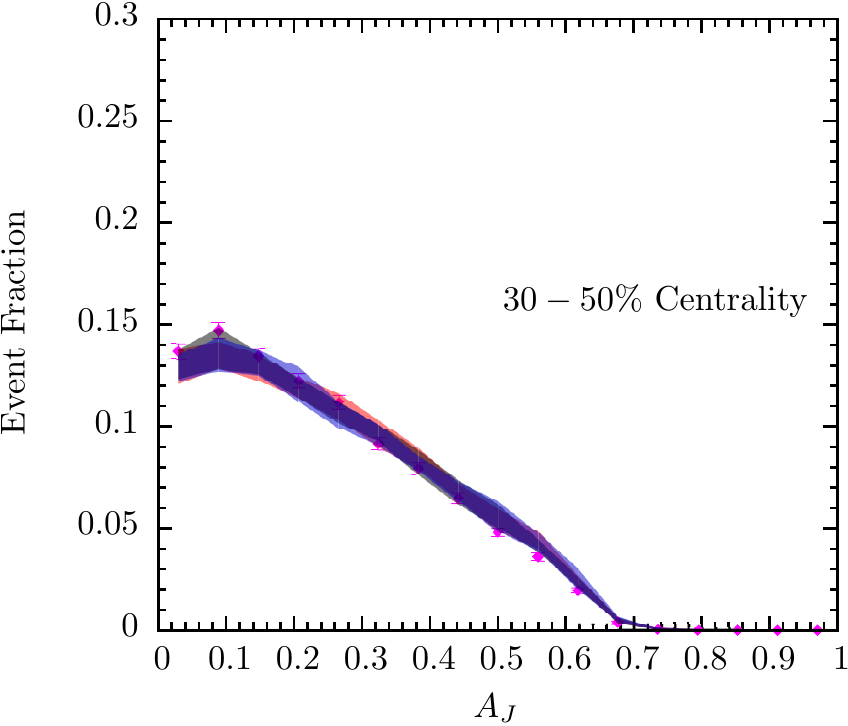}
\put(-88,88){\small{$\sqrt{s}=5.02~\rm{ATeV}$}}
\end{tabular}
\caption{\label{Fig:RAAcomp_prediction}  
Model dependence of dijet observables for Pb-Pb collisions with $\sqrt{s}=5.02$~ATeV.
Upper panel: Jet $\Raa$ as a function of $\pt$ for LHC collisions
in two different centrality bins for the three energy loss models from
Section~\ref{sec:hybridmodel}.
Lower panel: Dijet imbalance distribution in two different centrality 
bins for the three energy loss models. }
\end{figure}

The sensitivity of these predictions  to the form we assume for the rate of energy loss
is examined in Fig.~\ref{Fig:RAAcomp_prediction}. 
As at the lower collision energy,
this set of predictions shows 
little discriminating power to our choice among the three models for the rate of energy loss that we have investigated.
The strongly coupled form  (\ref{CR_rate}) for $dE/dx$
leads to a slightly large suppression and a slightly bigger dijet imbalance than the two  control models (\ref{Eloss_equations}),
bottom-left panel of Fig.~\ref{Fig:RAAcomp_prediction},
but the effect is small compared to current uncertainties. 
Some separation among models is observed with increasing centrality, top right panel of  Fig. \ref{Fig:RAAcomp_prediction}, although the largest centralities are more and more sensitive to the energy lost by 
energetic partons during
the hadronic phase of the collision, 
which we are neglecting.  

\begin{figure}[tbp]
\centering 
\begin{tabular}{cc}
\includegraphics[width=.5\textwidth]{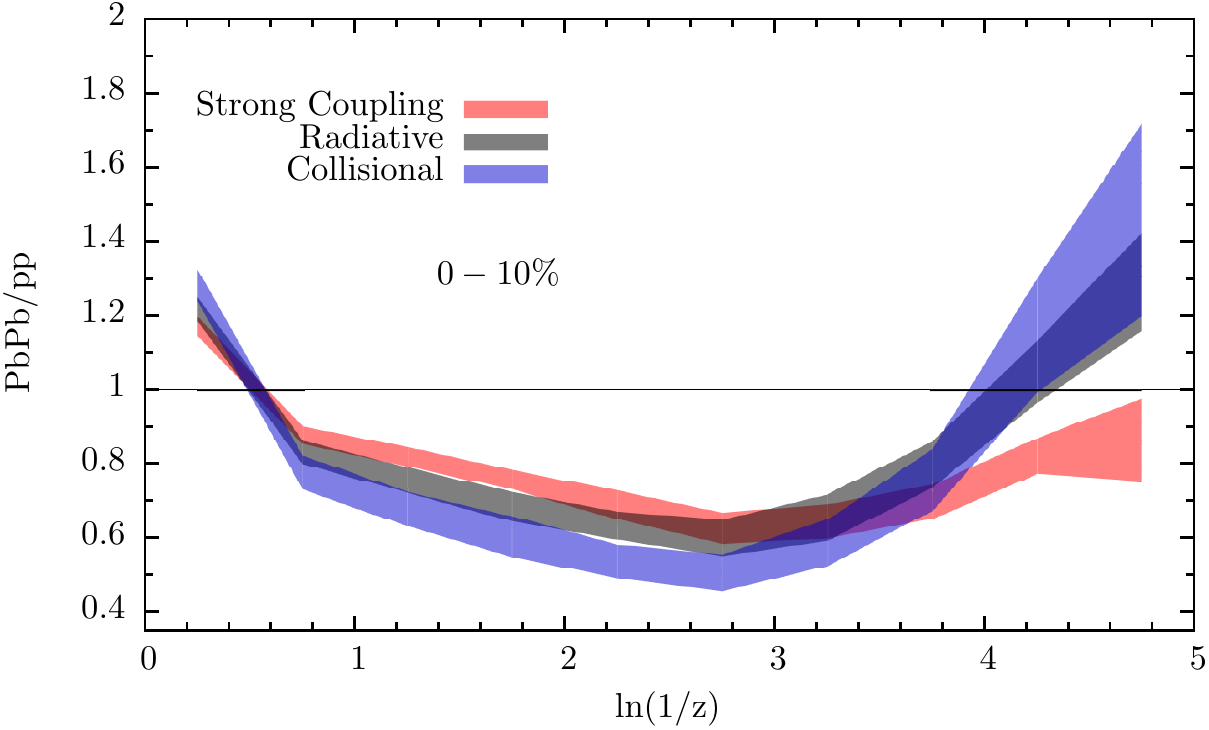}
\put(-90,108){\tiny{$\sqrt{s}=5.02~\rm{ATeV}$}}
&
\includegraphics[width=.5\textwidth]{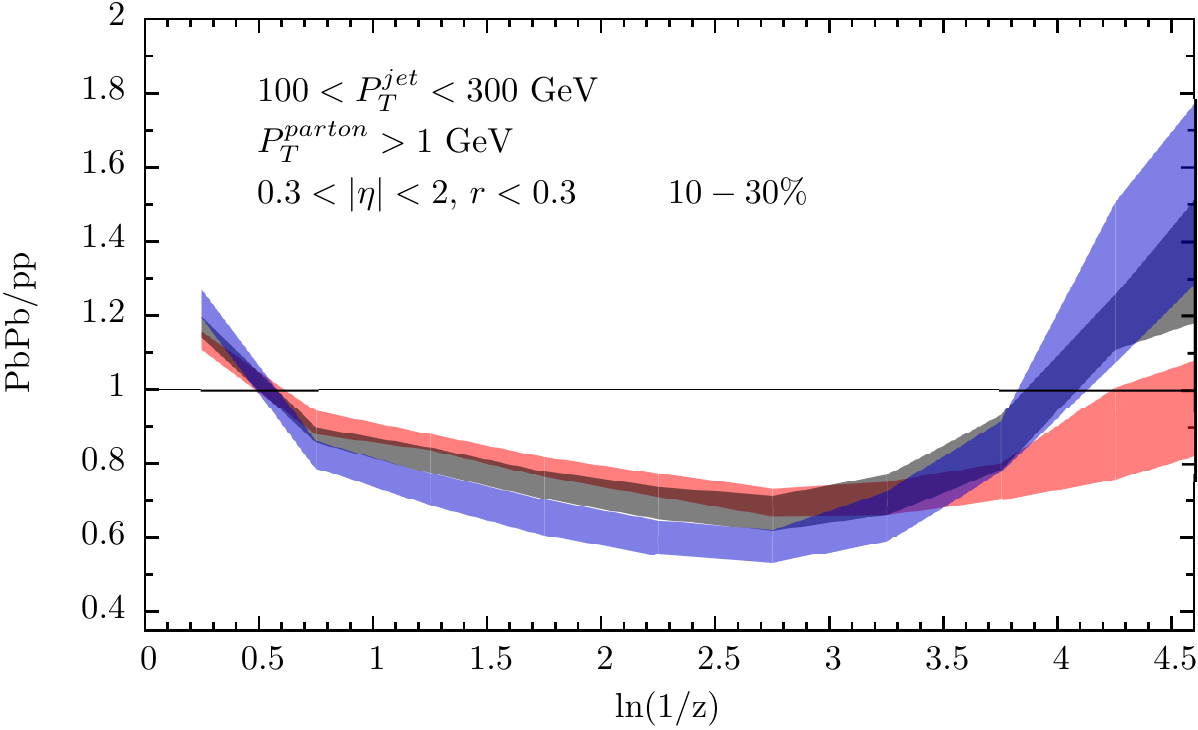}
\put(-97,108){\tiny{$\sqrt{s}=5.02~\rm{ATeV}$}}
\\
\includegraphics[width=.5\textwidth]{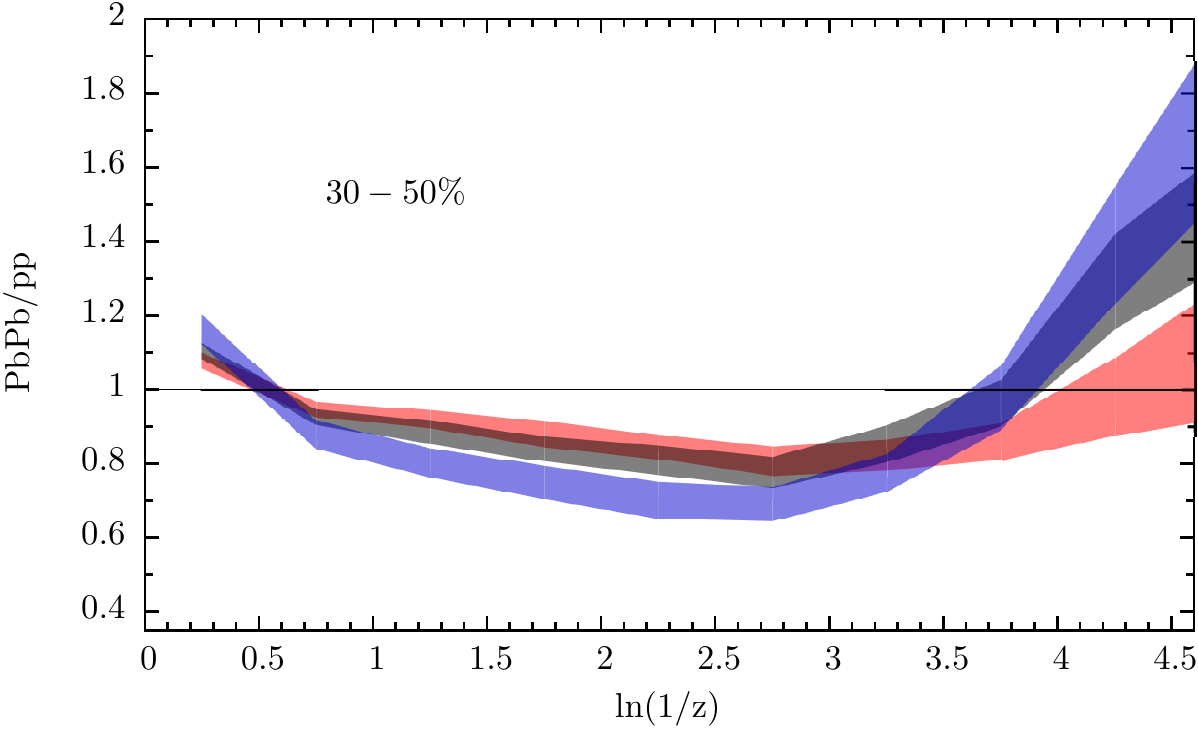}
\put(-90,108){\tiny{$\sqrt{s}=5.02~\rm{ATeV}$}}
&
\includegraphics[width=.5\textwidth]{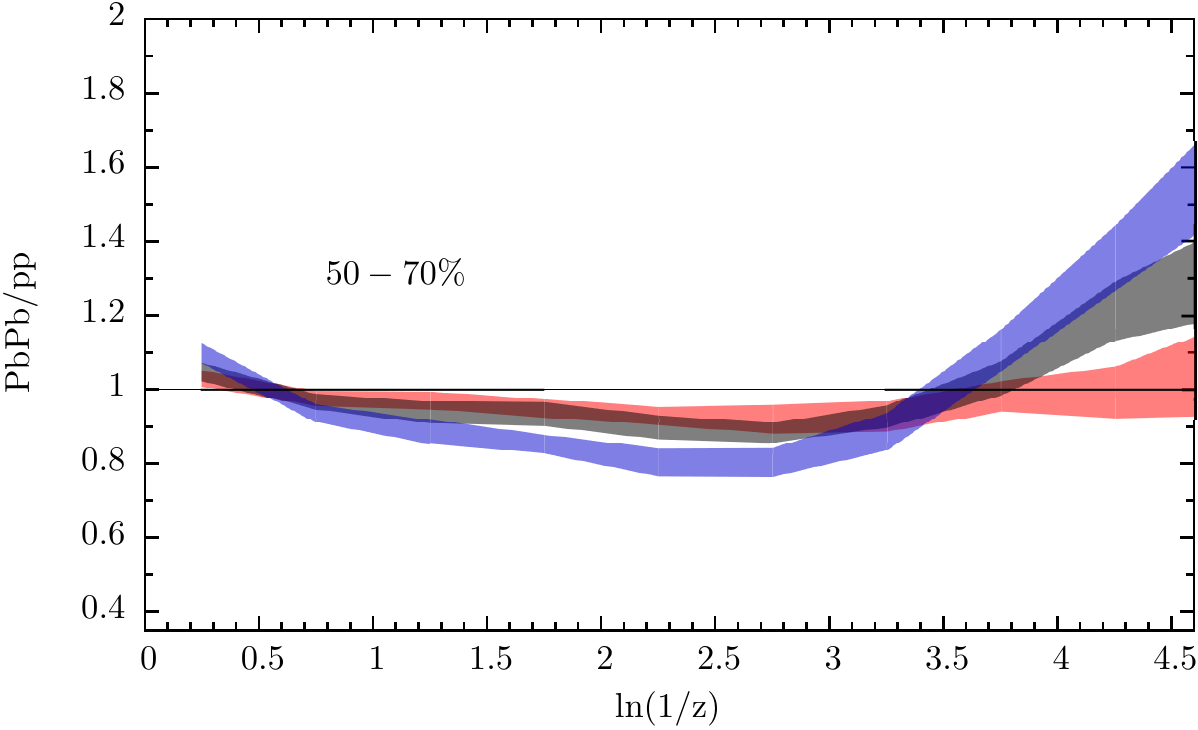}
\put(-90,108){\tiny{$\sqrt{s}=5.02~\rm{ATeV}$}}
\end{tabular}
\caption{\label{Fig:FF_prediction}  Partonic fragmentation functions for jets of $100<\pt^{\rm jet} <300$~GeV in heavy ion collisions
with $\sqrt{s}=5.02$~ATeV for three different models for the rate of energy loss and for four different centralities. 
Jets are reconstructed with the anti-$k_T$ algorithm with $R=0.3$. 
The jet fragments consist of final state partons within a cone of angle $r=0.3$ around the jet axes determined by the reconstruction algorithm. 
These partons are classified with respect to the longitudinal variable $z=p^\parallel/ p^{\rm jet}$ with $p^\parallel$ the momentum of the fragments along the jet axis.
}
\end{figure}

Finally, in Figure \ref{Fig:FF_prediction} we compare the predictions for partonic fragmentation functions 
obtained from the three models for the rate of energy loss.
These are, again, similar to one another. The modest  
separation between the model predictions
at intermediate values of $z$ observed at $\sqrt{s}=2.76$~ATeV is also observed at $\sqrt{s}=5.02$~ATeV. 
However, the model predictions separate most in the region of $\log z >3.5$ 
where they cannot be relied upon. In this softest region of the fragmentation function,
there is an additional contribution that is not included in the model: the backreaction of
the medium to the jet passing through it will result in additional soft particles in the jet cone.

\section{\label{Model_dep}Model Dependence of Boson-Jet Correlations}

In this Appendix, we study the sensitivity of the different photon-jet and Z-jet 
observables that we have considered in Sections~\ref{sec:photonjet_results} and \ref{sec:zjet_predictions} 
to the microscopic mechanism responsible for energy loss. 
To do so, we repeat our analysis with the strongly coupled form for the rate of energy loss (\ref{CR_rate})
that we use in our hybrid model replaced by one or other of the expressions (\ref{Eloss_equations})
that define our two control models, which are inspired by the radiative and collisional energy loss mechanisms. 
As in the strong coupling case, these expressions for $dE/dx$
also contain a single parameter which is fitted to the single-jet production rate at one transverse momentum and
centrality. 
The values of the parameters obtained from these fits are summarized in Table~\ref{alphatable}.

\begin{figure}[tbp]
\centering 
\begin{tabular}{cc}
\includegraphics[width=.5\textwidth]{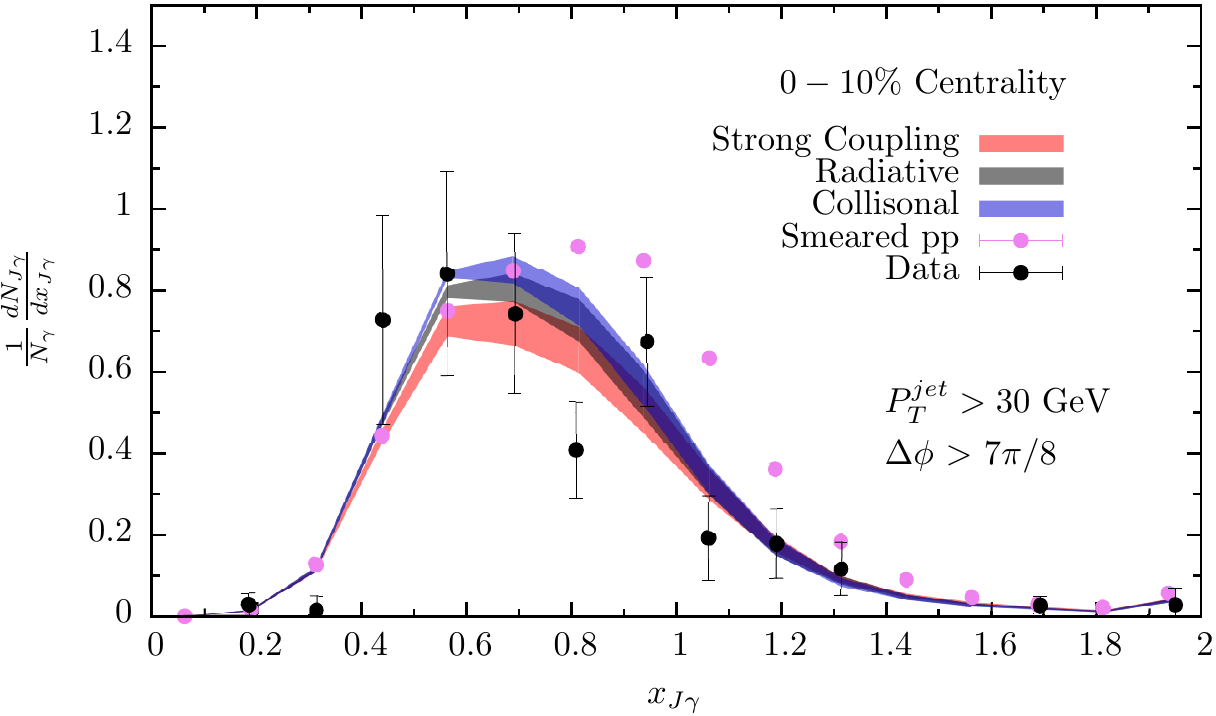}
\put(-180,110){\tiny{$\sqrt{s}=2.76~\rm{ATeV}$}}
&
\includegraphics[width=.5\textwidth]{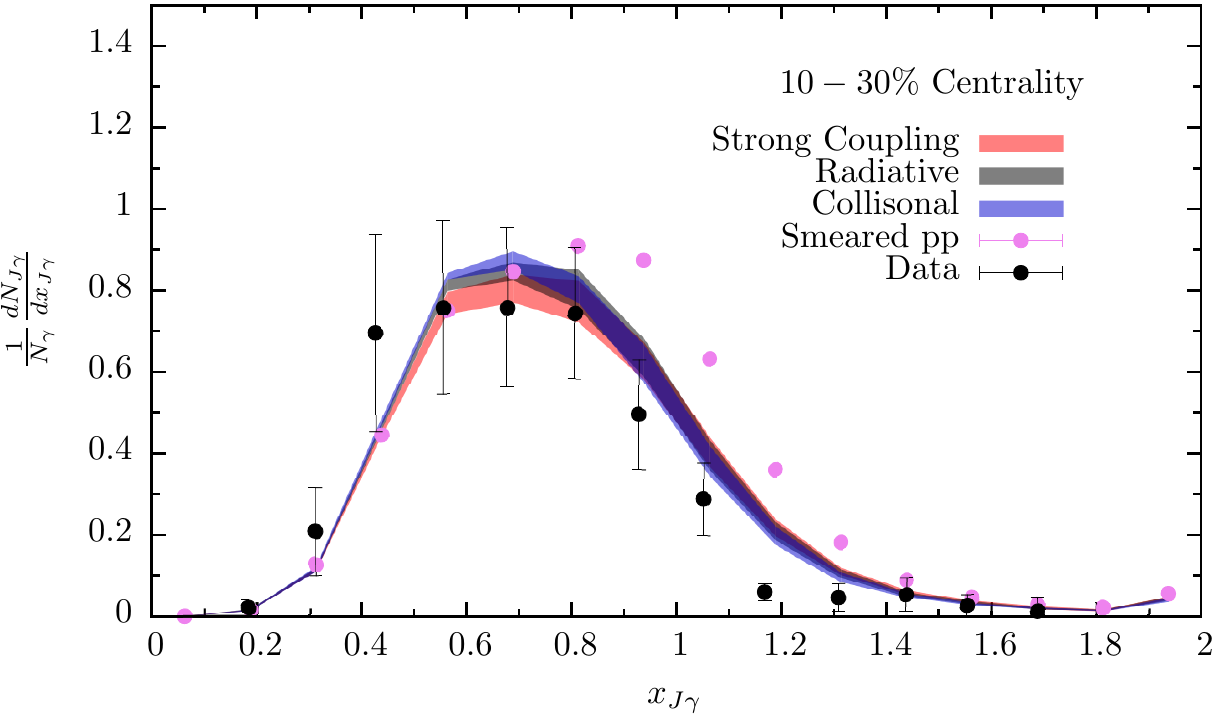}
\put(-180,110){\tiny{$\sqrt{s}=2.76~\rm{ATeV}$}}
\\
\includegraphics[width=.5\textwidth]{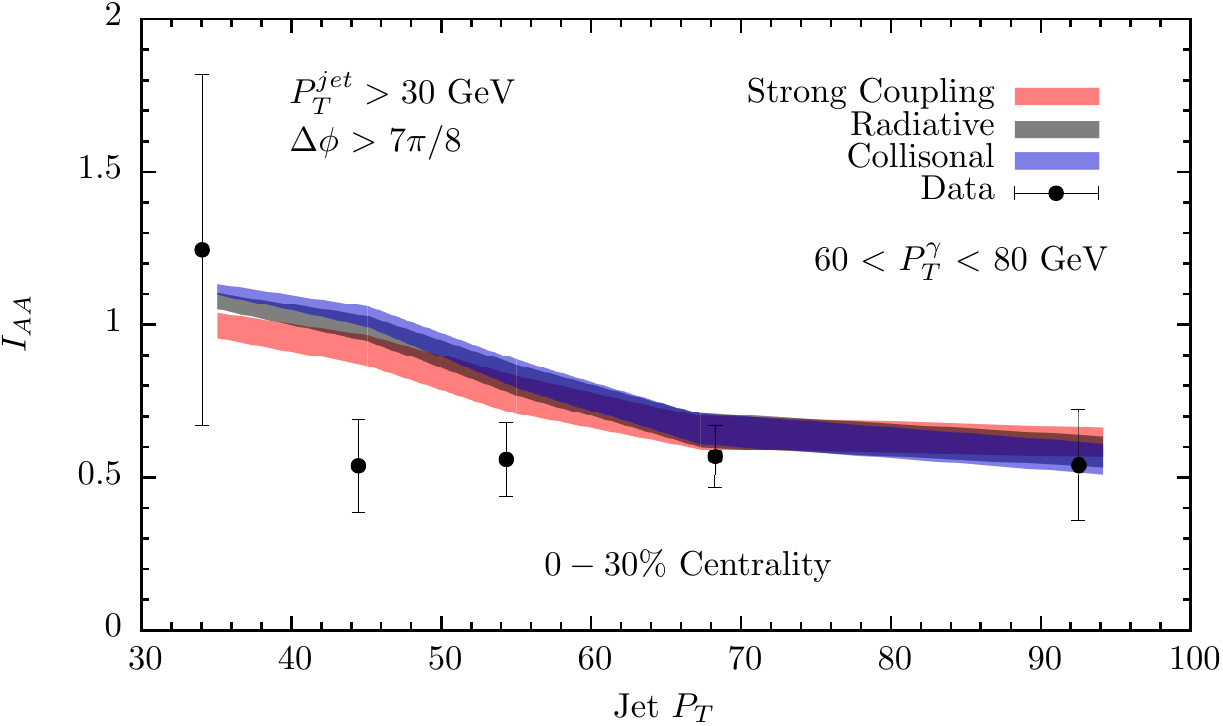}
\put(-165,90){\tiny{$\sqrt{s}=2.76~\rm{ATeV}$}}
&
\includegraphics[width=.5\textwidth]{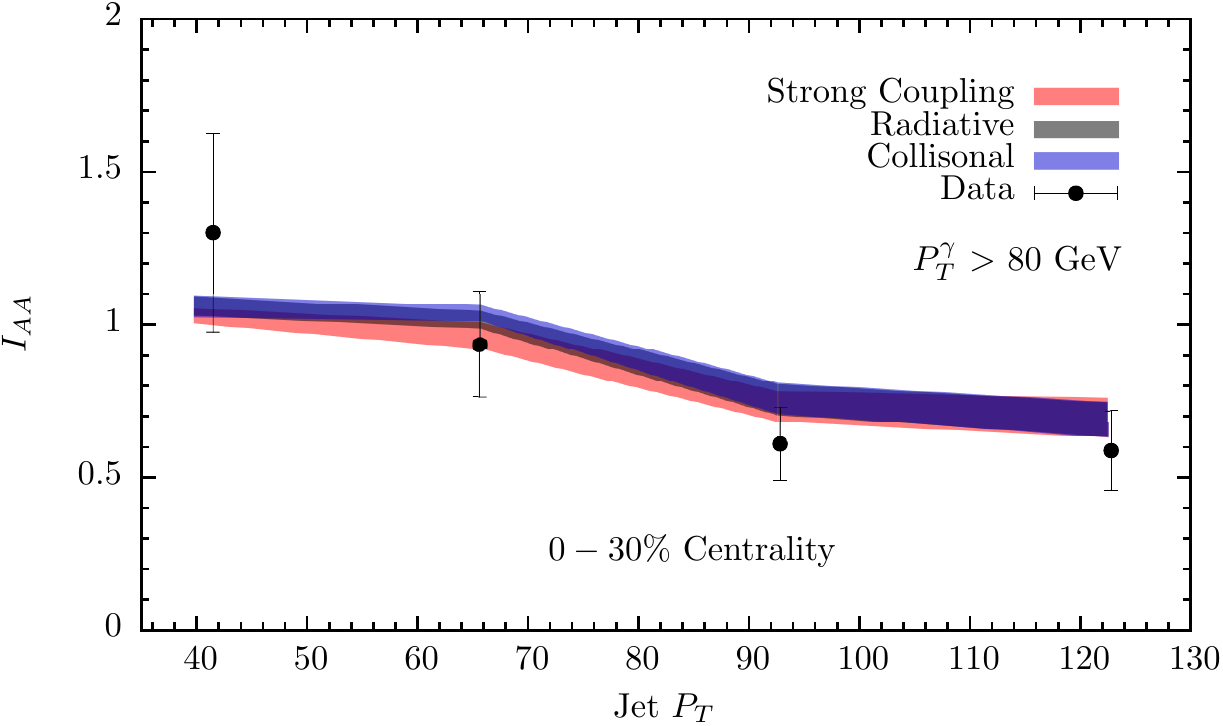}
\put(-165,90){\tiny{$\sqrt{s}=2.76~\rm{ATeV}$}}
\\
\multicolumn{2}{c}{\includegraphics[width=.5\textwidth]{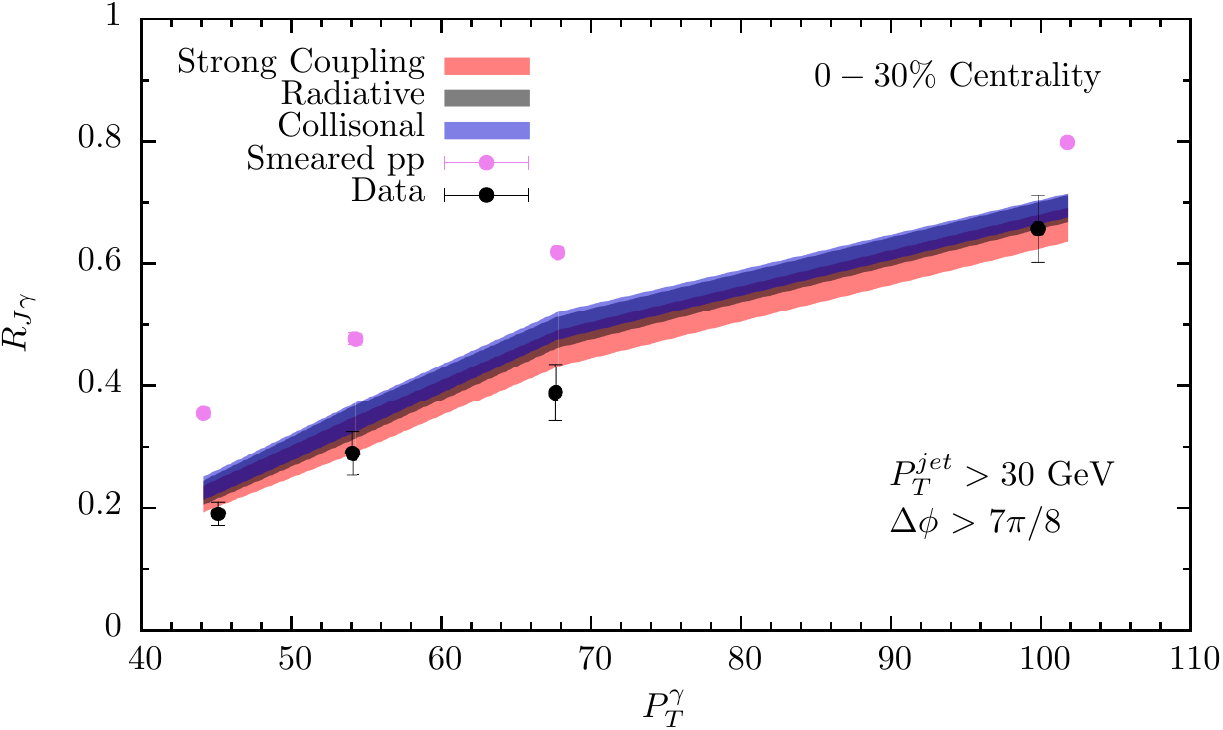}}
\put(-189,55){\tiny{$\sqrt{s}=2.76~\rm{ATeV}$}} 
\end{tabular}
\caption{\label{Fig:PJImodeldep}  
Computations of several photon-jet observables  using three different models of the energy loss mechanism for heavy ion
collisions with $\sqrt{s}=2.76$~ATeV. The distributions of the transverse momentum imbalance
of photon-jet pairs for two different centralities are displayed in the upper panels. 
The middle panel shows the ratio of the 
transverse momentum spectra of jets produced in association with an isolated photon in Pb-Pb collisions
to that in p-p collisions for two different centralities. The lower panel shows the fraction of isolated photons produced 
in association with a hard jet with $\pt^{\rm jet}> 30$~GeV at an azimuthal angle more than $7\pi/8$ away from
that of the isolated photon.
Data are taken from Ref.~\cite{CMS-HIN-13-006}.  
}
\end{figure}

In Fig.~\ref{Fig:PJImodeldep} we show the 
results that we have obtained from the hybrid strong/weak coupling model and our two control models for the
photon-jet observables that we have analyzed in heavy ion collisions with $\sqrt{s}=2.76$~ATeV. 
The procedure for determining each of the observables is identical for all models and  
is described in Section~\ref{sec:photonjet_results}. 
These plots make clear that the uncertainties in the present low-statistics data are too large
to make it possible to use
 the photon jet imbalance or the spectrum of jets produced in association with isolated photons or the fraction of isolated
 photons with an associated jet 
 to differentiate between microscopic models of the dynamics of energy loss.
 Within the current uncertainties, all the models agree 
 with the CMS data displayed in these plots. 
 The photon-jet imbalance in central collisions, top left panel of Fig.~\ref{Fig:PJImodeldep},
 does have modest power to discriminate between the strongly coupled model
 and the control models. This indicates that the higher statistics photon-jet data sets
 anticipated in LHC heavy ion Run 2 could shed light on the microscopic dynamics via which
 jets lose energy.

\begin{figure}[tbp]
\centering 
\begin{tabular}{cc}
\includegraphics[width=.5\textwidth]{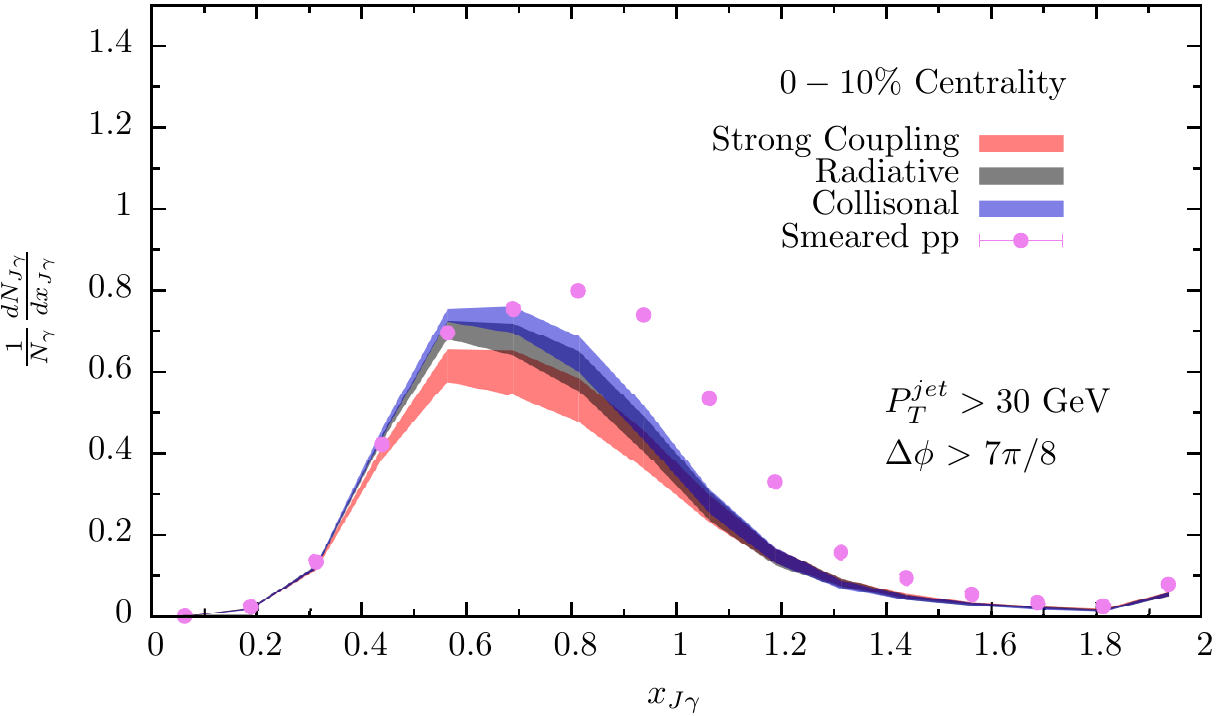}
\put(-180,110){\tiny{$\sqrt{s}=5.02~\rm{ATeV}$}}
&
\includegraphics[width=.5\textwidth]{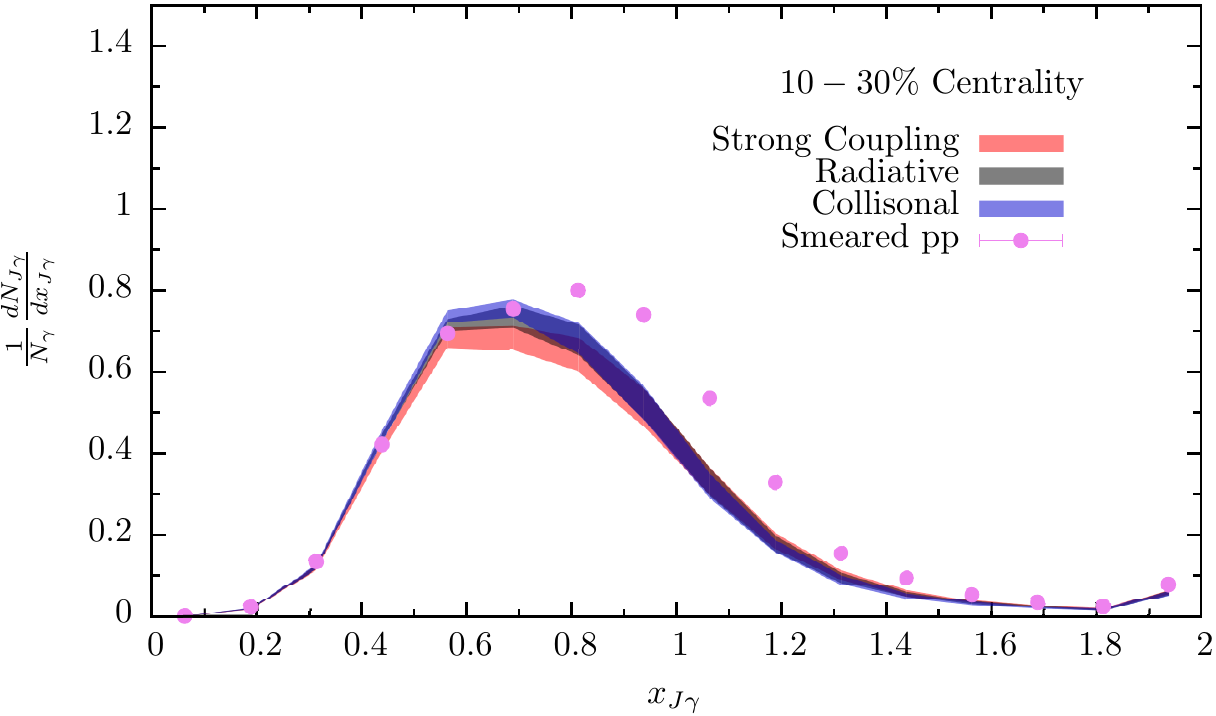}
\put(-180,110){\tiny{$\sqrt{s}=5.02~\rm{ATeV}$}}
\\
\includegraphics[width=.5\textwidth]{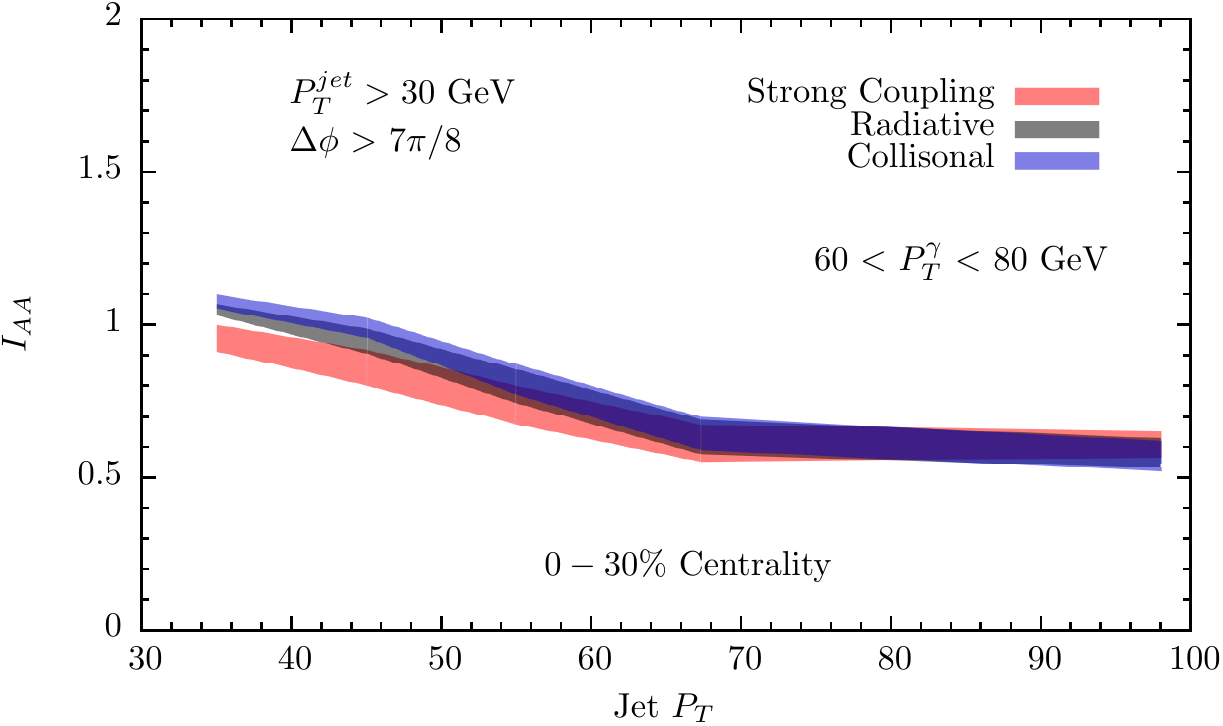}
\put(-165,90){\tiny{$\sqrt{s}=5.02~\rm{ATeV}$}}
&
\includegraphics[width=.5\textwidth]{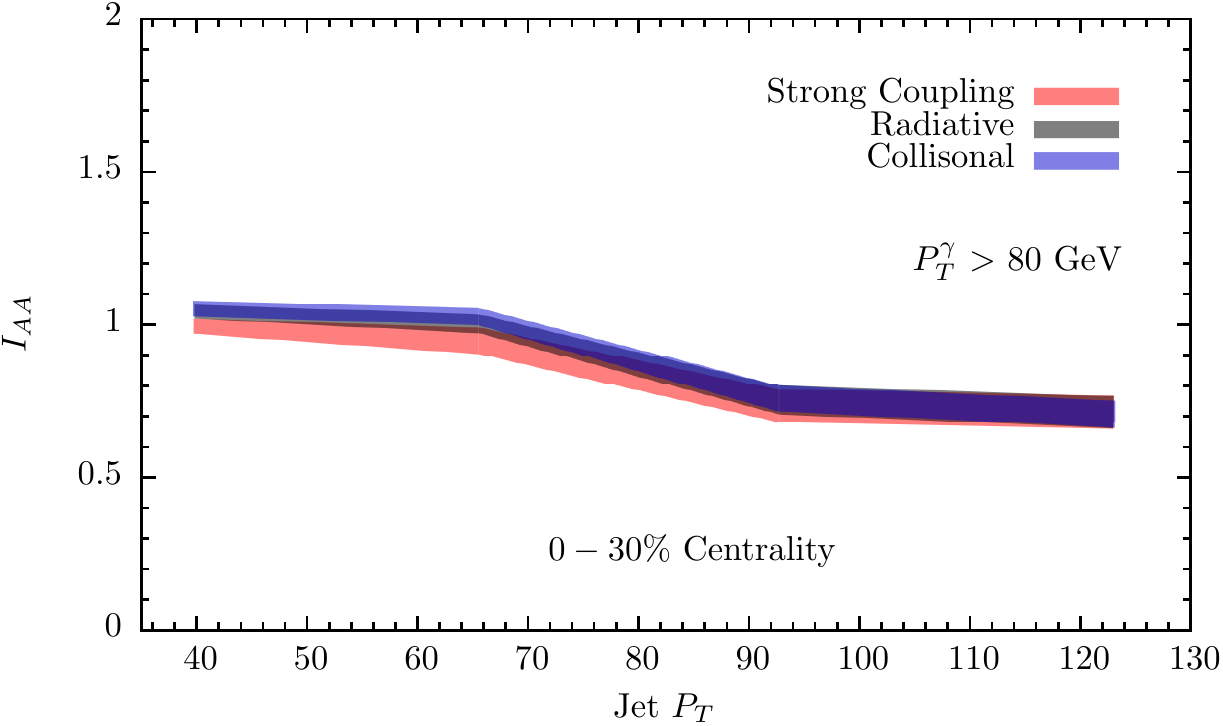}
\put(-165,90){\tiny{$\sqrt{s}=5.02~\rm{ATeV}$}}
\\
\multicolumn{2}{c}{\includegraphics[width=.5\textwidth]{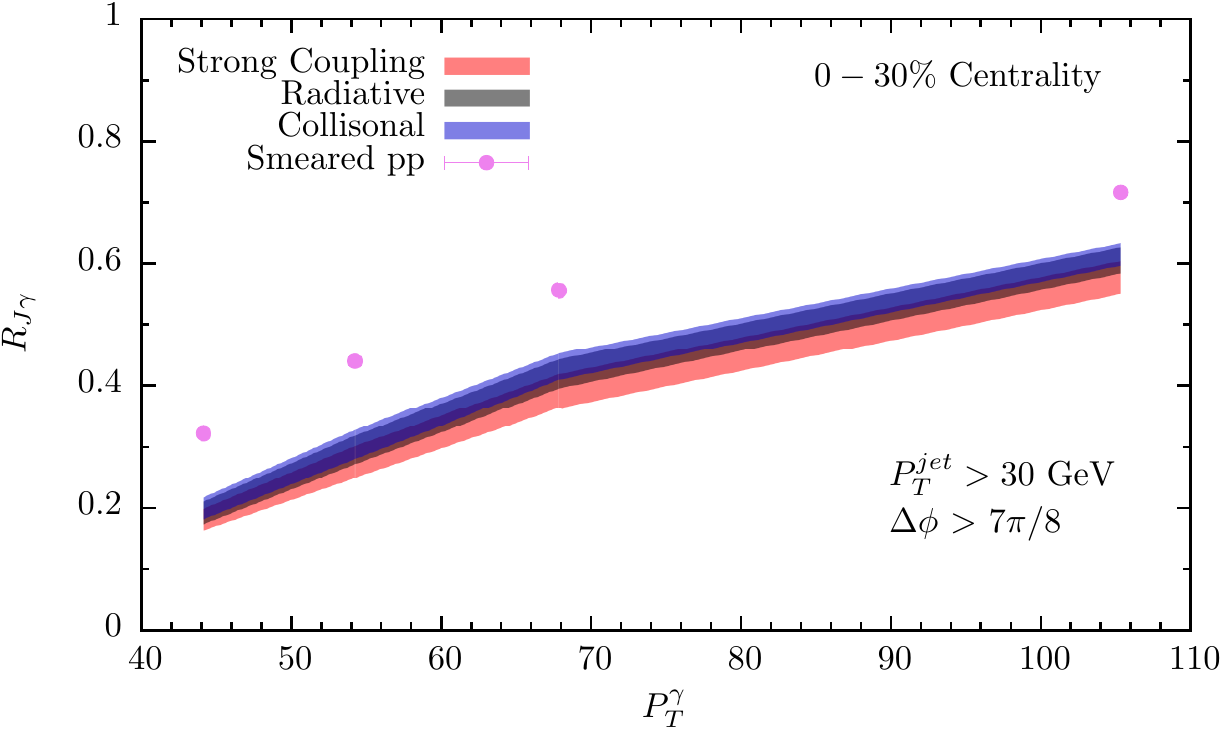}}
\put(-189,55){\tiny{$\sqrt{s}=5.02~\rm{ATeV}$}}
\end{tabular}
\caption{\label{Fig:PJI_prediction_modeldep}  
Predictions for  several photon-jet observables computed with three different models of the energy loss mechanism 
in heavy ion collisions with $\sqrt{s}=5.02$~ATeV. 
The distributions of the transverse momentum imbalance
of photon-jet pairs for two different centralities are displayed in the upper panels. 
The middle panel shows the ratio of the 
transverse momentum spectra of jets produced in association with an isolated photon in Pb-Pb collisions
to that in p-p collisions for two different centralities. The lower panel shows the fraction of isolated photons produced 
in association with a hard jet with $\pt^{\rm jet}> 30$~GeV at an azimuthal angle more than $7\pi/8$ away from
that of the isolated photon.}
\end{figure}

\begin{figure}[tbp]
\centering 
\begin{tabular}{cc}
\includegraphics[width=.5\textwidth]{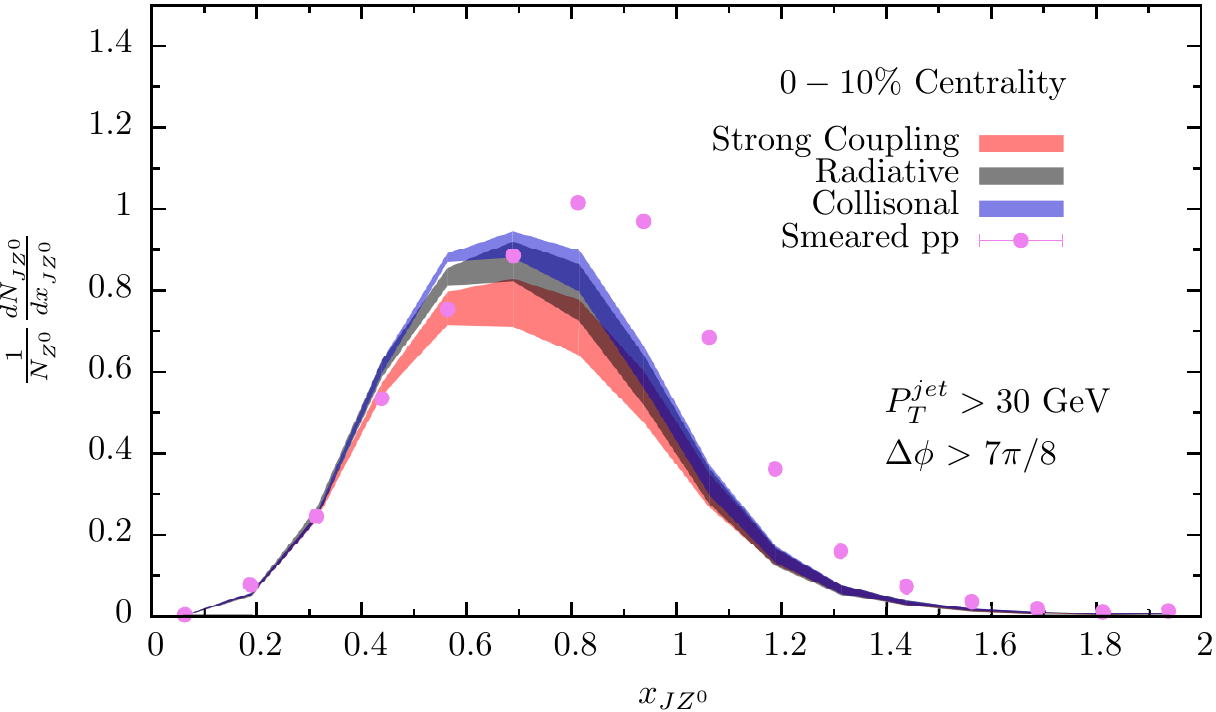}
\put(-180,110){\tiny{$\sqrt{s}=5.02~\rm{ATeV}$}}
&
\includegraphics[width=.5\textwidth]{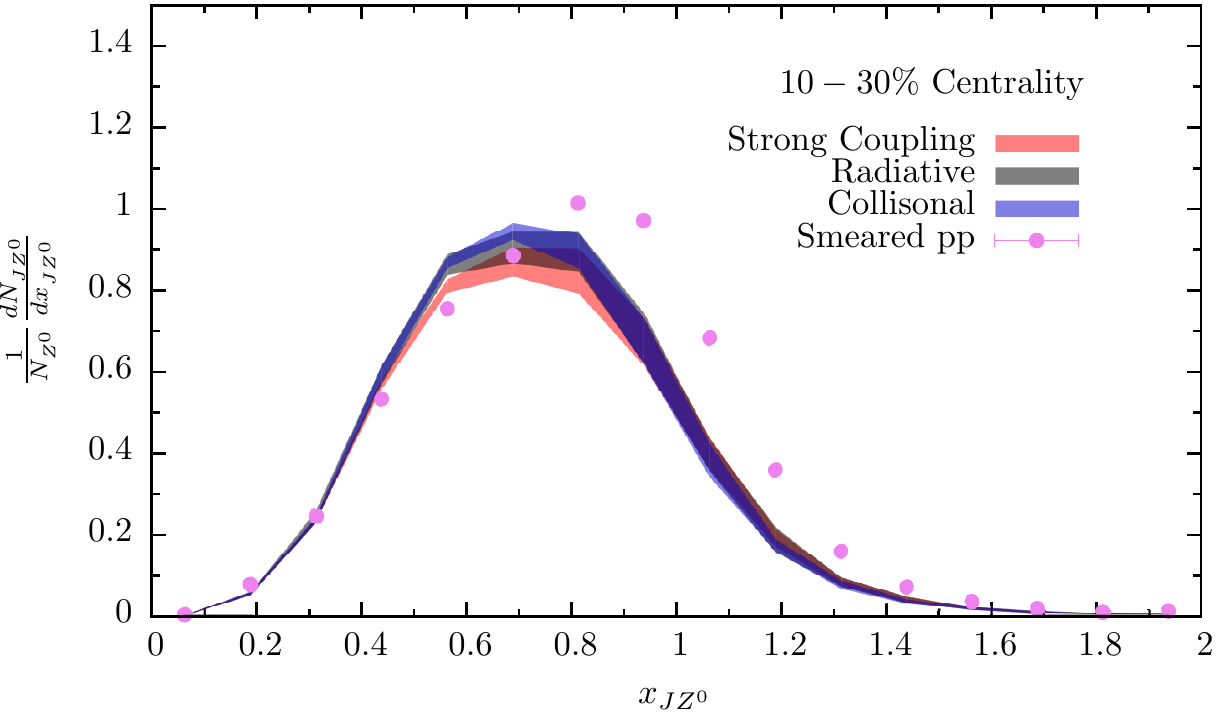}
\put(-180,110){\tiny{$\sqrt{s}=5.02~\rm{ATeV}$}}
\\
\includegraphics[width=.5\textwidth]{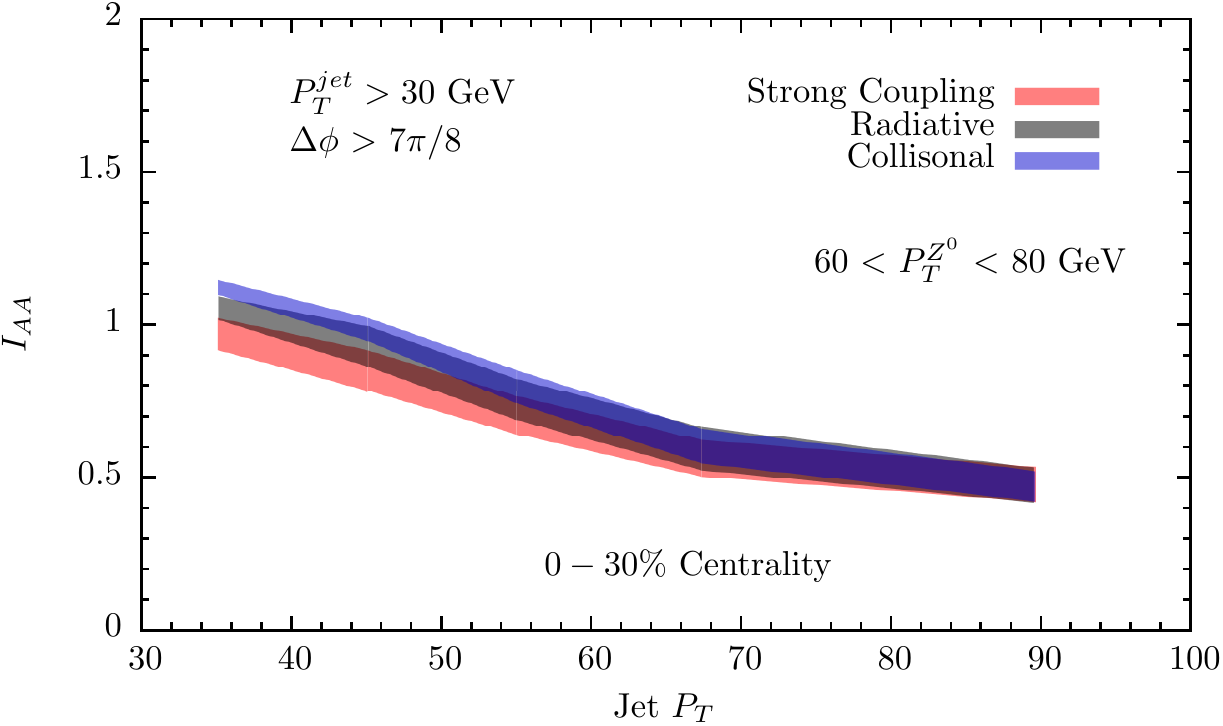}
\put(-165,90){\tiny{$\sqrt{s}=5.02~\rm{ATeV}$}}
&
\includegraphics[width=.5\textwidth]{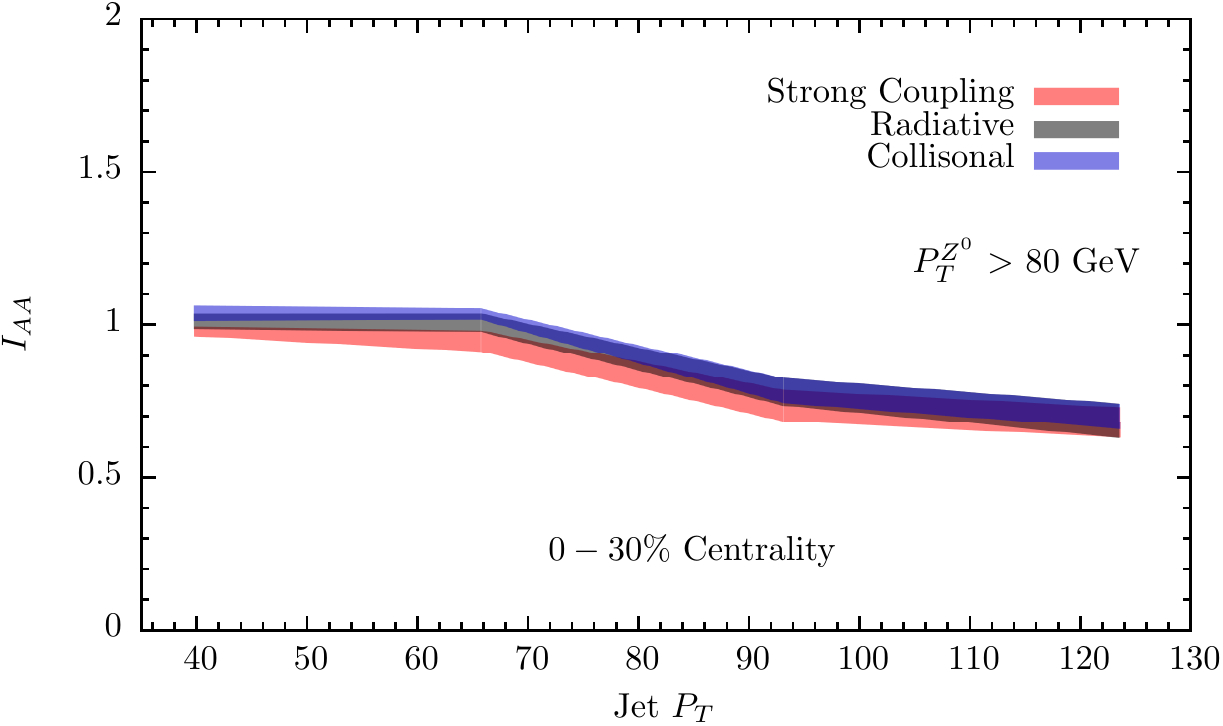}
\put(-165,90){\tiny{$\sqrt{s}=5.02~\rm{ATeV}$}}
\\
\multicolumn{2}{c}{\includegraphics[width=.5\textwidth]{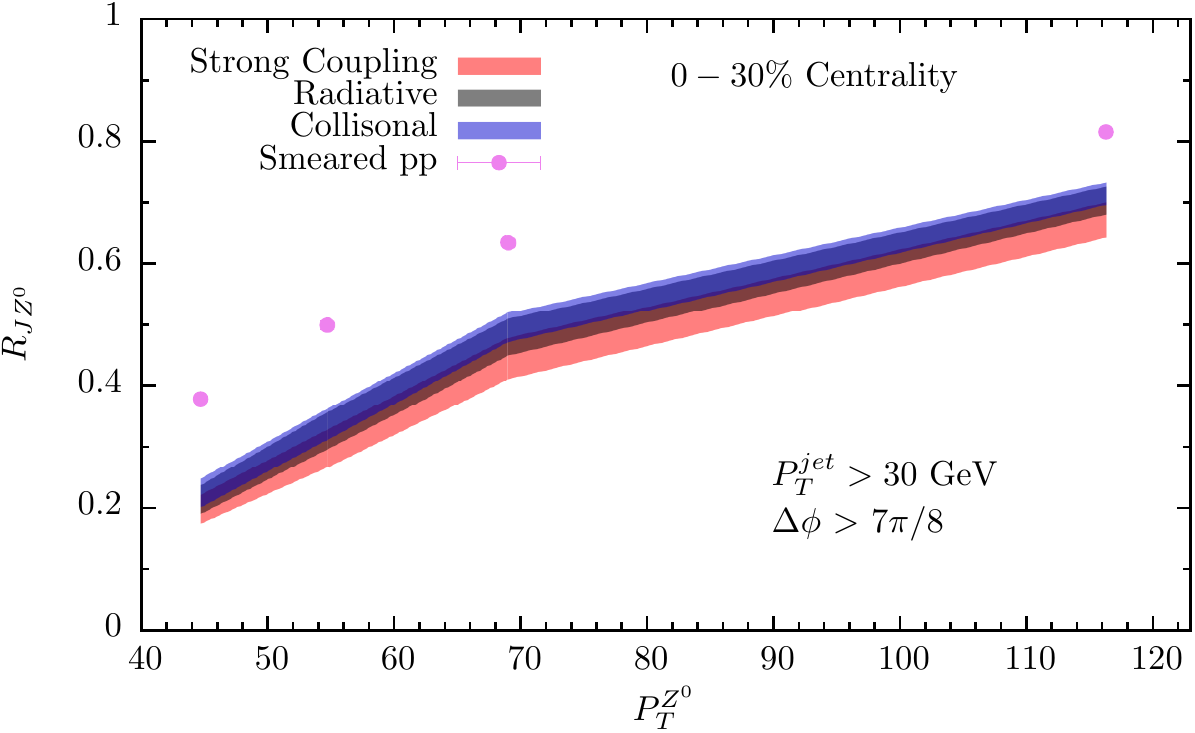}}
\put(-198,55){\tiny{$\sqrt{s}=5.02~\rm{ATeV}$}}
\end{tabular}
\caption{\label{Fig:ZJImodeldep}  
Predictions for  several Z-jet observables computed with three different models of the energy loss mechanism in
heavy ion collisions with $\sqrt{s}=5.02$~ATeV. 
The distributions of the transverse momentum imbalance
of Z-jet pairs for two different centralities are displayed in the upper panels. 
The middle panel shows the ratio of the 
transverse momentum spectra of jets produced in association with a Z-boson in Pb-Pb collisions
to that in p-p collisions for two different centralities. The lower panel shows the fraction of Z-bosons produced 
in association with a hard jet with $\pt^{\rm jet}> 30$~GeV at an azimuthal angle more than $7\pi/8$ away from
that of the isolated photon.
}
\end{figure}

To better compare our computations with future higher statistics data from LHC heavy ion Run 2, 
we also explore the model predictions of the different energy loss mechanisms 
for both photon-jet observables, displayed in Fig.~\ref{Fig:PJI_prediction_modeldep}, and Z-jet observables, displayed in Fig~\ref{Fig:ZJImodeldep}, 
in heavy ion collisions with $\sqrt{s}=5.02$~ATeV. As at the lower collision energy,  
little discriminating power is observed.
Again as at the lower collision energy,
there is some separation among the predictions of our hybrid model with its strongly coupled rate
of energy loss and the control models in the photon-jet and Z-jet momentum imbalance distributions in the most
central collisions, displayed in the upper-left panels of Figs.~\ref{Fig:PJI_prediction_modeldep} and \ref{Fig:ZJImodeldep}, respectively.

\end{document}